\documentclass[11pt]{article}
\usepackage{amsmath,amsfonts,amssymb,amsthm,bbm,graphicx,enumerate,times}
\usepackage{mathtools}
\usepackage{fullpage}
\usepackage[usenames,dvipsnames]{color}
\usepackage{hyperref}
\usepackage{comment}
\usepackage{graphicx}
\usepackage{tcolorbox}
\usepackage{float}
\usepackage{etoolbox}
\usepackage{comment}
\usepackage{accents}
\usepackage{stmaryrd}
\usepackage{cite}

\definecolor{jens}{rgb}{1,0,0}
\definecolor{alex}{rgb}{0.2,0.6,0}

\usepackage{dsfont}
\newcommand{\id}{\mathds{1}}

\newcommand{\tr}{\mathrm{tr}}

 \renewcommand{\i}{\,\ensuremath\mathrm{i}}

\newcommand{\ket}[1]{\left.\left|{#1}\right.\right\rangle}
\newcommand{\sket}[1]{\left.|{#1}\right.\rangle}

\newcommand{\bra}[1]{\left.\left\langle{#1}\right.\right|}

\newcommand{\sandwich}[3]
  {\left\langle  #1 \right| #2 \left| #3 \right\rangle}

\DeclareMathAlphabet{\mathpzcc}{OT1}{pzc}{m}{it}
\DeclareMathAlphabet{\mathpzc}{T1}{pzc}{m}{it}{\huge}

\newcommand{\m}{\operatorname{\gamma}}

\newcommand\Par[1][]{%
  \ifstrempty{#1}{%
    \mathcal P_\text{tot}
  }{
    \mathcal P_{#1}
  }
}

\def\and{\quad\text{and}\quad}

\definecolor{linkgray}{rgb}{0.3,0.3,0.45}

\newtcolorbox{infobox}[1][]{
  title=#1,
  colback=gray!15!white,
  colbacktitle=gray!15!white,
  coltitle=black,
  fonttitle=\bfseries,
  bottomrule=1pt,
  toprule=1pt,
  leftrule=2pt,
  rightrule=2pt,
  titlerule=0pt,
  arc=6pt,
  outer arc=6pt,
  colframe=black,
}

\title{Holographic tensor network models and quantum error correction: \\
A topical review}
\date{}
\author{Alexander Jahn and Jens Eisert \\
Dahlem centre for Complex Quantum Systems, \\
Freie Universit{\"a}t Berlin, 14195 Berlin, Germany}

\begin{document}
\maketitle
\begin{abstract}
    Recent progress in studies of holographic dualities, originally motivated by insights from string theory, has led to a confluence with concepts and techniques from quantum information theory. A particularly successful approach has involved capturing holographic properties by means of tensor networks which not only give rise to physically meaningful correlations of holographic boundary states, but also reproduce and refine features of quantum error correction in holography. This topical review provides an overview over recent successful realizations of such models. It does so by building on an introduction of the theoretical foundations of AdS/CFT and necessary quantum information concepts, many of which have themselves developed into independent, rapidly evolving research fields.
\end{abstract}

\tableofcontents

\newpage
\section{Introduction}

This topical review covers the intersection of two fields of physics that have recently been identified as being closely related to each other.
One of these two fields is high-energy physics, in form of the \emph{holographic principle} as realized by the \emph{anti-de Sitter/conformal field theory correspondence} (AdS/CFT), a striking duality conjecture that in itself brings together notions of gravity and of conformal field 
theories.
The other research field is that of quantum information, which has turned out to offer surprising new insights into the properties of holographic dualities.
This applies specifically to notions of \emph{quantum error correction} that have arisen in the context of quantum computing, connecting a topic of immense practical relevance to one developed from pure theory. 
An increasing number of connections between holography and quantum information are being unearthed as research within a steadily growing research community progresses, bridging the extensive theoretical foundation underlying research on AdS/CFT and quantum error correction.
In these endeavours, one particularly useful and concrete tool that captures key aspects of holography is that of tensor networks.
Originally born out of condensed matter and mathematical physics, they have taken centre stage in many questions at the heart of quantum information science due to their inherent relation to the concept of quantum entanglement, but they have also found wide-ranging applications that include numerical analysis, machine learning and probabilistic modelling.
In this review, we describe how tensor networks have developed from a practical tool for computing properties of low-dimensional quantum systems to an indispensable component of holographic models. However, the full range of such models would already extend the bounds of a \emph{topical} review due to the vast amount of research work produced in the last few years. For this reason, we focus here on those tensor networks exploring a particularly salient quantum information feature of AdS/CFT, that of quantum error correction.
While the sub-field of \emph{holographic quantum error correction} extends significantly beyond tensor network methods, these offer an especially transparent and practically computable framework in which to explore such questions.

Before exploring the state of current research in this field, Sec.\ \ref{SEC_FOUNDATIONS} begins with a  detailed introduction into the foundations of the field both on the high-energy and quantum information side.
We provide a lightning review of string theory and the specific setup of the AdS/CFT correspondence which served as the starting point of much of the vast amount of research on holography in the past twenty years.
Then we introduce the two topics of quantum information theory essential for this review: Tensor networks and quantum error correction, for both of which we sketch their original development and basic principles.  
In Sec.\ \ref{SEC_HOLOGRAPHY_IN_QI}, we then describe how quantum information and holography became connected, starting with the discovery of \emph{holographic entanglement entropy} and the role of entanglement in modern notions of quantum gravity. 
The inclusion of tensor networks and quantum error-correcting codes into the scope of holography is then laid out, laying the groundwork for the core topic of this review.
In Sec.\ \ref{SEC_FOUNDATIONS} and \ref{SEC_HOLOGRAPHY_IN_QI} we end each subsection with a brief referral to more detailed introductory work to each topic where such is available.
Bringing all the previously introduced topics together, Sec.\ \ref{SEQ_HOLO_QEC_TN} then discusses holographic tensor network models of quantum error correction. A large focus of this section is the class of holographic toy models known as \emph{HaPPY codes} which reproduce many of the features of continuum models of quantum error correction in AdS/CFT. 
Subsequently, we introduce the language of \emph{Majorana dimer states} that makes it possible to compute directly many of the boundary properties of the codes, which we discuss in detail.
We then describe ways in which the original HaPPY proposal can be extended to produce more general holographic models, and what different approaches exist for capturing holographic quantum error correction in tensor networks.
Sec.\ \ref{SEC_OUTLOOK} then closes with an outlook on the future of the field as well as acknowledgements.

\section{Foundations}
\label{SEC_FOUNDATIONS}

Before exploring the concrete duality proposals relevant for this topical review, we begin with the foundation upon which they all rest, the \emph{holographic principle}. Its origins can be traced back to the 1970s, when physicists began to consider black holes from a new vantage point: Based on the work of Stephen W.~Hawking and Jacob D.~Bekenstein, it has been realized that black holes are thermodynamic objects with a well-defined temperature and entropy \cite{Bekenstein:1973ur,Hawking:1974sw}. For a black hole with a given mass $M$, these are given by\footnote{During the remainder of this review, \emph{natural units} with $\hbar=c=1$ will be used. Newton's constant $G$ will be kept explicit, as it acts as a useful scale in AdS/CFT.}
\begin{align}
T_\text{H} &= \frac{\hbar c^3}{8\pi k_\text{B} G M} \ , &
\label{EQ_BEK_HAW}
S_\text{BH} &= \frac{4 G M^2}{\hbar c} = \frac{c^3 A_\text{hor}}{4 \hbar G} \ ,
\end{align}
commonly called the \emph{Hawking temperature} and \emph{Bekenstein-Hawking entropy} of a black hole, respectively. When expressed in terms of the area $A_\text{hor}$ of the event horizon, the latter equation, more succinctly written in natural units as $S_\text{BH} = A_\text{hor}/4 G$, contains a surprising insight: Rather than growing with its volume, as a conventional thermodynamical system, a black hole's entropy grows with its surface area! 
This has led to the suggestion that the information of a black hole's micro-states are \emph{holographically} encoded on its horizon.
The ``resolution'' of this encoding is on the order of the Planck scale, as we can see by writing the denominator of \eqref{EQ_BEK_HAW} in terms of the Planck length $l_\text{P}$ as $4 \hbar G / c^3 = 4 l_\text{P}^2$.
From this observation, Leonard Susskind and Gerardus 't Hooft declared that a consistent theory of quantum gravity would have to obey a \emph{holographic principle}: The dynamics of gravity in $3{+}1$-dimensional space-time in such a theory would have be reducible to an effective $2{+}1$-dimensional description \cite{Susskind:1994vu,tHooft:1993dmi}.
While the entropy scaling in terms of area rather than volume has appeared in gravitational settings other than black holes \cite{Bousso:1999xy}, the holographic principle has remained fundamentally vague: It has neither been specified \emph{which} theory of quantum gravity would produce such a holographic mapping between systems in different dimensions, nor \emph{how} this mapping would actually be implemented.

For this reason, the conjecture of the \emph{AdS/CFT correspondence}, a specific holographic duality between a $d{+}1$-dimensional gravitational theory and a $d$-dimensional quantum field theory by Juan M. Maldacena in 1997 \cite{Maldacena98} was met with a tremendous amount of research activity. It fundamentally changed the field of string theory, from which it was derived, and had repercussions in a wide range of research areas beyond the high-energy theory community.
The basic setup of AdS/CFT, along with the necessary string-theoretic concepts it has been built upon, will be reviewed in the next sections.

\subsection{String theory}

The development of quantum field theory in the second half of the 20th century led to a consistent and precise description of high-energy processes occurring in nature. With the \emph{Standard Model} of particle physics, quantum field theory unified electromagnetic, weak and strong interactions into one formulation.
Quantum field theory, however, is an inherently effective theory. The fields of the Standard Model require \emph{renormalization}: Their naive formulation leads to diverging physical quantities, requiring the introduction of a regulating energy or length scale (similar to the lattice scale in solid state models) and leading to physical observables such as coupling ``constants'' depending on the energy at which the system is probed. This implies that as higher and higher energies are considered, the behaviour of the theory changes; a quantum field theory valid at lower energies may need to be replaced by a more complicated one at higher energies, e.g.,\ by introducing new intermediate particles. 
While the Standard Model with its finite parameters describes the three aforementioned forces in a manner that can in principle be extended to arbitrarily high energies, this is not true for the fourth fundamental force: Gravity. 
A quantum field theory of gravity which replaces the metric of space-time by a dynamical quantum field (with excitations known as \emph{gravitons}) can be easily constructed at low energies. However, at high energies the process of renormalization requires the introduction of increasingly many parameters to cancel out divergences, making the theory useless for actual predictions.
This suggests that a naive field quantization of gravity is only an effective theory for a more fundamental theory of \emph{quantum gravity} appearing at exceedingly high energies.

One candidate for such a theory is given by string theory. Rather than the fundamental point-like particles appearing in quantum field theory, this approach proposes the quantization of one-dimensional objects called strings.
Similar to how the trajectories of point particles correspond to worldlines in space-time, a string traces out a two-dimensional \emph{worldsheet} $X^\mu(\tau,\sigma)$ parametrized by two coordinates $\tau$ and $\sigma$. Note that the $D$-vector $X^\mu$ can describe a point in a \emph{target space-time} of arbitrary dimension $D>2$.
Extending the action of a point particle in special relativity to a two-dimensional object, one arrives at the \emph{Nambu-Goto} action\footnote{Named after Yoichiro Nambu and Tetsuo Goto, though no formal publication of theirs introduces it. In practice, a reformulation in terms of the equivalent \emph{Polyakov action} \cite{Polyakov1981} is more conveniently used.}
\begin{equation}
\label{EQ_NG}
S_\text{NG} = \frac{-1}{2\pi \sqrt{l_\text{s}}} \int \text{d}\tau\text{d}\sigma\, \sqrt{\left( \frac{\partial X^\mu_{\phantom{\mu}}}{\partial \tau} \frac{\partial X_\mu^{\phantom{\mu}}}{\partial \sigma} \right)^2 - \left( \frac{\partial X^\mu_{\phantom{\mu}}}{\partial \tau} \frac{\partial X_\mu^{\phantom{\mu}}}{\partial \tau} \right) \left( \frac{\partial X^\mu_{\phantom{\mu}}}{\partial \sigma} \frac{\partial X_\mu^{\phantom{\mu}}}{\partial \sigma} \right) } \ ,
\end{equation}
where $X_\mu = \eta_{\mu \nu} X^\nu$ (with the Minkowski metric $\eta_{\mu \nu}$) and the constant $l_\text{s}$ is known as the \emph{string length} (often replaced by a coefficient $\alpha^\prime = l_\text{s}^2$).

Solutions to the action \eqref{EQ_NG} can be either \emph{open} or \emph{closed} strings; in the former case, this means that the string endpoints need to be associated with Dirichlet or Neumann boundary conditions.
It has later been realized that in the first case, the dynamics of the endpoints are related to higher-dimensional objects known as \emph{Dirichlet-branes} or \emph{D-branes} for short.

The spectrum of possible excitations on strings can be identified with particles of mass $M$. For the purely bosonic action \eqref{EQ_NG}, the vacuum state of both open and closed strings leads to unphysical \emph{tachyons} with $M^2 < 0$.
The first excited states, however, become massless if the target space dimension is chosen as $D=26$. 
These states can be identified with gauge bosons for open strings and gravitons for closed strings. In addition, closed string excitations contain scalar \emph{dilatons} and an antisymmetric tensor field.

Interactions in string theory are considerably more constrained than in regular quantum field theory, where coupling constants are usually free parameters to be determined by experiment. In contrast, interactions of strings follow directly from geometrical considerations: For example, by pinching together two points of a closed string it is split into two new ones. Similarly, open strings can turn into closed ones through the joining of endpoints. The effective string coupling $g_\text{s}$ is determined by the vacuum expectation value of the dilaton field.

While twenty-six dimensions could be reduced to our familiar four by compactification of the remaining dimensions to small scales, leading to new effective lower-dimensional fields, the problem of a tachyonic ground state is not easily circumvented. 
However, after extending the bosonic action to a \emph{supersymmetric} one containing both bosonic and fermionic degrees of freedom (see Summary 1), the tachyonic states can be removed through the \emph{GSO projection} \cite{Gliozzi1977}. This projection removes states of even fermionic parity, including the unphysical vacuum. 
In the case of supersymmetric string theory, the critical number of dimensions necessary to produce massless states is reduced to ten.
Different projections led to different ten-dimensional superstring theories. For closed superstrings, due to different possible choices of worldsheet (anti-)periodicity of the left- and right-moving modes, two consistent models known as \emph{type IIA} and \emph{type IIB} superstring theory emerge. In addition, by separately placing bosonic and supersymmetric modes in the left- and right-moving sector, another consistent solution known as \emph{heterotic} string theory is recovered.\footnote{Due to different possible symmetries of this construction, there are actually two heterotic string theories: $SO(32)$ and $E_8\times E_8$.}
Finally, another possibility is given by \emph{type I} superstring theory, which contains both open and closed \emph{unoriented} strings.
This web of consistent string theories has later been found to be connected by dualities that map from one theory to another.
Furthermore, it has been speculated that these theories might be related to a unique eleven-dimensional theory called \emph{M theory} \cite{Witten:1995ex}.

For the purposes of the AdS/CFT correspondence, we are mostly interested in type IIB superstring theory. In the low-energy limit, only the lowest string excitions are relevant, leading to an effective quantum field theory known as type IIB \emph{super-gravity}. In addition to the graviton, this theory contains a number of additional fields that preserve supersymmetry.
Interestingly, this theory allows for non-perturbative (solitonic) objects known as D-branes \cite{Dai:1989ua} that fill out some of the ten dimensions. Beyond containing endpoints of open strings, as mentioned earlier, D-branes can themselves carry masses and charges and perturb the metric around them. 
These two perspectives on D-branes are essential for the construction that led to AdS/CFT.
For more information on string theory, refer to one of the several textbooks on the subject available both at the introductory (undergraduate) \cite{ZwiebachStrings04,LustTheisenStrings89} and advanced level \cite{GreenSchwarzWittenString2012,PolchinskiStrings1998}.

\begin{figure*}
\begin{infobox}[Summary 1: Supersymmetry]
Classifying quantum field theories by their symmetries has driven much of the development of the Standard Model over the late 20th century.
While bosons and fermions in the Standard Model are intimitely related by gauge symmetries, it is possible to extend its field content to allow for a direct symmetry between bosonic and fermionic fields, called \emph{supersymmetry}.
Following Noether's theorem, supersymmetry implies the existence of \emph{supercharges} $Q$. These act as operators on fields that change their spin by $1/2$, turning bosonic fields into fermionic ones and vice-versa.
The number $\mathcal{N}$ of possible distinct supercharges is subject to physical constraints. Models of \emph{extended supersymmetry} with $\mathcal{N}>1$ preserve chiral symmetry and are thus incompatible with the Standard Model.
In $3{+}1$ dimensions, a theory with spins $s\leq 1$ can have $\mathcal{N}=4$ at most. A particular example of such a theory is $\mathcal{N}=4$ \emph{super-Yang-Mills} theory, which also possesses conformal symmetry.
Supersymmetries with $\mathcal{N}>8$ imply fields with spin $s>2$.
\end{infobox}
\end{figure*}

\subsection{The AdS/CFT proposal}
\label{SS_ADS_CFT}

The original AdS/CFT setup \cite{Maldacena98} is based on type IIB superstring theory in $D{=}10$ space-time dimensions. 
This theory supports non-perturbative D-brane solutions. We consider a stack of $N$ parallel D3-branes filling out three of the nine spatial dimensions. This setup has the following parameters: The string coupling $g_\text{s}$, the string length $l_\text{s}$, the distance $d$ between the branes and their number $N$. 

As has been mentioned in the previous section, D-branes serve as endpoints of open strings and carry fields.
We consider the low-energy limit of vanishing string length $l_\text{s} \to 0$, i.e., we ``zoom out'' to scales where all excitations (of order $1/l_\text{s}$) beyond the ground state become negligible. To keep the mass of string modes between different branes constant, $U=d/l_\text{s}^2$ is fixed (so that $d \to 0$, as well).
Considering the open strings as small perturbations (in $N g_\text{s}$) of the branes, one finds an effective $U(N)$ Yang-Mills theory with coupling constant $g_\text{YM}^2 = 2\pi g_\text{s}$ in the low-energy limit. Specifically, this theory has $\mathcal{N}=4$ supercharges (half of the original type IIB theory, broken by the D-branes) and is known as $\mathcal{N}=4$ super-Yang-Mills (SYM) 	theory.
The limit $l_\text{s} \to 0$ also removes interactions between the open strings on the branes and the closed strings in the type IIB background, so that the $U(N)$ theory decouples.
The remaining closed strings can be treated by the low-energy limit of type IIB strings, given by type IIB super-gravity.

\begin{figure}
\centering
\includegraphics[width=0.4\textwidth]{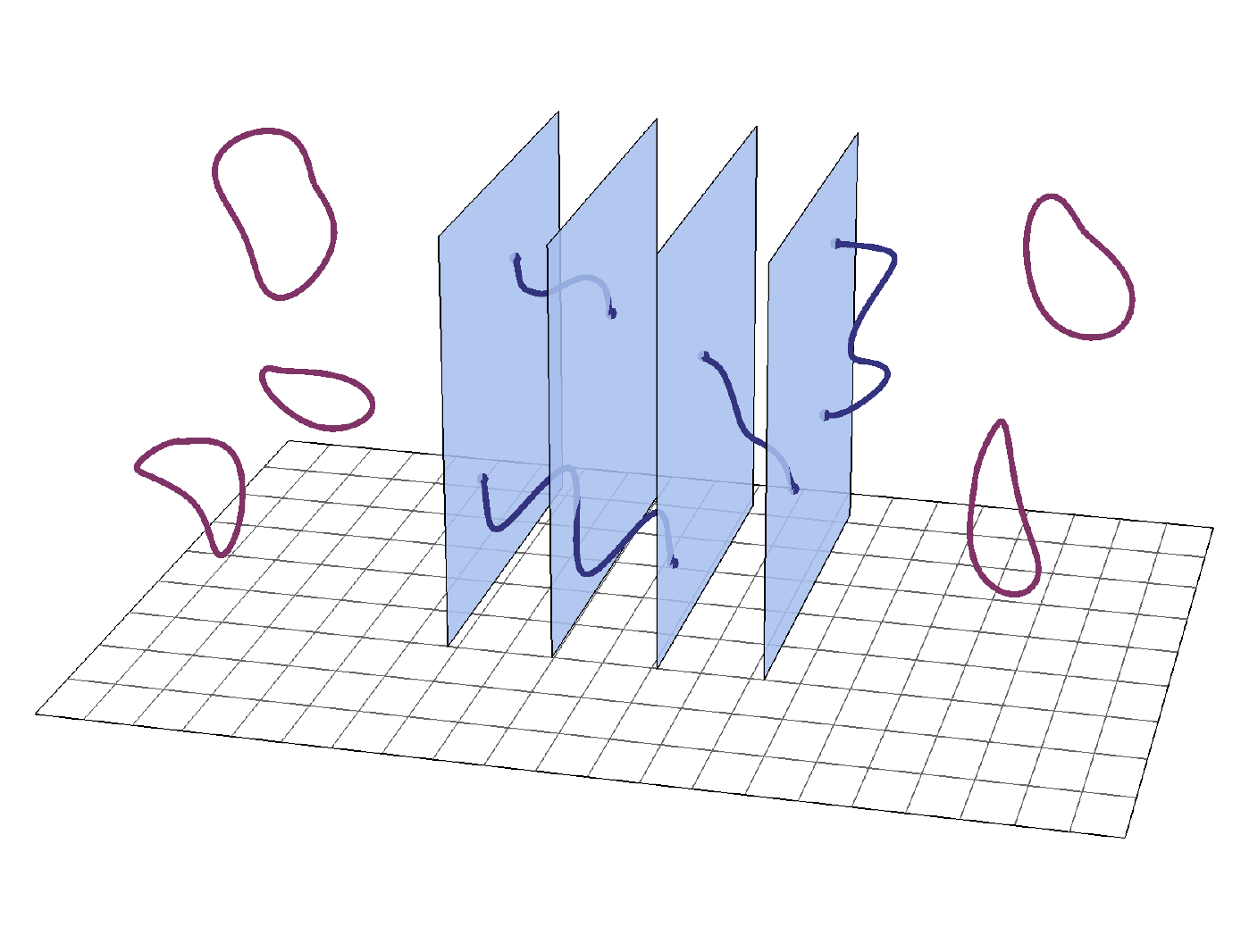}
\hspace{0.07\textwidth}
\includegraphics[width=0.4\textwidth]{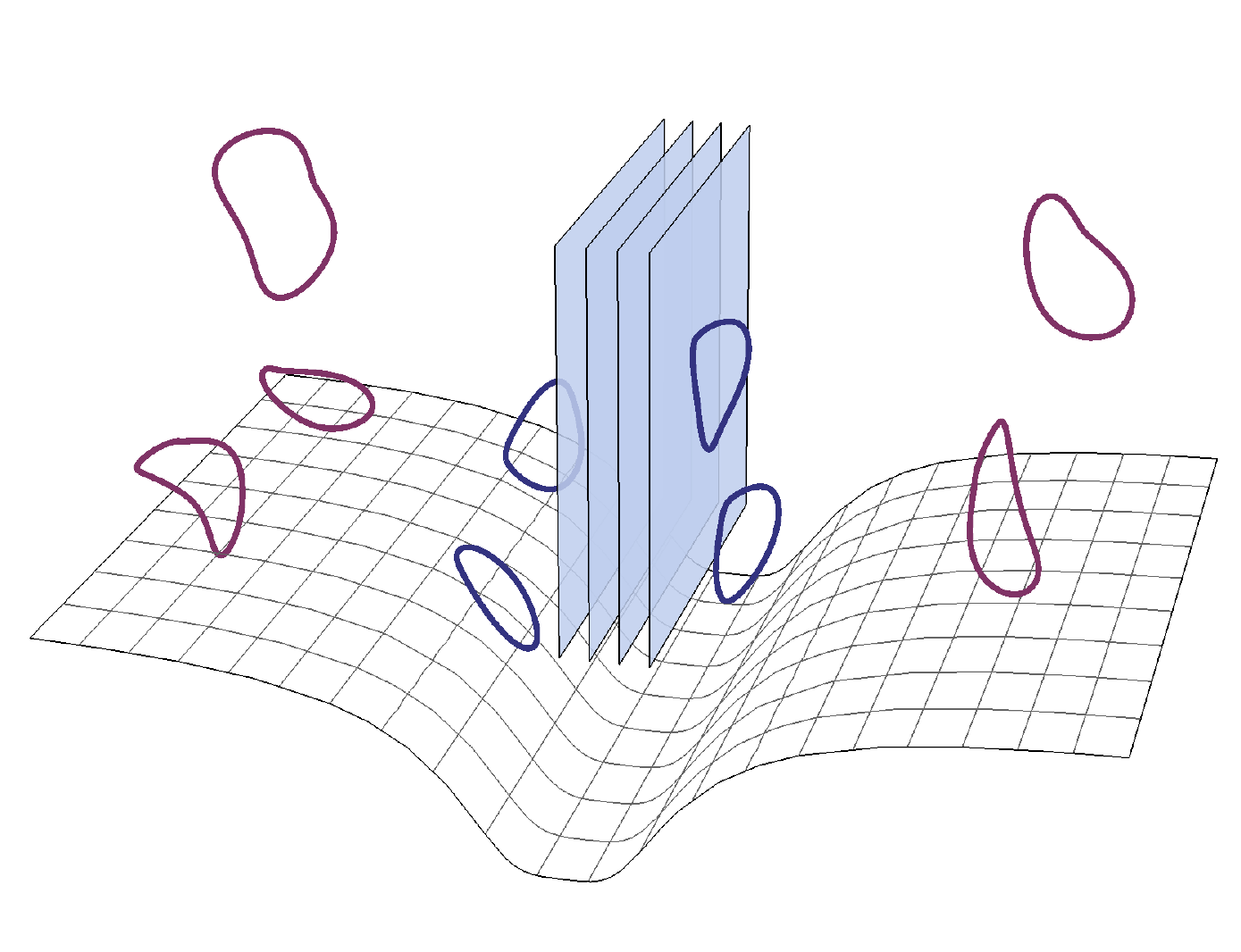}
\caption{Visualization of two perspectives on the AdS/CFT setup. \textsc{Left:} At small coupling $N g_\text{s}$, open strings between $N$ D-branes form an effective $U(N)$ gauge theory in the low-energy limit, with decoupled closed strings described by IIB super-gravity in a flat $\mathbb{R}^{9,1}$ background.
\textsc{Right:} At large $N g_\text{s}$, the D-branes deform the space-time background filled with closed strings.
At low energies, strings near to and far away from the D-branes decouple; both are approximated by IIB super-gravity, with the metric described by AdS$_5 {\times} $S$^5$ and flat $\mathbb{R}^{9,1}$, respectively.
}
\label{FIG_ADS_CFT}
\end{figure} 

Alternatively, we can look at the branes as a massive perturbation of the background of the closed strings in the type IIB string theory. 
The metric around $N$ D3-branes is given by \cite{Horowitz:1991cd}
\begin{equation}
\text{d}s^2 = \frac{\text{d}x^2}{\sqrt{1 + 4 \pi N g_\text{s} (l_\text{s}/r)^4 }} + \sqrt{1 + 4 \pi N g_\text{s} (l_\text{s}/r)^4 } \left( \text{d}r^2 + r^2 \text{d}\Omega_5^2 \right) \ ,
\end{equation}
where $x$ are the space-time coordinates along the branes and $r$ is the radial coordinate away from the D-branes.
At large $r$, this metric simply describes flat 10-dimensional space.
However, at small $r$ we can rewrite it as 
\begin{equation}
\label{EQ_ADS_S5_POINCARE}
\text{d}s^2 = \frac{r^2}{\alpha^2} \text{d}x^2 + \frac{\alpha^2}{r^2} \text{d}r^2 + \alpha^2 \text{d}\Omega_5^2 \ ,
\end{equation} 
where we defined $\alpha = (4 \pi N g_\text{s})^{1/4}\, l_\text{s}$. The metric \eqref{EQ_ADS_S5_POINCARE} describes $4{+}1$-dimensional \emph{anti-de Sitter} (AdS) space-time in the \emph{Poincar\'e coordinates} $(x,r)$ in addition to the angular coordinates $\Omega_5$ of the 5-sphere S$^5$. This combination is denoted as AdS$_5{\times}\text{S}^5$.
The AdS radius $\alpha$ is also the radius of the 5-sphere in this setup. AdS space-time has constant negative curvature of the same magnitude as the positive curvature of S$^5$.
As explained in Summary 2, the branes at $r=0$ form a horizon that is infinitely spatially separated from the remaining space-time.

\begin{figure*}
\begin{infobox}[Summary 2: Anti-de Sitter space-time]
A particularly symmetric class of $D$-dimensional space-times are those with constant \emph{scalar curvature} $R$ at all points. For vanishing $R$, we find flat Minkowski space-time $\mathbb{R}^{D-1,1}$. The cases $R>0$ and $R<0$ are known as \emph{de Sitter} (dS) and \emph{anti-de Sitter} (AdS) space-times, respectively. The latter case can be expressed in different metrics; most commonly used are \emph{global coordinates} $(\tau,\rho,\Omega_1,\dots,\Omega_{D-2})$ with
\begin{equation}
\label{EQ_ADS_GLOBAL}
\text{d}s^2 = -\left(1 + \frac{\rho^2}{\alpha^2} \right)\text{d}\tau^2 + \frac{\alpha^2}{\alpha^2 + \rho^2} \text{d}\rho^2 + \rho^2 \text{d}\Omega_{D-2}^2 \ ,
\end{equation}
and \emph{Poincar\'e coordinates} $(t,r,x_1,\dots,x_{D-2})$ with 
\begin{equation}
\label{EQ_ADS_PDISK}
\mathrm{d}s^2 = \frac{r^2}{\alpha^2} \left( -\text{d}t^2 +  \text{d}\vec{x}^2 \right) + \frac{\alpha^2}{r^2} \text{d}r^2 \ .
\end{equation}
The \emph{AdS radius} $\alpha$ determines the scalar curvature $R=-D(D-1)/\alpha^2$.
Characteristic of AdS space-time is a horizon at spatial infinity ($\rho\to\infty$ or $r\to 0$) that no timelike geodesics can reach. The metric at this horizon is given by flat $D{-}1$-dimensional Minkowski space-time $\mathbb{R}^{D-2,1}$.
AdS space-time has $SO(D-1,2)$ symmetry, the same symmetry as a $D{-}1$-dimensional conformal field theory (CFT), an important cornerstone of the AdS/CFT correspondence.
\end{infobox}
\end{figure*}

We again consider the low-energy limit of this setup: As $l_\text{s} \to 0$, the closed strings at both large and small $r$ are described by type IIB super-gravity while decoupling from one another. At large $r$, the space-time background is flat and we find the same super-gravity theory as in the previous setup where we considered the open string dynamics between branes.
However, at small $r$ we find a theory of super-gravity on an AdS background, rather than the $U(N)$ theory resulting from the previous analysis using open strings. 
Assuming that both descriptions of the D-brane setup are equally valid across the whole range of couplings $N g_\text{s}$, it appears that both theories should be equivalent, as well. This leads to the following duality:
\begin{equation*}
\mathcal{N}=4\; SU(N) \text{ SYM theory on } \mathbb{R}^{3,1} \quad\equiv\quad
\text{Type IIB superstring theory on AdS$_5$$\times$S$^5$} \ .
\end{equation*}
Note that we changed the gauge group from $U(N)$ to $SU(N)$, as a set of $U(1)$ modes on the boundary is non-dynamical \cite{Witten:1998wy}.
The coupling constants are related via $g_\text{YM}^2 = 2\pi g_\text{s}$ and $2 N g_\text{YM}^2 = (\alpha/l_\text{s})^4$.
This leads to a remarkable property.
The effective coupling constant in the SYM theory is given by
\begin{equation}
\lambda = N g_\text{YM}^2 = \frac{\alpha^4}{2 l_\text{s}^4} \ .
\end{equation}
As we are working in the $l_\text{s} \ll \alpha$ limit, $\lambda$ is large and the SYM theory is thus \emph{strongly} coupled. 
However, if we are also taking the $N \to \infty$ limit, the string coupling $g_\text{s} = \lambda / (2\pi N)$ is \emph{weak} and the type IIB superstring theory can be studied perturbatively.\footnote{The large $N$ limit at fixed $\lambda$ is usually called the \emph{'t Hooft limit}, after an earlier observation that Yang-Mills theory in this limit has a perturbation series similar to that of a quantized string \cite{tHooft1974}.}
This duality between supersymmetric gauge theory and super-gravity (or gauge/gravity duality, for short) is thus often called a strong/weak  duality.
The name \emph{AdS/CFT correspondence} comes from a particular property of the $SU(N)$ SYM theory: It possesses \emph{conformal} invariance (see Summary 3) and is thus belongs to the class of \emph{conformal field theories} (CFTs).

\begin{figure*}
\begin{infobox}[Summary 3: Conformal field theory]
As stated by the Coleman-Mandula theorem \cite{PhysRev.159.1251}, it is generally not possible to combine internal symmetries of quantum fields with space-time symmetries in any nontrivial way.
One exception to this theorem is supersymmetry, based on a graded Lie algebra beyond the scope of Coleman-Mandula, which directly relates bosonic and fermionic fields (see Summary 1). \cite{Haag:1974qh} 
As the theorem is based on the properties of the $S$-matrix describing scattering between asymptotic particles, it also breaks down in theories without a length scale. This includes scale-invariant and \emph{conformally} invariant models.
Conformal transformations $g_{\mu \nu}(x) \to \Omega^2(x) g_{\mu \nu}(x)$ (with positive $\Omega^2(x)$) preserve local angles but not lengths. 
This greatly restricts the physical properties of a conformal field theory (CFT). Correlations generally depend polynomially on distances, with the form of two- and three-point functions fixed by symmetry.
Extending $D$-dimensional Poincar\'e symmetries (translations, rotations and Lorentz transformations) with conformal symmetry leads to the \emph{conformal group} $SO(D,2)$.
While higher-dimensional CFTs are hard to study analytically, in $1{+}1$ dimensions many examples (such as the critical \emph{Ising model} \cite{Ising1925}) are exactly solvable.
Further requiring supersymmetry leads to \emph{superconformal} theories, a class that also includes $\mathcal{N}=4$ super-Yang-Mills (see Summary 1).
\end{infobox}
\end{figure*}

The range of applicability of the AdS/CFT correspondence appears to be much larger than the specific example just given: Rather than a relationship between super-gravity in $4{+}1$-dimensional AdS$_5$ and a $3{+}1$-dimensional CFT$_4$ ($\mathcal{N}=4$ SYM), similar AdS$_{D+1}$/CFT$_D$ dualities can be constructed for different $D$.\footnote{Other examples for $D=1,2,3$ and $6$ were already proposed in Maldacena's original work \cite{Maldacena98} using different D-brane setups and compactifications.}
This is consistent with the symmetries of both theories: The space-time symmetries of AdS$_{D+1}$ are given by $SO(D,2)$, as it can be embedded onto a hyperbola in flat $\mathbb{R}^{D,2}$ space-time. This exactly matches the space-time and conformal symmetries of a CFT$_D$, which taken together also form $SO(D,2)$. The observation that the algebra of AdS$_{D+1}$ symmetry generators turns into the $D$-dimensional conformal algebra at the asymptotic boundary of AdS$_{D+1}$ space-time was already observed in the $D=2$ case long before the AdS/CFT correspondence \cite{Brown:1986nw}.
Supersymmetry on the field theory side corresponds to the additional compact dimensions of matching symmetry on the gravity side: For the AdS$_5$/CFT$_4$ case, the $\mathcal{N}=4$ supersymmetry corresponds to an $SU(4)$ symmetry that again matches the $SO(6) \sim SO(4)$ symmetry of $S^5$.
This relationship between supersymmetry and additional compact dimensions suggests the existence of some non-supersymmetric duality between gravity in AdS$_{D+1}$ and CFT$_D$. 
Similarly, one may speculate about the validity of AdS/CFT at small 't Hooft coupling $\lambda$, where a weakly coupled CFT in the above construction appears to be related to a strongly interacting --- i.e., non-perturbative --- theory of quantum gravity.
While examples in both directions have been constructed, the general validity of AdS/CFT remains unknown. 
This is intimately tied to the problem that in a strong/weak duality, one of the two sides of the duality will always be hard to treat analytically. 
For this reason, a fundamental motivation for the focus of this review
is the construction of simpler models that can be more directly studied with analytical and numerical tools.

\begin{figure}
\centering
\includegraphics[width=0.4\textwidth]{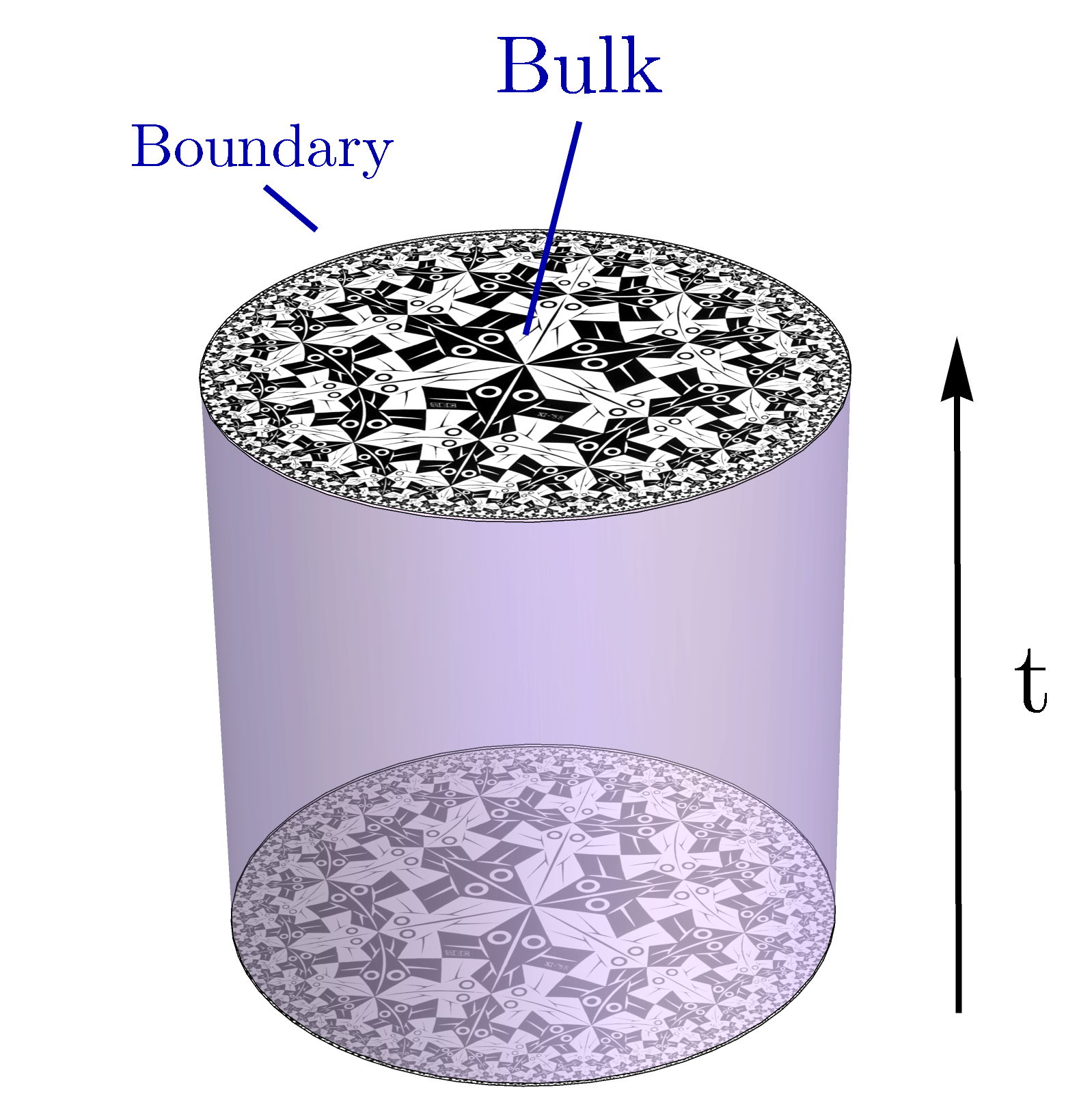}
\caption{Bulk/boundary relation in the AdS/CFT correspondence: The bulk AdS$_{D+1}$ space-time (shaded cylinder) has a flat asymptotic boundary at spatial infinity. Each time-slice $t\,{=}\,\text{const}$ is a hyperbolic space with negative curvature.
}
\label{FIG_BULK_BOUNDARY}
\end{figure} 

As a duality between theories, AdS/CFT furnishes a concrete \emph{dictionary} between degrees of freedom on both sides. First, note that the flat background space-time of the gauge theory side is associated with the location of the D-brane stack, which lies at the asymptotic boundary $r=0$ of the AdS space-time \eqref{EQ_ADS_S5_POINCARE}. 
As explained in Summary 2, this is a natural identification, as AdS$_{D+1}$ space-time indeed has a flat $\mathbb{R}^{D-1,1}$ horizon at spatial infinity.
It is customary to refer to the AdS$_{D+1}$ space-time as the \emph{bulk} and to the asymptotic $\mathbb{R}^{D-1,1}$ as the \emph{boundary}. The relation between the two is visualized in Fig.\ \ref{FIG_BULK_BOUNDARY}.
The bulk/boundary mapping effected by AdS/CFT generally relates fields $\phi$ in the AdS background to operators $\mathcal{O}$ in the boundary conformal field theory.
The dynamics on both sides are equivalent; in the language of partition functions,
\begin{equation}
Z_\text{bulk} [\phi] = Z_\text{boundary} [\mathcal{O}] \ .
\end{equation}
This relationship has first been proposed by Edward Witten \cite{Witten:1998qj}. Concretely, consider a CFT operator $\mathcal{O}$ with scaling dimension $\Delta$, i.e., with two-point correlations
\begin{equation}
\langle \mathcal{O}(x) \mathcal{O}(y) \rangle_\text{CFT} \propto \frac{1}{|x-y|^{2\Delta}} \ ,
\end{equation}
between two boundary points $x$ and $y$. Assume that the AdS/CFT dictionary relates $\mathcal{O}$ to a dual field $\phi$ with boundary values $\phi^0$. The bulk configuration of $\phi$ is determined as a boundary value problem from a given $\phi^0$, so that the bulk action can be expressed purely in terms of $\phi^0$.\footnote{As the fields $\phi$ are technically divergent at the boundary, one generally defines $\phi^0 = \lim_{r \to 0} (r^{\Delta} \phi)$, where $r$ is the radial AdS coordinate from \eqref{EQ_ADS_S5_POINCARE} and $\Delta$ the scaling dimension of its dual field. The dependence of the bulk fields $\phi$ on the boundary fields $\phi^0$ can be written in a diagrammatic expansion known as \emph{Witten diagrams}.} 
The bulk action then follows from a simple coupling between $\mathcal{O}$ and $\phi^0$ as
\begin{equation}
\label{EQ_ADS_CFT_Z}
Z_\text{bulk} [\phi] \equiv Z_\text{bulk} [\phi^0] = \left\langle \exp \int\text{d}^D x\, \phi^0\, \mathcal{O} \right\rangle_\text{CFT} \ .
\end{equation}
The boundary operator $\mathcal{O}_k$ thus acts as a source term for the boundary bulk field $\phi_k^0$. Conversely, for expectation values on the boundary, $\phi_k^0$ acts as a source term for the operator $\mathcal{O}_k$.
Remarkably, \eqref{EQ_ADS_CFT_Z} leads to a direct relationship between the mass $m$ of a massive bulk field and the scaling dimension $\Delta$ of its dual operator, given by \cite{Witten:1998qj}
\begin{equation}
\Delta = \frac{1}{2} \left( D + \sqrt{D^2 + 4 m^2} \right) \ .
\end{equation}
AdS/CFT thus implies a concrete relationship between asymptotic bulk fields and boundary operators in a conformal field theory.\footnote{Note that the \emph{operator/state correspondence} allows each CFT state to be characterized by a single, local operator, as scale invariance allows us to effectively project the path integral evolution of any state onto a point. Specifying CFT states and operators is thus equivalent.}

While the AdS/CFT correspondence is still a conjecture, many specific examples of the AdS/CFT dictionary with applications from high-energy to condensed matter physics have been found, with its impact on quantum information theory being a particular focus of this review. With now more than fifteen thousand citations, Juan Maldacena's original work has led to a vast amount of research whose end is nowhere in sight.
Beyond technical work, a number of introductory texts to AdS/CFT have been written, from formal textbooks \cite{Natsuume:2014sfa,NastaseIntro,AmmonErdmengerIntro} to notes that are freely available online \cite{Aharony:1999ti,Nastase:2007kj,McGreevy:2009xe,Ramallo:2013bua,Harlow:2018fse}, occasionally by the same authors. Due to the breadth of current AdS/CFT research, each of the citation listed has its own target audience, with numerous more specialized introductions available.

\subsection{Tensor networks}
\label{S_TN}

Hilbert spaces of physical systems are generally huge in their dimension. While most problems in classical mechanics can be reduced to a small parameter space that is approachable with efficient analytical and numerical techniques, the state spaces of quantum mechanics rarely offer such a relief. 
Beyond perturbative methods that can describe problems close to one of the few analytically solvable, usually \emph{non-interacting} ones, only approximate numerical techniques are available.
To see how the size of Hilbert spaces becomes a fundamental problem in this approach, consider a simple system of $N$ quantum mechanical degrees of freedom each corresponding to an $M$-level system (e.g.,\ spins for $M=2$). To describe a single pure quantum state $\psi$ in this system, we use a basis representation
\begin{equation}
\label{KET_PSI}
\ket\psi = \sum_{k_1,k_2,\dots,k_N = 1}^M T_{k_1,k_2,\dots,k_N} \ket{k_1,k_2,\dots,k_N}  \ ,
\end{equation}
where each basis state can be expressed as a direct product of local state vectors
\begin{equation}
\ket{k_1,k_2,\dots,k_N} \equiv \ket{k_1} \otimes \ket{k_2} \otimes \dots \otimes \ket{k_N} \ .
\end{equation}
The state \eqref{EQ_BEK_HAW} is thus expressed by the $M^N$ amplitudes $T_{k_1,\dots,k_N} \in \mathbb{C}$. We can view $T$ as a complex-valued rank $N$ tensor. The dimension of each index, often called the \emph{bond dimension} $\chi$, is given by $\chi=M$.
A fundamental problem of any numerical method to tackle a quantum-mechanical problem --- e.g., finding the ground state of a Hamiltonian --- is that describing a quantum state and optimizing over its components takes an exponential amount of memory. A spin chain of only fifty sites already requires $2^{50} \approx 1.126 \cdot 10^{15}$ complex numbers to store, which in the \texttt{C++} type \texttt{complex<double>} corresponds to around 18 petabytes of data, slightly less than the 30 petabytes that the entire LHC experiment produces every year.\footnote{See \url{https://home.cern/resources/faqs/facts-and-figures-about-lhc}.}
When performing classical algorithms on such a gigantic state vector, even operations scaling linearly in the number of components require extreme computational resources.

A naturally occurring question is thus: Do we really \emph{need} to be able to describe the full Hilbert space in most practical problems? If we already know certain physical properties that our quantum states have to fulfill, can we simply ignore the part of the Hilbert space that contains states irrelevant to our problem?
One of the properties for which the answer appears to be yes is \emph{area-law entanglement} \cite{AreaReview}.
Pure state entanglement in many-body systems is commonly quantified by the \emph{entanglement entropy} $S_A$, defined for a subdivision of the entire physical space into a subsystem $A$ and its complement $A^\text{C}$. Specifically, for a total system specified by a density matrix $\rho$,
\begin{align}
S_A &= - \tr (\rho_A \log \rho_A) \ , &
\rho_A &= \tr_{A^\text{C}} (\rho) \ ,
\end{align}
where $\rho_A$ is the \emph{reduced density matrix} on the subsystem $A$ of which the von Neumann entropy is taken. In a precise sense, the entanglement entropy uniquely characterizes the entanglement content of a pure quantum state.
We say that a system's entanglement entropy follows an area law if $S_A$ scales with the size of the boundary $\partial A$ of $A$; in particular, for an area law in $d=1$ dimensions $S_A$ is constant.
Entanglement area laws are characteristic of ground states of Hamiltonians that are \emph{local}, i.e.\ contain only coupling terms over a  distance that does not grow with the total system size, and \emph{gapped}, meaning that even in the continuum limit a separation between the ground state and the first excitation exists. 
These area law conditions have been proven rigorously in $1{+}1$ dimensions \cite{Hastings2007} and for non-interacting gapped bosonic (and fermionic) systems in arbitrary dimensions
\cite{Cramer:2005mx,PhysRevLett.71.666}.
%\cite{Bombelli:1986rw,PhysRevLett.71.666,Latorre:2003kg,Cramer:2005mx}. 
As an energy gap $\Delta E$ induces an energy scale, these conditions exclude scale-invariant (and by extension, conformally invariant) quantum systems.

Having identified a class of interesting states with characteristic properties, how do we restrict the Hilbert space to states than conform to these? For the case of area-law states, we clearly need a state description that encodes the (potential) entanglement entropy between subsystems, so we can discard states with large long-range entanglement.
This description is afforded by \emph{tensor networks}.
A tensor network is an ansatz for the state amplitudes $T_{k_1,k_2,\dots,k_N}$ in \eqref{KET_PSI} in terms of a \emph{contraction} of complex-valued tensors.
For example, for $N=3$ we may write 
\begin{equation}
\label{EQ_T_CONTR_EX}
T_{k_1,k_2,k_3} = \sum_{j_1,j_2,j_3=1}^{\chi_j} U_{k_1,j_3,j_1} V_{k_2,j_1,j_2} W_{k_3,j_2,j_3} \ ,
\end{equation}
where $U,V,W$ are rank $3$ tensors and $\chi_j$ is the bond dimension (number of possible values) of the $j$ indices.
Note that this form of contraction is simply a sum over pairs of indices, and a rank $r$ tensor an $r$-dimensional collection of complex numbers; there are no notions of co- and contravariance here that the words \emph{tensor} and \emph{contraction} imply in the context of relativistic theories.
As the name suggests, tensor networks can be represented as a graph, with nodes and edges representing tensors and their indices, respectively. 
For example, we can represent the rank $3$ tensor $U$ as 
\begin{equation}
U_{k,j_1,j_2} = 
\begin{gathered}
\includegraphics[height=0.09\textheight]{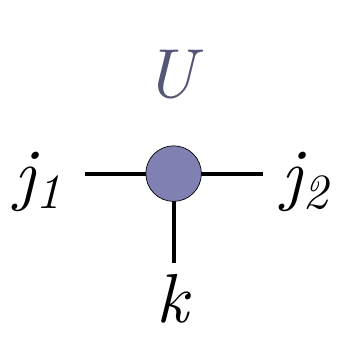}
\end{gathered} \ .
\end{equation}
The full contraction \eqref{EQ_T_CONTR_EX} is represented graphically as
\begin{equation}
\label{EQ_T_CONTR_EX_G}
T_{k_1,k_2,k_3} = 
\begin{gathered}
\includegraphics[height=0.1\textheight]{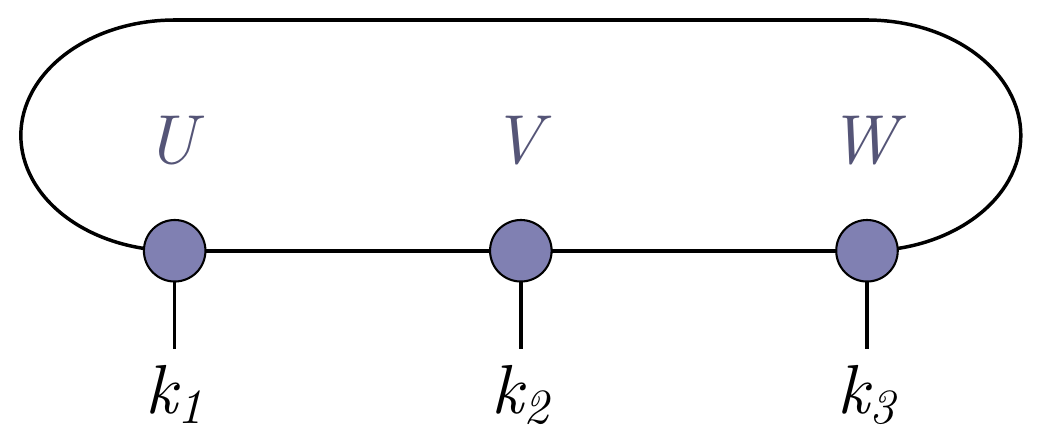} 
\end{gathered} \ ,
\end{equation}
where the contracted indices $j_1,j_2,j_3$ correspond to connected edges between nodes (tensors) without being labeled explicitly. 
The uncontracted edges are often referred to as \emph{open legs} or \emph{free indices}. Equivalently, a tensor of rank $r$ is often called an $r$-leg tensor.
The tensor labels on each node are often suppressed in larger networks for clarity.

The ansatz \eqref{EQ_T_CONTR_EX} may not appear as a particularly a smart one: Assuming a bond dimensions $\chi_k$ for the $k$ indices, we have expressed $\chi_k^3$ coefficients on the left-hand side through $3\chi_k \chi_j^2$ on the right-hand side. Unless $\chi_k > \sqrt{3} \chi_j$, the tensor network does not reduce the number of coefficients required to describe a state. 
However, this changes dramatically when looking at larger tensor networks. The generalization of \eqref{EQ_T_CONTR_EX} and \eqref{EQ_T_CONTR_EX_G} to $N$ sites, known as a \emph{matrix product state} (MPS)\footnote{The terminology becomes clear when rewriting each three-leg tensor $U_{k_n,j_{n-1},j_n}$ in the chain as a matrix $(U_{k_n})_{j_{n-1},j_n}$, as the tensor contraction can then be written as $\tr [U_{k_1} U_{k_2} \dots U_{k_N}]$.}, consists of a chain of $N$ 3-leg tensors each contracted with two neighbours (excepting possible non-periodic boundary conditions).
Such a tensor network describes $\chi_k^N$ coefficients by $N \chi_k \chi_j^2$ ones, thus allowing for a representation exponentially smaller in $N$. Of course, this means that unless $\chi_j$ is chosen to depend exponentially on $N$, only a subset of the full $N$-site Hilbert space can be reached by the MPS.
Fortunately, the this ansatz is sufficient to describe ground states of one-dimensional local Hamiltonians as long as they are gapped \cite{Verstraete2006mps,PerezGarcia2006,Hastings2007},
%, or more generally, exhibit an %exponential decay of correlations %\cite{Brandao:2014ppa}. 
which implies an exponential decay of correlations \cite{HastingsKoma06}.
In fact, in one spatial dimension, this condition is sufficient to lead to an area law, which is indeed reproduced with the MPS ansatz: Entanglement entropies $S_A$ of connected subsystems $A$ are constant in the subsystem size $|A|$ \cite{Hastings2007}. What is more, the converse is also true, and MPS deliver good approximations of any quantum state that satisfies a suitable area law \cite{Schuch_MPS} in terms of the \emph{R\'enyi entropy} \cite{Renyi1960}, a quantity generalizing entanglement entropy.

\begin{figure}
\centering
\includegraphics[width=0.5\textwidth]{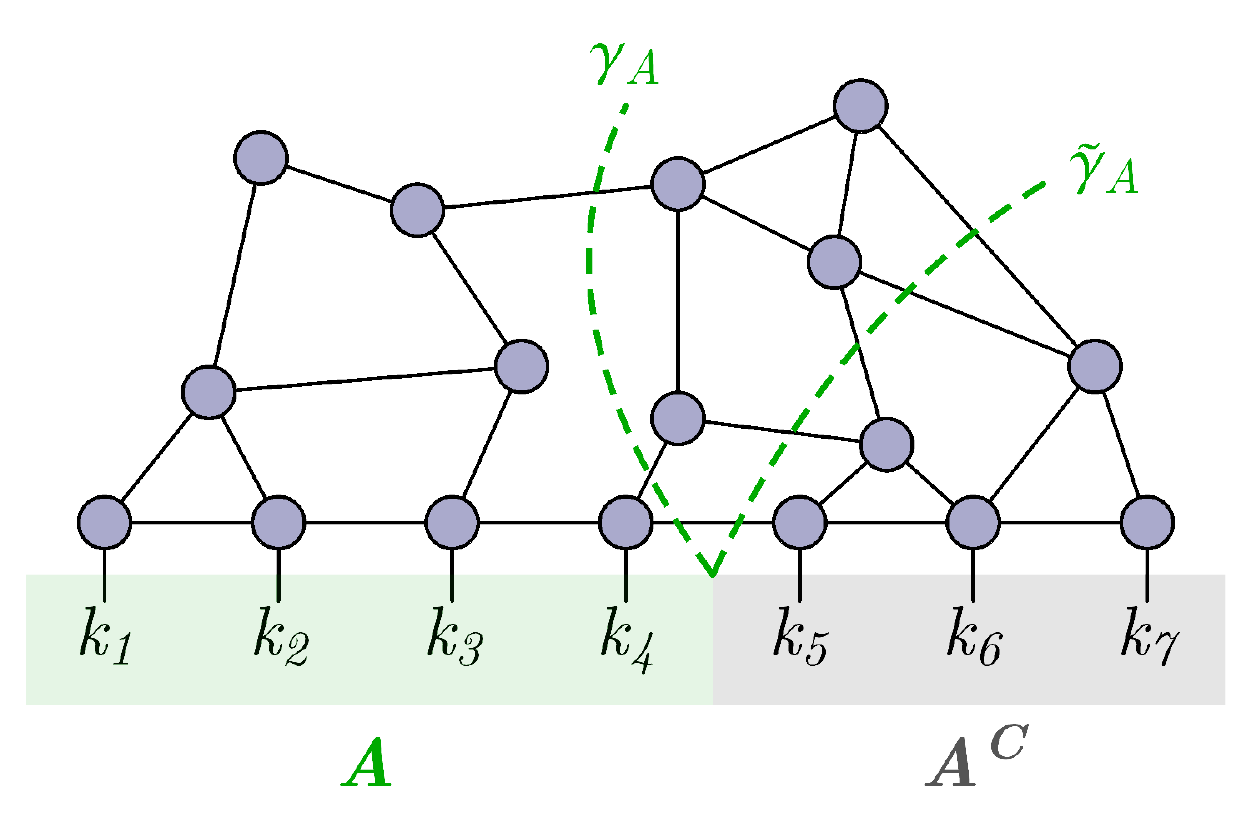}
\caption{Cuts $\gamma_A$ and $\tilde{\gamma}_A$ through a tensor network, with their endpoints on the boundary between the subsystem $A$ and its complement $A^\text{C}$. $\gamma_A$ is the minimal cut, passing through the fewest legs between tensors.
}
\label{FIG_TN_CUTS}
\end{figure}

Which tensor network geometries are needed to reproduce more complicated entanglement? To answer this question, it is necessary to understand how boundary subsystems and tensor network subregions are related. When computing a reduced density matrix $\rho_A$ from whose spectrum entanglement entropies are gleaned, we decompose our system in two parts: $A$ and its complement $A^\text{C}$. If no entanglement between both parts exist, the total state vector can be written as a tensor product of two parts in each subsystem, 
\begin{equation}
\label{EQ_PRODUCT_STATE}
\ket{\psi} = \ket{\psi}_A \otimes \ket{\psi}_{A^\text{C}} \ ,
\end{equation}
so that $\rho_A = \ket{\psi}_{A^\text{C}}\bra{\psi}_{A^\text{C}}$ is pure, leading to $S_A=0$.
For a generic entangled state, we need the more general form known as the \emph{Schmidt decomposition}
\begin{equation}
\label{EQ_SCHMIDT_DECOMP}
\ket{\phi} = \sum_k \lambda_k  \ket{\psi_k}_A \otimes \ket{\psi_k}_{A^\text{C}} \ ,
\end{equation}
where the $\ket{\psi_k}_A$ and $\ket{\psi_k}_{A^\text{C}}$ each form an orthogonal set of state vectors on $A$ and $A^\text{C}$, respectively.
The number of terms in the Schmidt decomposition increases with the amount of entanglement between both subsystems: If the Schmidt values $\lambda_k$ are normalized so that $\sum_k |\lambda_k|^2=1$, we find an entanglement entropy
\begin{equation}
\label{EQ_SA_SCHMIDT_DECOMP}
S_A = - \sum_k |\lambda_k|^2 \log |\lambda_k|^2 \ .
\end{equation}
For a single nonzero term with $\lambda_k = \delta_{k,1}$, we recover \eqref{EQ_PRODUCT_STATE} and $S_A=0$.
Given an MPS, the Schmidt decomposition for any connected subsystem $A$ is exactly the contraction of the block of tensors corresponding to $A$ with the remaining tensors, that is, the contraction over two legs (one at each endpoint of $A$). Thus \eqref{EQ_SCHMIDT_DECOMP} and \eqref{EQ_SA_SCHMIDT_DECOMP} can contain at most $\chi_j^2$ terms, where $\chi_j$ is again the bond dimension of all internal contracted legs. $S_A$ becomes maximal if all Schmidt values are identical, i.e., $|\lambda_k|^2 = 1/\chi_j^2$.
Thus the entanglement entropy is generally bounded as
\begin{equation}
S_A \leq 2 \log \chi_j \ .
\end{equation}
This equation can be generalized to tensor networks of arbitrary geometry: Any cut through the network ending on the boundary $\partial A$ between $A$ and $A^\text{C}$ can be associated with a decomposition \eqref{EQ_SCHMIDT_DECOMP}. As shown in Fig.\ \ref{FIG_TN_CUTS}, such a cut effectively associates two smaller tensor networks to each subsystem $A$ and $A^\text{C}$. The contraction between them over the chosen cut entangles both boundary regions, with the total entanglement entropy bounded by the length of the cut. This bound is tightest for the \emph{minimal cut} $\gamma_A$ over the fewest legs. Again assuming constant bond dimension $\chi_j$ on all internal legs, we thus arrive at the well-known bound
\begin{equation}
\label{EQ_TN_ENT_BOUND}
S_A \leq |\gamma_A| \log{\chi_j} \ ,
\end{equation}
where $|\gamma_A|$ is the length of $\gamma_A$ counted as the number of cut legs ($|\gamma_A|=2$ for the example in Fig.\ \ref{FIG_TN_CUTS}). 
In order to describe states with more entanglement than a simple area law, it is thus necessary to increase either the bond dimension $\chi_j$ or the lengths $|\gamma_A|$ of minimal cuts through the tensor network, i.e., consider geometries more complicated that the geometry of the physical sites themselves.

A particularly interesting class of states with entanglement not following an area law is afforded by \emph{critical} or \emph{gapless states}, which naturally follow from systems with conformal invariance: In a system without characteristic length scale $l$, there also cannot exist a characteristic energy $\epsilon$, such as $\epsilon=\hbar c / l$. 
In $1{+}1$-dimensional conformal field theory, for example, the entanglement entropy of a subsystem $A$ of length $\ell=|A|$ generally follows a logarithmic scaling \cite{Cardy:1988tk,Holzhey:1994we,CalabreseReview}
\begin{equation}
\label{EQ_SA_CRIT}
S_A = \frac{c}{3} \log \frac{\ell}{a} \ ,
\end{equation}
where $c$ is the central charge and $a$ is a lattice regulator. Note that in the continuum limit $a \to 0$, $S_A$ is infinite; entanglement on all scales provides contributions that diverge as infinitely small scales are included.

\begin{figure}
\centering
\begin{minipage}[c]{0.52\textwidth}
\includegraphics[width=0.9\textwidth]{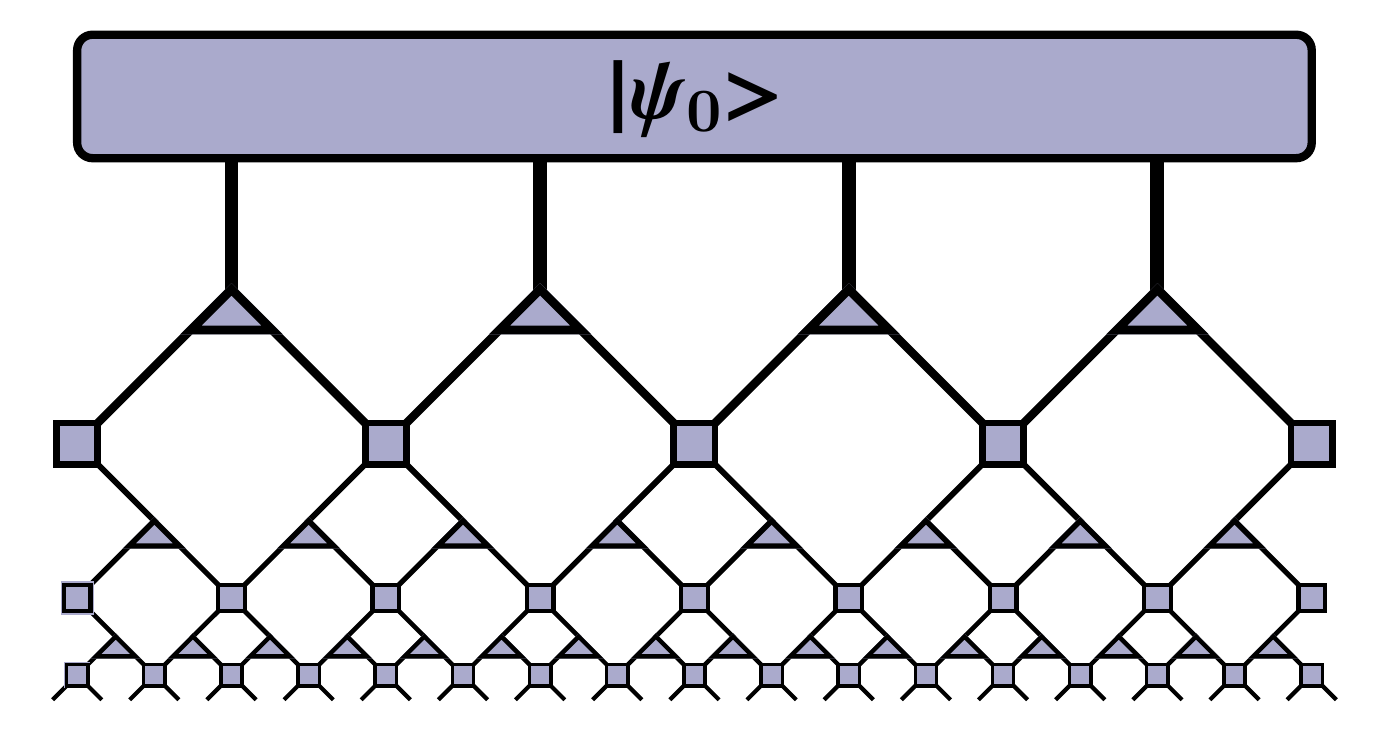}
\end{minipage}
\begin{minipage}[c]{0.45\textwidth}
\hspace{0.1\textwidth}
\begin{align*}
\begin{gathered}
\includegraphics[height=0.1\textheight]{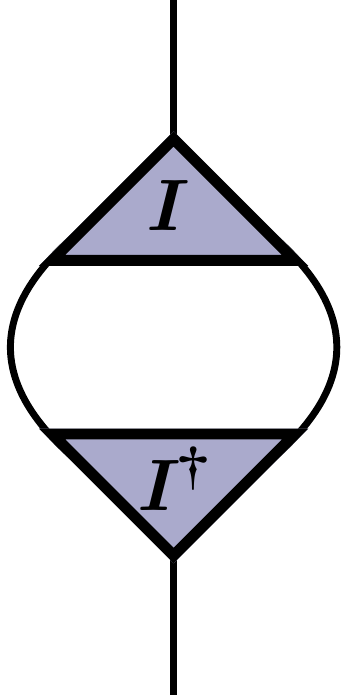}
\end{gathered}
\;
&=
\begin{gathered}
\includegraphics[height=0.1\textheight]{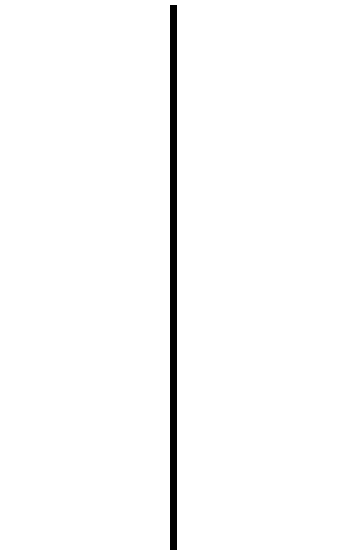}
\end{gathered} \ , &
\begin{gathered}
\includegraphics[height=0.1\textheight]{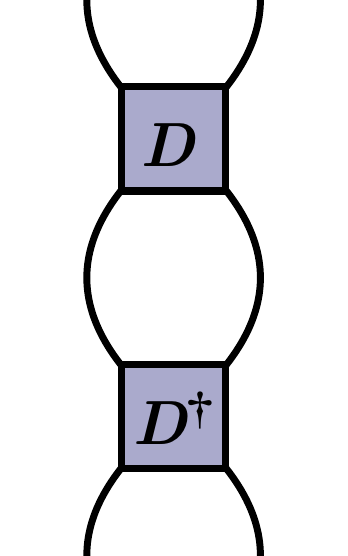}
\end{gathered}
&=
\begin{gathered}
\includegraphics[height=0.1\textheight]{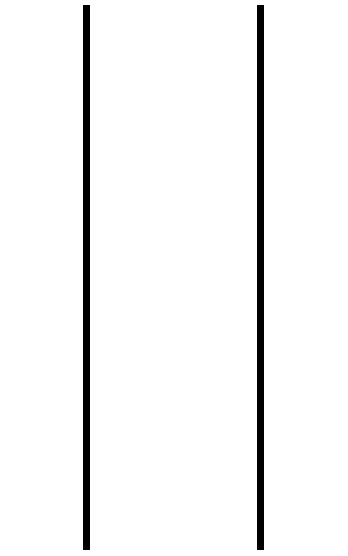}
\end{gathered}
\end{align*}
\end{minipage}
\caption{A MERA tensor network composed of isometries (triangles) and disentanglers (squares). \textsc{Left:} The full network acting on an initial coarse-grained state vector $\ket{\psi_0}$. \textsc{Right:} Identities of the isometries and disentanglers for contractions of each tensor with its Hermitian conjugate over two legs.
}
\label{FIG_TN_MERA}
\end{figure}

To reproduce states with an entanglement following \eqref{EQ_SA_CRIT}, an MPS would require a bond dimension that scales linearly in $\ell/a$ and would thus fail at reproducing entanglement of sufficiently large subsystems.
A more effective approach is to replace the MPS by a tensor network whose geometry automatically reproduces a relation $|\gamma_A| \propto \log \ell/a$. Such a tensor network indeed exists and is known as the \emph{multi-scale entanglement renormalization ansatz} (MERA), first introduced by Guifr\'e Vidal \cite{Vidal:2008zz}. As visualized in Fig.\ \ref{FIG_TN_MERA}, it consists of two types of tensors, unitary \emph{disentanglers} and directional \emph{isometries}, arranged in a tree-like structure. Due to special properties of each of these two tensor types, evaluating local observables is significantly more efficient than an evaluation of the entire tensor network would be in principle.
The tree-like structure ensures that a shortest cut associated with a boundary region $A$ generally protrudes deeper into the ``bulk'' geometry as $|A|$ is increased, leading to the desired entanglement entropy bounds.

The MERA also has a further interesting property: It can be interpreted as an \emph{entanglement renormalization} \cite{Vidal:2007hda} that transforms a coarse-grained state to a fine-grained one (or vice-versa), as shown in Fig.\ \ref{FIG_TN_MERA} for a coarse-grained state vector $\ket{\psi_0}$. 
Each layer in the MERA network thus has an associated length or energy scale at which it produces entanglement in the output state.
As we will see later, these properties of the MERA are reminiscent of AdS/CFT, of which it has been proposed as an early toy model \cite{PhysRevD.86.065007}, further explored in Ref.\  \cite{PhysRevD.97.026012}. The connection between between MERA and hyperbolic geometries as featuring in the AdS/CFT correspondence has been made plausible also in terms of the causal structure emerging from a MERA ansatz \cite{Beny:2011vh}. Much of the mindset of MERA also carries over to \emph{hyperinvariant tensor networks} \cite{PhysRevLett.119.141602,Steinberg:2020bef}.
The usefulness of tensor networks in understanding and modelling properties of AdS/CFT is the main motivation of a large part of the work presented in this topical review, extending beyond the tensor network approaches presented in this introduction.
Reviews for a broader introduction to tensor network methods include Refs.\ \cite{EisertTensors,Orus-AnnPhys-2014,Bridgeman2017hw,PEPSRMP}.

\subsection{Quantum error correction}

Information storage and transmission are susceptible to errors. Even a purely classical system --- digital or analog --- is affected by corruption of the physical medium carrying the information. 
In the case of information storage, this includes corruption of bits on hard drives or SSDs, while in the case of transmission, noise in the conducting material can affect the signal from which the data is later read.

As it is impossible to preclude errors completely, error correction becomes necessary. This means that information is encoded such that recovery of the logical data is still possible after small errors have occurred.
The simplest way of achieving such resilience is by simply storing or transmitting multiple copies of the original data in what is called a \emph{repetition code}. For example, one may transfer the bit sequences \texttt{000} and \texttt{111} in place of the logical bits \texttt{0} and \texttt{1}. If one of the bits becomes corrupted, the remaining two still allow the reconstruction of the original logical bit (e.g.,\ \texttt{1} from \texttt{101}).

Classical codes are often categorized by the notation $[n,k,d]$, which denotes an encoding of $k$ logical bits in $n$ physical ones, with a \emph{Hamming distance} $d$. The latter is the minimal number of physical single-bit errors required to map one logical state (sequence of bits) to another. If we think of all possible physical bit-strings for a given code block as nodes in a graph, and of single-bit errors as edges connecting them, $d$ becomes the minimal graph distance between the bit-strings corresponding to logical states, showcasing the notion of code distance.
In the given notation, an $n$-fold repetition code for a single logical bit is denoted as an $[n,1,n]$ code, as $n$ physical bits need to be flipped in order to change the bit-string \texttt{00\dots 0} to \texttt{11\dots 1} and vice-versa.

Classical codes in practical use are much more complicated that simple repetition codes, but rely on the same concept of spreading out the information of logical bits over larger bit-strings. For example, the popular class of \emph{Reed-Solomon codes} interprets $k$ logical values as coefficients in a polynomial function whose result is mapped onto $n>k$ physical ones \cite{ReedSolomon1960}, leading to an $[n,k,n-k+1]$ code.

The appearance of errors and methods for their correction are fundamentally different for quantum systems.
When interacting with an environment, isolated quantum systems exhibit \emph{decoherence}, i.e., the breakdown of quantum superposition and in turn, entanglement. As entanglement is a necessary resource for any quantum computation,\footnote{Though necessary, more entanglement does not automatically make a quantum system more useful for computations \cite{PhysRevLett.102.190501}.} its breakdown must be avoided if computational power beyond classical limits is desired.

Methods of \emph{quantum error correction} are thus required to store and manipulate quantum information with a certain resilience to coupling with an environment. The most useful approach in classical error correction, the duplication of information, is impossible for quantum systems due to the \emph{no-cloning theorem}: No unitary operator, and thus no physical time evolution, can duplicate an arbitrary quantum state \cite{Wootters:1982zz,DIEKS1982271}.
Quantum error correction thus requires other approaches. The most popular and relevant for this review is the use of \emph{stabilizer codes}, first introduced by Daniel Gottesman in his PhD thesis \cite{Gottesman1997}, extending earlier approaches to the problem by Peter Shor and Andrew Steane \cite{PhysRevA.52.R2493,Steane:1995vv}.
Stabilizer codes can be represented as an encoding of quantum information in ground states of Hamiltonians 
\begin{equation}
H_\mathcal{S} = - \sum_{i=1}^m S_i
\end{equation}
that are given by the sum of orthogonal operators $S_i$, called the \emph{generators} of the \emph{stabilizer} 
\begin{equation}
\mathcal{S} = \{S_1, S_2,\dots,S_m\}.
\end{equation}
The generators are chosen to commute with one another and act as independent ``parity checks'' on different parts of the Hilbert space, i.e., have eigenvalues $\pm 1$.
The space of ground states of $H_\mathcal{S}$, given an $n$-qubit system, is thus $2^{n-m}$-dimensional and contains all states that are in the ${+}1$-eigenspace of each generator. 
For qubits, it is convenient to choose stabilizer generators that are 
%direct 
tensor products of the Pauli operators $\sigma_x,\sigma_y,\sigma_z$ and the identity $\id$, as well as using them as a basis set for operators that represent local errors.
This ensures that any product of such errors either commutes or anti-commutes with each generator.
The errors thus flip the eigenvalue of one or more of the generators, leading to a measured pattern or \emph{syndrome} from which the type of error can be deduced and reversed. 

Stabilizer codes are generally denoted as $[[n,k,d]]$ codes, in a generalization of the notation for classical codes introduced above.
Here $n$ and $k$ again denote the number of physical and logical sites, respectively, usually qubits. The code distance $d$, however, has a slightly more nuanced meaning than the classical Hamming distance. Consider, for example, a single logical qubit encoded in a basis of states $\bar{0}$ and $\bar{1}$ (read as ``logical zero'' and ``logical one'') as
\begin{equation}
\sket{\bar{\psi}} = \alpha \ket{\bar{0}} + \beta \ket{\bar{1}} \ ,
\end{equation}
where $|\alpha|^2+|\beta|^2=1$. The quantum analogon of classical bit flip errors is a basis flip $\bar{0} \leftrightarrow \bar{1}$, expressed by an operator $\mathcal{O}_\text{b}$ that interchanges the basis as
\begin{equation}
\mathcal{O}_\text{b} \sket{\bar{\psi}} = \beta \ket{\bar{0}} + \alpha \ket{\bar{1}} \ .
\end{equation}
Clearly this operator fulfills the condition $\mathcal{O}_\text{b}^2 = \id$, which followed directly from our expression of errors in terms of products of Pauli operators. However, there exists another type of error which fulfills this condition as well; these \emph{phase flip} errors, expressed by an operator $\mathcal{O}_\text{p}$ act on a logical qubit basis as
\begin{equation}
\mathcal{O}_\text{p} \sket{\bar{\psi}} = \alpha \ket{\bar{0}} - \beta \ket{\bar{1}} \ .
\end{equation}
Note that this type of error maps the $\bar{0}$ basis state onto itself, but adds a phase $e^{-\i \pi}$ to $\bar{1}$. 
This implies that when calculating the error distance $d$, we have to count the minimal number of fundamental error operations (local Pauli operators) that not only map logical basis states to other eigenstates of the stabilizer Hamiltonian, but also include errors than produce basis-dependent phases. Note that in a classical system, an operation of arbitrary complexity that maps each bit string to itself produces no effective error, a simplification that no longer applies for code states in a quantum superposition under a given basis.
As in classical codes, to increase $d$ one generally needs to increase $n$, the number of physical sites, as well. This is quantified by the \emph{quantum Hamming bound} \cite{Ekert:1996pg}, which can be derived from the following argument: 
The full $n$-qubit Hilbert space can contain $2^n$ orthogonal states, $2^k$ of which are logical states. If the code distance is $d$, then $\lfloor \frac{d-1}{2} \rfloor$ errors can be corrected, i.e., lead to distinct orthogonal states.\footnote{Note that we can \emph{detect} $d-1$ errors, but may not be able to identify the original logical state.} There are $3n$ possible local errors, one for each physical qubit and Pauli operator, and  $3^m {n\choose m}$ possibilities of applying exactly $m$ non-trivial errors.
As all errors applied to the logical states must be distinct and contained in the physical Hilbert space, we arrive the following bound 
\begin{equation}
\sum_{m=0}^{\lfloor \frac{d-1}{2} \rfloor} 3^m {n\choose m} \leq \frac{2^n}{2^k} 
\end{equation}
for an $[[n,k,d]]$ code.
For a single logical qubit, the quantum Hamming bound leads to the requirement of $n \geq 5$ physical sites. Indeed, a $[[5,1,3]]$ code that can correct an arbitrary Pauli-type error on one logical qubit exists, often simply called the ``5-qubit code''. This code, which will be highly relevant troughout this work, is explained in more detail in Summary 4.

\begin{figure*}
\begin{infobox}[Summary 4: The 5-qubit code]
The $[[5,1,3]]$ quantum error correcting code \cite{Bennett:1996gf,Laflamme1996} is built from the stabilizers
\begin{equation}
\mathcal{S}_5 = \{\, \sigma_x \sigma_z \sigma_z \sigma_x \id ,\;\; \id \sigma_x \sigma_z \sigma_z \sigma_x,\;\; \sigma_x \id \sigma_x \sigma_z \sigma_z,\;\; \sigma_z \sigma_x \id \sigma_x \sigma_z \,\} \ .
\end{equation}
Note that all generators are cyclic permutations of one another, and that multiplying all of them yields a fifth generator $\sigma_z \sigma_z \sigma_x \id \sigma_x$ that is precisely the missing permutation. 
This code is optimal in a variety of ways: It saturates the quantum Hamming bound \cite{PhysRevA.54.1862} as well as the \emph{quantum Singleton bound} \cite{Gottesman1997}, which follows from conditions on reconstructability after erasures \cite{Knill:1996ny} and is given by
\begin{equation}
n \geq 2(d-1) + k \ ,
\end{equation}
for an arbitrary $[[n,k,d]]$ code. The two local eigenstates $\bar{0}$ and $\bar{1}$ of the 5-qubit code can be distinguished by the total parity $\sigma_z^{\otimes 5}$, which thus acts as the logical parity operator $\bar{\sigma}_z$. Similarly, $\bar{\sigma}_x=\sigma_x^{\otimes 5}$ and $\bar{\sigma}_y=\sigma_y^{\otimes 5}$ act as the remaining logical Pauli operators.
Conveniently, a \emph{Jordan-Wigner transformation} maps $\bar{0}$ and $\bar{1}$ to fermionic states that are Gaussian, i.e., can be expressed as ground states of a Hamiltonian that is only quadratic in fermionic operators.
\end{infobox}
\end{figure*}

A widely used class of stabilizer codes are \emph{Calderbank-Shor-Steane (CSS) codes} \cite{Calderbank:1995dw,Steane:1995vv}, built from a combination of two classical codes. Each is mapped onto stabilizer generators containing, up to local identities, only $\sigma_x$ or only $\sigma_z$ operators, respectively, which makes it easier to realize such codes in practice; indeed, they were the first quantum codes to be realized experimentally \cite{PhysRevLett.81.2152}.
Recently, stabilizer-based \emph{topological codes} \cite{Kitaev_1997} like the \emph{surface code} \cite{Dennis:2001nw} and the \emph{color code} \cite{PhysRevLett.97.180501} have enjoyed large popularity for potential quantum error correction in large systems of well-protected logical qubits (see, e.g.,\ Ref.\ \cite{Litinski_2019}).
Rather than full fault-tolerant quantum computation, a goal still intensely pursued on many fronts, current stabilizer code implementations are restricted to small-scale \emph{quantum error detection} which provides some limited benefits \cite{Vuillot2017,Andersen2019}.
An experimental realization of the 9-qubit \emph{Bacon-Shor code}, a \emph{subsystem code} that extends stabilizer codes by a notion of gauge transformations \cite{PhysRevLett.95.230504}, has recently been achieved in trapped ions \cite{egan2020faulttolerant}.
However, as the codes considered in the context of holography are usually simple stabilizer codes, no more details on the fascinating subject of practical realizations of quantum error correction will be given here.

As the general reference for all things quantum information, the book by Michael A.\ Nielsen and Isaac L.\ Chuang \cite{nielsen_chuang_2010} contains a broad introduction to quantum error correction, as does chapter 7 of the lecture notes by John Preskill \cite{PreskillNotes1998}.
The previously mentioned PhD thesis by Daniel Gottesman \cite{Gottesman1997} provides a more extensive introduction to the field of quantum error correction. For more recent overviews of the field, consult Refs.\ \cite{RevModPhys.87.307,Devitt_2013}.

\section{Holography in quantum information}
\label{SEC_HOLOGRAPHY_IN_QI}

In recent years, viewing the AdS/CFT correspondence through the lens of quantum information theory has led to surprising connections between both fields. These ``holographic'' descriptions of quantum information concepts deepen our understanding of holography itself, but also offer potential approaches to problems that were not originally thought to be associated with AdS/CFT.

\subsection{Holographic entanglement entropy}
\label{SS_HOLO_EE}

\begin{figure}
\centering
\includegraphics[height=0.2\textheight]{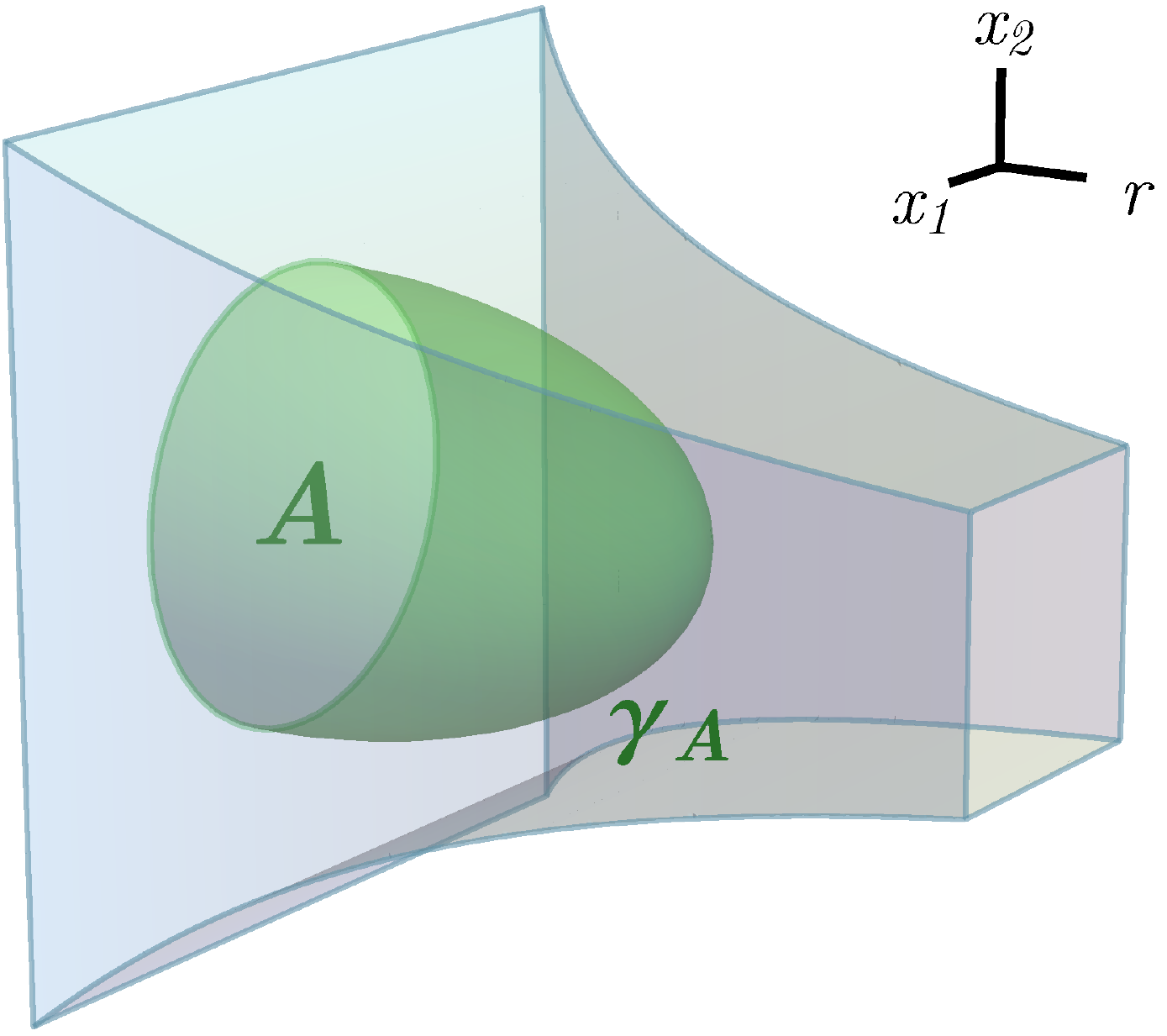}
\hspace{2.2cm}
\includegraphics[height=0.2\textheight]{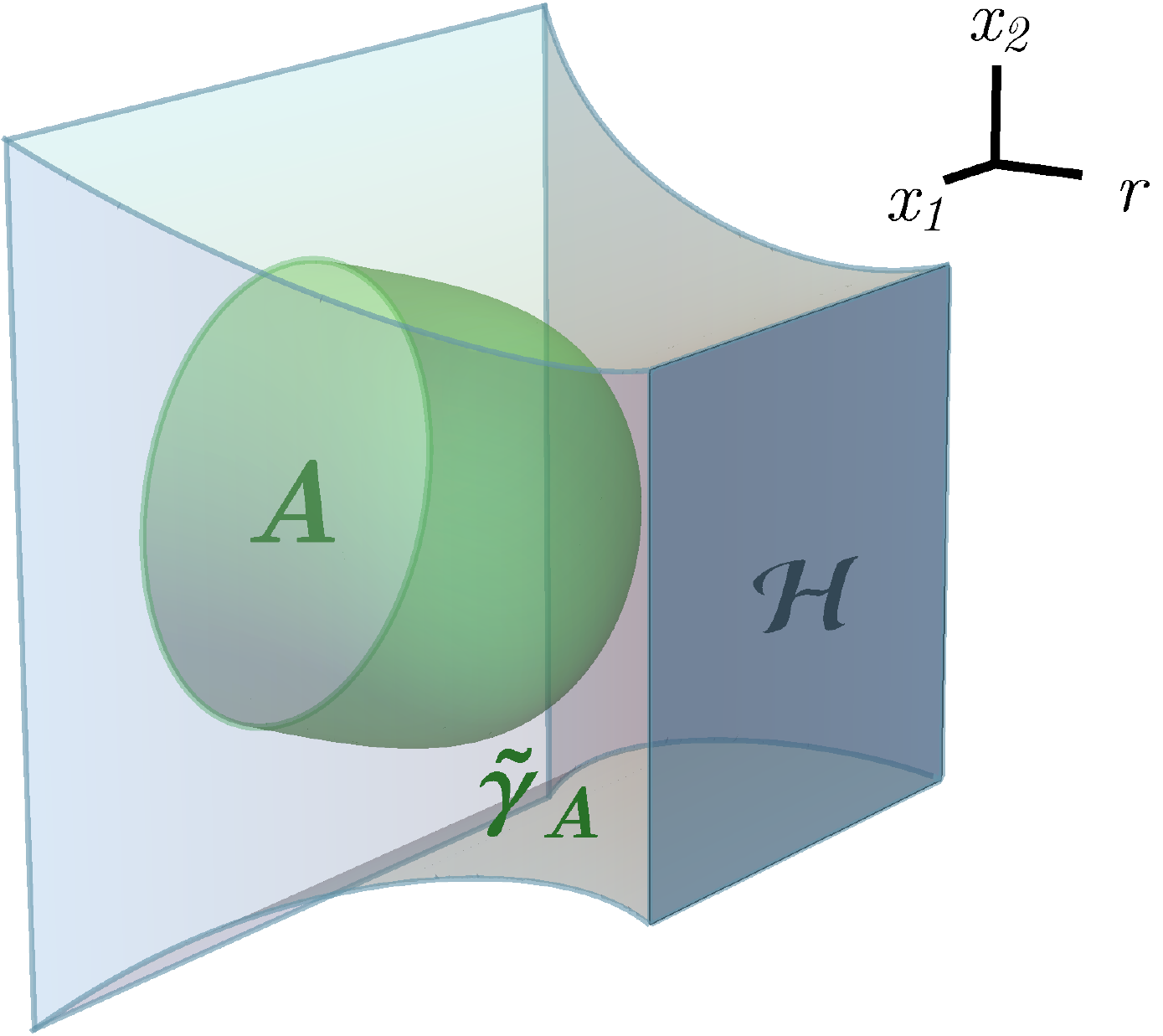}
\caption{\textsc{Left:} Minimal surface $\gamma_A$ homologous to a boundary region $A$ in an AdS time-slice (blue-shaded throat region).
\textsc{Right:} Deformed minimal surface $\tilde{\gamma}_A$ in an AdS geometry with black hole horizon $\mathcal{H}$.  
}
\label{FIG_RT}
\end{figure}

Probably the first connection between AdS/CFT and quantum information has been introduced by Shinsei Ryu and Tadashi Takayanagi, when they considered the following question: What is the dual AdS$_{d+1}$ bulk description of the entanglement entropy $S_A$ of a boundary subsystem $A$ in a holographic CFT$_d$? The answer is provided by a startling generalization of the black hole entropy formula \eqref{EQ_BEK_HAW} which relates the black hole entropy to its horizon area. It turns out that the bulk quantity dual to $S_A$ is the area of a $d{-}1$-dimensional minimal surface $\gamma_A$ homologous to $A$, i.e., with $\partial A = \partial \gamma_A$ (see Fig.\ \ref{FIG_RT}, left). This is quantified by the \emph{Ryu-Takayanagi (RT) formula} \cite{PhysRevLett.96.181602}
\begin{equation}
\label{EQ_RYU_TAKAYANAGI}
S_A = \frac{|\gamma_A|}{4 G} \ ,
\end{equation}
where $|\gamma_A|$ is the area of $\gamma_A$ and $G$ is the gravitational constant in the $d{+}1$-dimensional bulk space-time. Note that this formula has no dependence on the actual holographic model, e.g., the degree of supersymmetry and the bulk structure.
Given an excited CFT with a dual bulk geometry other than ``pure'', undeformed AdS, the shape and area of $\gamma_A$ change, reflecting entanglement produced or destroyed by the excitation.
For example, consider a thermal CFT, whose bulk dual is given by an AdS black hole geometry (see Fig.\ \ref{FIG_RT}, right). The horizon deforms the minimal surface towards the AdS boundary, increasing its area and thus reproducing the thermal entanglement associated with a finite-temperature CFT \cite{PhysRevLett.96.181602}. 
If we start growing the subsystem $A$ until it encompasses the entire boundary, the minimal surface starts wrapping around the black hole horizon, as this horizon is itself extremal. In that limit, \eqref{EQ_RYU_TAKAYANAGI} becomes the Bekenstein-Hawking formula \eqref{EQ_BEK_HAW} for the classical entropy $S$, showing the intimate connection between both formulae. 

Strictly speaking, the definition of $\gamma_A$ is only properly coordinate-independent if we restrict the boundary system to the time-slice of a static space-time geometry. For more general space-like boundary regions $A$, we have to consider a space-like bulk surface $\gamma_A$ that is \emph{extremal}, leading to a generalized form of the RT formula is often called the \emph{Hubeny-Rangamani-Takayanagi (HRT) formula} \cite{Hubeny:2007xt}. Such an extremal surface is minimal with regard to space-like variations and maximal with regard to time-like ones \cite{Wall:2012uf}.

After the initial proposal of the RT formula it was quickly found to be consistent with the property of \emph{strong subadditivity} of entanglement entropy \cite{Headrick:2007km}, providing an important consistency check. It has later also been proven within AdS/CFT, first only for $1{+}1$-dimensional CFTs \cite{Faulkner:2013yia,Hartman:2013mia} and shortly afterwards in the more general case \cite{Lewkowycz:2013nqa}.
It was soon understood that \eqref{EQ_RYU_TAKAYANAGI} only holds in the AdS/CFT limit of large $G$ (classical bulk gravity) and $N \to \infty$, where $N$ is the rank of the gauge group $SU(N)$ of the boundary CFT, and that quantum corrections lead to additional terms constant in $G$ and $N$ that can be interpreted as entanglement between bulk regions \cite{Faulkner:2013ana}. 
For an introduction to holographic entanglement entropy, the extended version of 
Shinsei Ryu and Tadashi Takayanagi's original work \cite{Ryu:2006ef} is a good starting point. There also exists a book on the topic \cite{Rangamani:2016dms}, a shortened version of which is available online \cite{Rangamani:2016dms}.

\subsection{Gravity and entanglement}
\label{SS_GRAV_ENT}

A particularly fascinating feature of AdS/CFT is that it relates a theory with (quantum) gravity to one without it. This has led to the suggestion that gravity, whose failure at consistent quantization on arbitrarily small scales has been one of the driving motivations behind the development of string theory, is indeed not a fundamental force at all, but rather holographically emergent from quantum degrees of freedom.
More concretely, Mark Van Raamsdonk suggested that the connectivity between regions of space-time could be a consequence of entanglement between them \cite{VanRaamsdonk:2010pw}: First, he interpreted an earlier setup relating a maximally entangled AdS black hole space-time to two copies of a CFT  \cite{Maldacena:2001kr} as an example of a non-entangled system being spatially separated.
An entangled system, on the other hand, would be characterized by a nonvanishing \emph{mutual information} (MI)
\begin{equation}
I(A:B) = S_A + S_B - S_{A \cup B} \ ,
\end{equation}
between two regions $A$ and $B$. The behaviour of MI in holography is well understood, and associated with a phase transition as the distance between $A$ and $B$ is varied \cite{Headrick:2010zt,Molina-Vilaplana:2013xqa}.
MI also serves as an upper bound to two-point correlation functions between both regions \cite{Wolf:2007tdq}; in a holographic theory, where such correlators are expected to decay exponentially with geodesic distance through the bulk, an increase in entanglement would thus imply a closer spatial bulk distance.

\begin{figure}
\centering
\includegraphics[height=0.25\textheight]{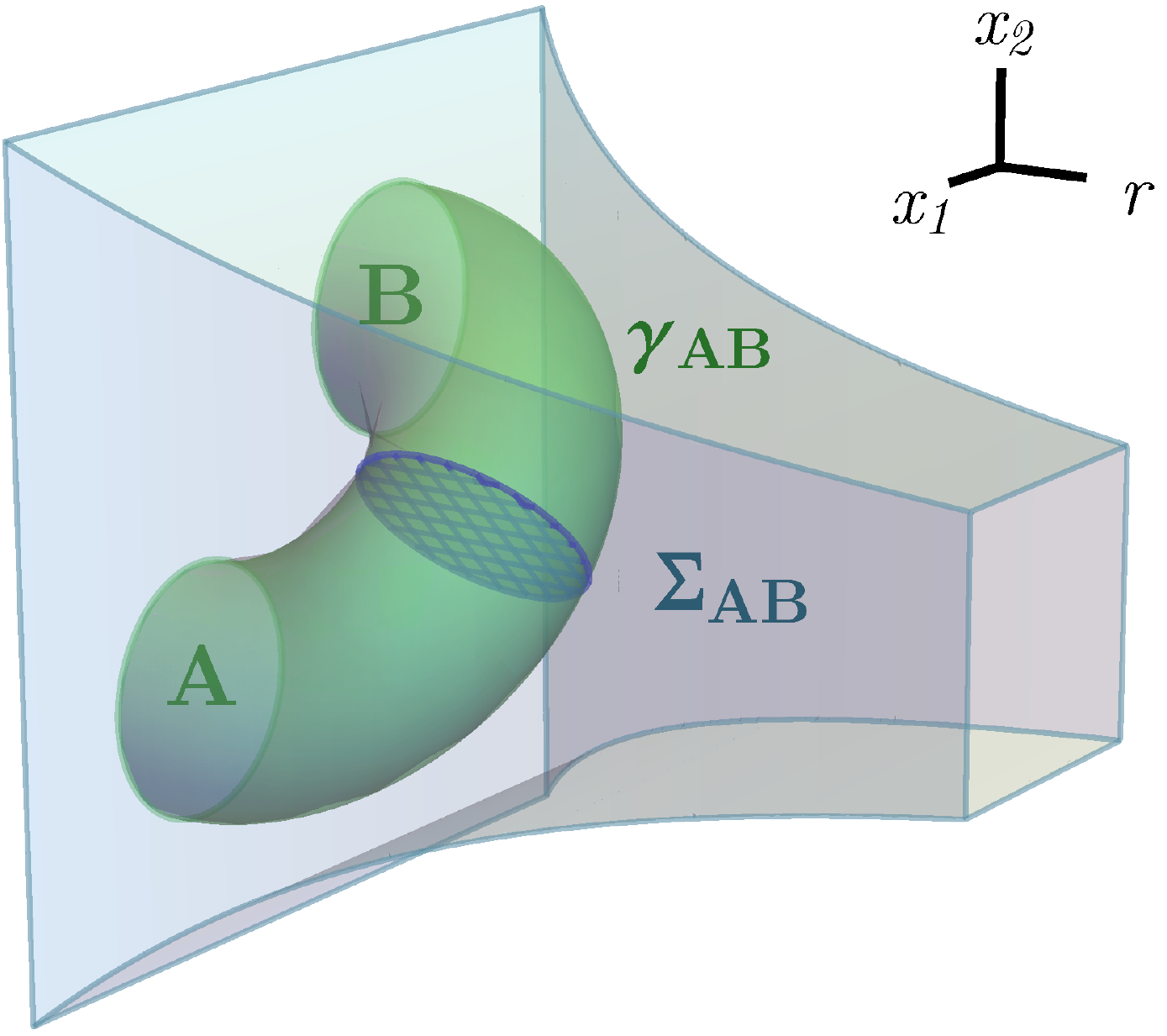}
\caption{Holographic prescription of the entanglement of purification $E_P$ for a disjoint boundary region $A\cup B$: The minimal (Ryu-Takayanagi) surface $\gamma_{AB}$ connects both boundary regions when $A$ and $B$ are close together. The area of the minimal \emph{entanglement wedge cross-section} $\Sigma_{AB}$ is conjectured to be related to the entanglement of purification via $E_P = |\Sigma_{AB}^\text{min}|/4 G_N$ \cite{Takayanagi:2017knl}.
}
\label{FIG_HEOP}
\end{figure}

While this conjectured connection between entanglement and the emergence of gravity remains far from being understood, many similar ideas have appeared throughout the decade since its proposal. For example, a similar behaviour has been found in the holographic description of quantities other than entanglement entropy. This includes
the \emph{entanglement of purification}
\cite{Terhal:2002eop}, 
an entanglement measure for mixed states related holographically to the minimal cross-section of the throat-like RT surface of a two-component region \cite{Takayanagi:2017knl,Caputa:2018xuf} (see Fig.\ \ref{FIG_HEOP} for details), again relating spatial connectivity and entanglement. Another example is the suggestion that the \emph{state complexity} of a quantum state, quantifying the number of local operations or quantum gates required to construct it from a reference state \cite{Nielsen:2006mn2}, may possess a holographic dual either in terms of a bulk volume \cite{PhysRevD.90.126007}
(``complexity equals volume'')
or an action evaluated on as specific bulk region \cite{Brown:2015bva} 
(``complexity equals action''). The validity of either proposal is still hotly debated.

As most of these proposals are motivated around conceptual issues of quantum gravity, it may be surprising that a holographic description of gravity may have a bearing on actual qubit experiments: It was been proposed that certain experimental setups for \emph{quantum teleportation}, the transmission of quantum states via entanglement, may be effectively described by a holographic construction involving traversable wormholes \cite{Gao:2016bin,Brown:2019hmk,Nezami:2021yaq}. Though no such experiments have yet been conducted, this emphasizes the potential of bridging the fields of quantum computation and fundamental physics.
Many of the early ideas of connecting gravity and entanglement are surveyed in Mark Van Raamsdonk's lecture notes on the same topic \cite{VanRaamsdonk:2016exw}.

\subsection{Holographic tensor networks}

The form of the Ryu-Takayanagi formula \eqref{EQ_RYU_TAKAYANAGI} bears a striking resemblance to the entanglement entropy bound \eqref{EQ_TN_ENT_BOUND} in generic tensor networks, both involving a minimal surface through a geometry than extends the direct geometry of the boundary state.
This leads to a straightforward question: Can the time-slice of an AdS space-time, which we consider in the RT formula, be expressed as a tensor network? The first proposal in this direction has been made by Brian Swingle, who suggested the MERA tensor network for this particular interpretation \cite{PhysRevD.86.065007}. This identification is tempting, because the MERA implements entanglement renormalization between discretized quantum systems at different scales. Similarly, we expect a time-slice in the AdS metric \eqref{EQ_ADS_PDISK} at fixed radius $r$ from the AdS boundary to describe an increasingly fine-grained system as $r$ is decreased. 
Furthermore, the gapless states produced by the MERA resemble those expected in the conformal boundary theories of AdS/CFT, though being ground states of much simpler critical Hamiltonians that usually feature neither supersymmetry nor non-Abelian gauge symmetries.
While the MERA produces boundary states with conformal symmetries, its tensor network geometry does not exactly match an AdS time-slice --- the hyperbolic \emph{Poincar\'e disk} --- and leads to inconsistencies when treated as such \cite{Bao:2015uaa}.
Alternatively, the MERA geometry has been interpreted both as a time-like surface in positively curved de Sitter (dS) space-time \cite{Beny:2011vh} and as a path integral discretization of an AdS light-like surface \cite{Milsted:2018san}. 
In either case, using the MERA is a discrete realization of AdS/CFT requires abandoning the simple time-slice picture in which the RT formula has been derived, leading to a setting whose relationship to holography is not yet fully understood.

In principle, the indices of any tensor network can be seperated into two sets between which it acts as a linear map on states. Clearly, labeling these two sets ``bulk'' and ``boundary'' and expectating the map to show any holographic features is pointless for most setups. 
Invoking the RT formula, we at minimum desire an ansatz that gives the correct entanglement entropy scaling for minimal cuts through the network. Choosing a tensor network whose geometry discretizes the Poincar\'e disk is not generally sufficient; we also need the entanglement entropy bound \eqref{EQ_TN_ENT_BOUND} to saturate for \emph{any} choice of subsystem.
The right choice of tensors is thus crucial. Surprisingly, choosing random tensors already reproduces many of the expectated properties, such as polynomially decaying correlation functions, as long as the bond dimension is large \cite{Hayden2016,Qi:2018shh}. Rather than a mapping between individual states, such a construction considers averages of random bulk configurations, leading to a bulk partition function that is in fact equivalent to the classical Ising model when evaluating R\'enyi entropies 
\begin{equation}
S^{(n)}_A =  \frac{1}{1-n}\log(\tr{\rho_A^n})
\end{equation} 
for $n=2$. These random tensor network models produce many holographic properties in the limit of large bond dimension, which is consistent with the semiclassical limit of AdS/CFT. However, the scaling of $S^{(n)}_A$ with $n$ as expected for CFT ground states is not reproduced in these models, even at large bond dimension.
Which other choices of tensors are possible? Conditions to constrain suitable tensors, as we will see in the next section, are found by considering quantum information quantities beyond simple entanglement measures. 
While a broad introduction to tensor network holography remains to be written, the initial proposal by Brian Swingle \cite{PhysRevD.86.065007} contains many of the key ideas that various implementations over the past decade have been based on.

\subsection{Holographic codes}

As a duality between bulk and boundary, AdS/CFT implies a complicated mapping of quantum information between both sides of the duality.
Considering subregions of bulk or boundary, the question arises if the information encoded in local bulk regions is contained in a local region on the boundary and vice-versa.
In general, this does not appear to be the case. Reconstructing a bulk field $\phi(x)$ at a point $x$ generally requires information about a boundary region that increases in size as $x$ is moved further into the bulk; in other words, information about fields close to the boundary can be recovered from a small boundary region, while information deep in the bulk is ``smeared out'' over the boundary \cite{Hamilton:2006az}.
More precisely, given a region $A$ on the boundary there exists a wedge extending into the bulk space-time whose information content can be reconstructed purely from $A$. 
An early candidate for such a bulk region has been the \emph{causal wedge} \cite{Czech:2012bh,Bousso:2012sj,Hubeny:2012wa}, which is the bulk region causally connected to the boundary \emph{domain of dependence} of $A$, which is the diamond-shaped boundary region containing the causal past and future of $A$.
It has later been found that an even larger bulk region, the \emph{entanglement wedge}, can be reconstructed from $A$ \cite{PhysRevLett.117.021601,Bao:2016skw}. This wedge is the bulk region enclosed between $A$ and its Ryu-Takayanagi surface $\gamma_A$ (as well as its bulk domain of dependence).

\begin{figure}
\centering
\includegraphics[height=0.28\textheight]{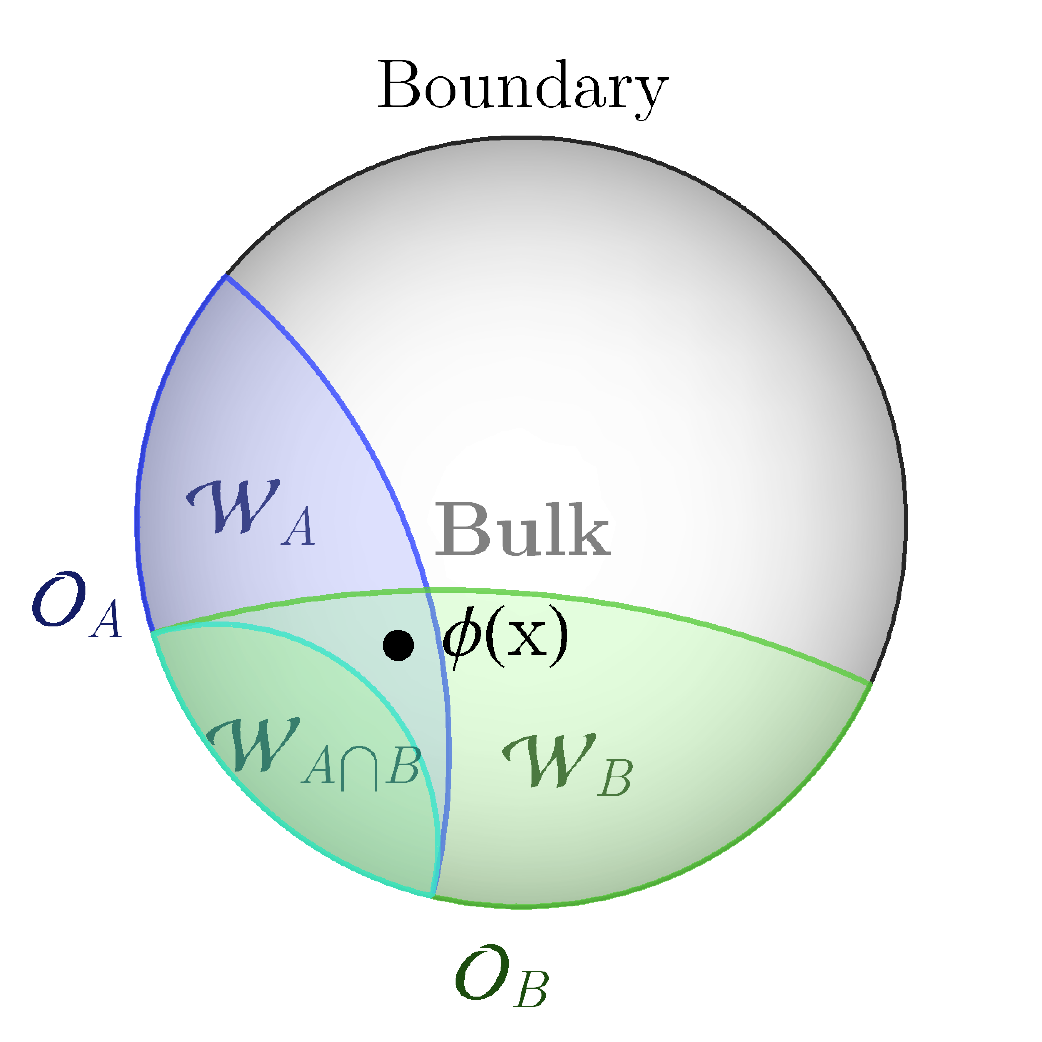}
\hspace{0.8cm}
\includegraphics[height=0.28\textheight]{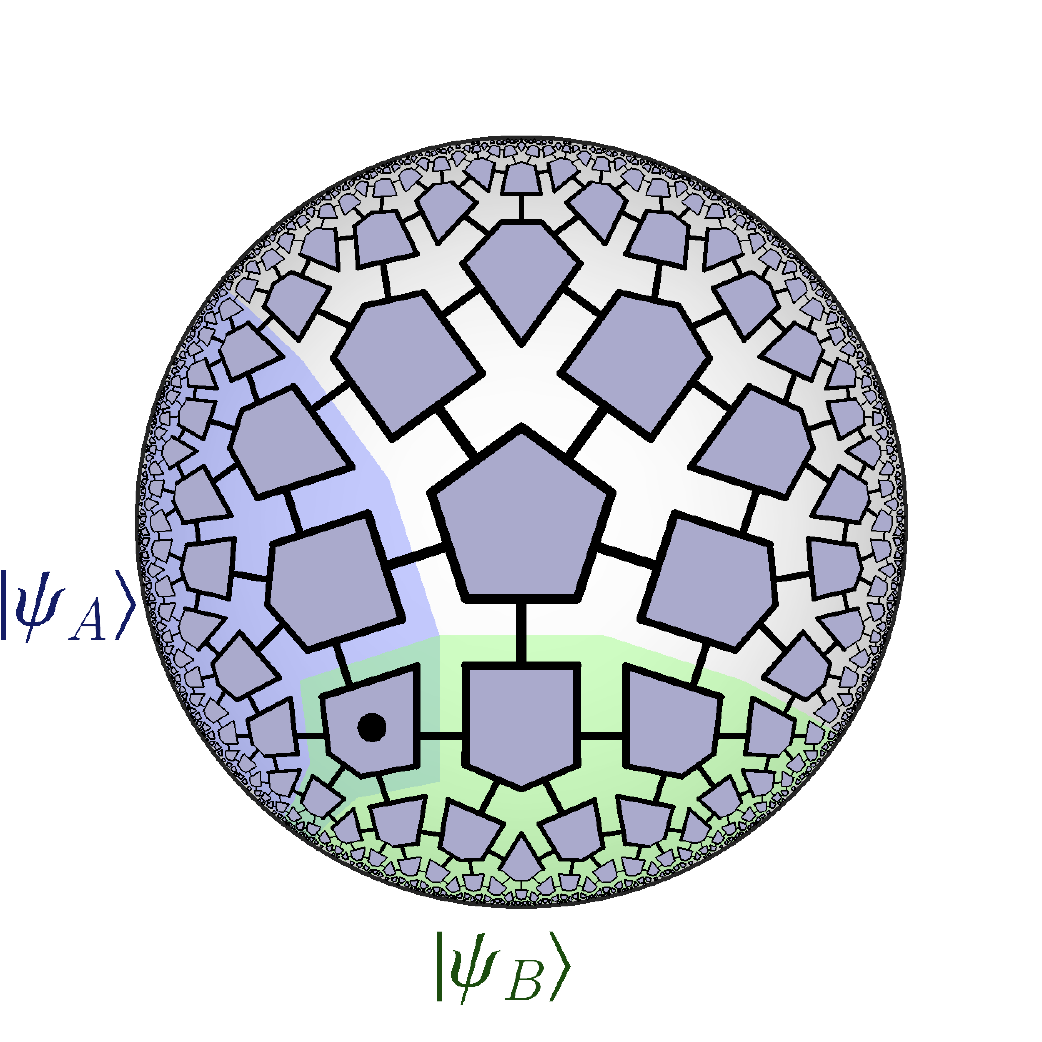}
\caption{\textsc{Left:} (Entanglement) wedges $\mathcal{W}_A$ and $\mathcal{W}_B$ in an AdS time-slice (Poincar\'e disk), corresponding to two regions in which a bulk field $\phi(x)$ can be reconstructed as two operators $\mathcal{O}_A$ and $\mathcal{O}_B$ with support on boundary regions $A$ and $B$. The wedge $\mathcal{W}_{A \cap B}$ of the intersection of $A$ and $B$ is insufficient.
\textsc{Right:} The discrete form of causal/entanglement wedges in the hyperbolic pentagon code. The logical state on the marked pentagon can be reconstructed from either of the two boundary state
vectors $\ket{\psi_A}$ and $\ket{\psi_B}$.
}
\label{FIG_WEDGES}
\end{figure}

As different boundary regions correspond to sometimes overlapping wedges in the bulk, local bulk information can in fact be reconstructed on \emph{different} boundary regions \cite{Morrison:2014jha}.
This leads to a conundrum: Consider, as shown in Fig.\ \ref{FIG_WEDGES} (left), two wedges $\mathcal{W}_A$ and $\mathcal{W}_B$ containing a point $x$. If we can reconstruct the bulk field $\phi(x)$ in both of the corresponding boundary regions $A$ and $B$, does this imply that its information is encoded in $A \cap B$? 
This conclusion cannot be correct, as $x$ can be chosen so that it is not contained within the wedge $\mathcal{W}_{A \cap B}$ and hence the information in $A \cap B$ must be insufficient to reconstruct $\phi(x)$.

The only resolution to this problem --- other than assuming that reconstructed operators are all trivial, acting as an identity --- is to conclude that $\phi(x)$ can be represented as \emph{different} equivalent operators on different boundary regions, an insight transparently captured in work by 
Ahmed Almheiri, Xi Dong, and Daniel Harlow \cite{Almheiri:2014lwa}.
Its information of $\phi(x)$ is thus stored redundantly, as removing parts of the boundary required for the reconstruction along one wedge does not prevent its recovery via another; in other words, bulk information is stored on the boundary in the manner of a quantum error-correcting code. 
Bulk quantum information stored at a point in the centre is protected against the erasure of large parts of the boundary, as the wedge of the erased region usually does not penetrate far enough into the bulk to affect the centre.
In contrast, bulk information at a point near the boundary is completely lost by an erasure of a small region whose wedge contains it. 
Properties of quantum error correction, in the form of a \emph{subregion duality} between boundary and bulk subsystems, thus appear to be a generic feature of AdS/CFT and can be shown to hold even in deformed holographic theories \cite{Lewkowycz:2019xse}.

The behaviour of holographic codes can be reproduced in an exceedingly simple discrete toy model first proposed in Ref.\ \cite{Almheiri:2014lwa} and extended in Refs.\ \cite{Latorre:2015xna,Harlow:2016vwg}: This model relies on the \emph{3-qutrit code} \cite{Cleve:1999qg} which encodes a logical qutrit (a three-level quantum state) in three physics ones. It contains the three logical basis states
\begin{align}
    \ket{\bar{0}} &= \frac{\ket{0,0,0} + \ket{1,1,1} + \ket{2,2,2}}{\sqrt{3}} \ , \\
    \ket{\bar{1}} &= \frac{\ket{0,1,2} + \ket{1,2,0} + \ket{2,0,1}}{\sqrt{3}} \ , \\
    \ket{\bar{2}} &= \frac{\ket{0,2,1} + \ket{2,1,0} + \ket{1,0,2}}{\sqrt{3}} \ .
\end{align}
This code enables \emph{quantum secret sharing}: To reconstruct a logical state vector (some superposition of the logical basis states), one requires access to the state on any two of the physical qutrits from which it can be recovered via a suitable unitary transformation. Conversely, the reduced density matrix of any single physical qutrit is maximally mixed, and no information about the logical state can be recovered from it at all.
The holographic interpretation of this code now comes about if we identify the logical qutrit with the bulk and the three physical ones with the boundary: We interpret the need for two physical qutrits for reconstruction of the logical one as the requirement that only large boundary regions have a wedge large enough to recover information far in the bulk.
This model reproduces a number of quantum error correction properties expected from the earlier AdS/CFT discussion: First, a bulk local operator (an operator acting on the logical state space) commutes with all local boundary operators (acting on the physical qutrits). This follows directly from the erasure property of the 3-qutrit code: If any single physical qutrit can be erased without affecting the logical state then such single-site operators must commute with the logical ones.
Second, there exists a subregion duality between any two physical qutrits on the boundary and the bulk information (the logical state): Any logical operator can be represented as an operator acting on only two boundary qutrits.
Third, the entanglement entropies of boundary subregions have a holographic interpretation as well: For a mixed logical state encoded in the density matrix $\bar{\rho}$, the 1- and 2-site entanglement entropies are
\begin{align}
    S_\text{1-site} &= \log 3 \ , & 
    S_\text{2-site} &= \log 3 + S(\bar{\rho})  \ .
\end{align}
We can identify the $\log 3$ contribution as an ``area term'' that corresponds to a minimal cut through the geometry over a single bond of dimension $\chi=3$. For boundary regions large enough to reconstruct the bulk logical state, we further find a ``bulk entropy term'' that depends on mixing within the logical state. In continuum AdS/CFT, such terms appear as higher-order corrections to the Ryu-Takayanagi formula \cite{Faulkner:2013ana} and appear to be deeply related to properties of quantum error correction \cite{Harlow:2016vwg}.
This 3-qutrit model can even be interpreted as containing a notion of a black hole: For any state in the code subspace we can reconstruct the logical qutrit and thus the full ``bulk geometry'', but any state outside of the code subspace makes it inaccessible, as if hidden behind a black hole horizon.

Clearly a model of three qutrits does not really describe the bulk geometry of an AdS space-time, but the underlying ideas were the starting point of the more elaborate construction presented in Ref.\ \cite{Pastawski2015} which will we explore in great detail in the next section.
The connection between quantum error correction and holography has only recently been established and is the subject of much ongoing research. While a formal review does not yet exist, a short introduction by Beni Yoshida, which also introduces the tensor network realization of a holographic code presented in the next section, is available on Caltech's \emph{Quantum Frontiers} blog \cite{YoshidaHQEC2015}.

\section{Holographic quantum error correction in tensor networks}
\label{SEQ_HOLO_QEC_TN}
\subsection{The HaPPY code}

In its continuum formulation with infinitely many degrees of freedom, the code picture of AdS/CFT is difficult to treat with the language usually used in the context of quantum information. Can we instead build a discrete toy model, based on simple quantum error-correcting codes that are already familiar to us? 
A class of such models has indeed been constructed and are now known as \emph{HaPPY codes} \cite{Pastawski2015}, an acronym containing the authors' initials. 
Before defining the entire class of codes, first consider a particularly simple instance of this code built on the previously mentioned five-qubit code: This \emph{hyperbolic pentagon code} \cite{Pastawski2015} has the geometry of a regular hyperbolic discretization built from pentagons, each pentagon being identified with one $[[5,1,3]]$ code and each pentagon edge with a physical qubit (or pairs thereof for each edge shared by two neighbouring pentagons). Representing the code as a six-leg tensor that maps between the logical and physical qubits, the five ``physical legs'' are then contracted following the adjacency of edges in the discretized geometry.
In this construction, visualized in Fig.\ \ref{FIG_WEDGES} (right), the reconstruction along different bulk wedges discussed in the previous section follows directly from the $[[5,1,3]]$ code's properties: Starting from a set of uncontracted physical sites on the (asymptotic) boundary, we can reconstruct the logical states on the near-boundary pentagons from just three physical sites. Recovering the physical state on the remaining edges, we then use these as inputs on the next layer of pentagons, reconstructing their logical states in turn. Through this procedure we gradually recover the logical states in the bulk from a boundary region, building up a discretized wedge until we can no longer find three physical sites around the same pentagon, i.e., until the boundary of the wedge is no longer concave.
This process, known as the \emph{greedy algorithm}, can be applied to any given boundary region, the state of which is determined by all logical states within the wedge. Conversely, the logical state on a single pentagon affects all physical boundary states in subsystems whose wedges include it. The hyperbolic pentagon code thus gives a concrete mapping between bulk and boundary states with the quantum error-correcting features of AdS/CFT.

The hyperbolic pentagon code is only a special case of a large class of tensor networks with similar properties: Their crucial ingredient is that of \emph{perfect} tensors, which act as \emph{isometries} between any bipartition of its indices as long as the number of output indices is at least as large as the number of input indices. For example, a perfect five-index tensor $T_{i,j,k,l,m}$ would fulfill constraints such as
\begin{equation}
\sum_{i,k,m} T^\star_{i,j,k,l,m} T_{i,n,k,o,m} \propto \delta_{j,n} \delta_{l,o} \ .
\end{equation}
Perfect tensors can be identified with \emph{absolutely maximally entangled} states \cite{Helwig:2012nha} by means of exploiting an equivalence with pure multi-partite quantum states, and in turn give rise to instances of stabilizer codes \cite{Mazurek:2019xph}.
Tensor networks built from such tensors allow for a variant of the greedy algorithm to be applied and thus lead to Ryu-Takayanagi-like entanglement scaling. While the property to be a perfect tensor makes perfect sense from the perspective of quantum error correction and renders the analysis of the resulting holographic code very transparent, from a physical perspective this is a rather strong property. In fact, random tensors drawn from a suitable probability measure are with high probability close to such perfect tensors \cite{Hayden2016} so that much of the analysis of Ref.\ \cite{Pastawski2015} carries over to the case of \emph{random tensor networks}.

We can think of the resulting codes as being akin to an omnidirectional MERA: The MERA tensor constraints (compare Fig.\ \ref{FIG_TN_MERA}) reduce the computation of local observables to a problem of evaluating a localized part of the tensor network, as most of the contractions simply reduce to identities. Similarly, an operator applied to a HaPPY code boundary can only affect the result of contractions within the wedge obtained from the boundary by application of the greedy algorithm. Unlike the MERA, HaPPY codes have no inherent directionality, as a regular hyperbolic tilings has the same geometrical structure around any given tensor.
These codes can be defined on any such tiling; they are generally labeled by the \emph{Schl\"afli symbol} $\{n,k\}$ denoting an $n$-gon tiling where $k$ $n$-gons meet at each vertex (or equivalently, whose \emph{dual tiling} that replaces vertices and $n$-gon centres yields a $k$-gon tiling). By considering the angles of such tilings one finds that hyperbolic tilings require $f(n,k) = n k - 2(n+k)$ to be positive (if $f(n,k)$ is zero the tiling is flat, and the integers satisfying $f(n,k)<0$ characterize regular polyhedra). 
The standard hyperbolic pentagon code is thus also refered to as a $\{5,4\}$ holographic code.
Any $[[n,m,d]]$ code can in principle be embedded into an $n$-gon tiling, with those whose encoding isometry can be representated as a perfect tensors inheriting the HaPPY properties.

Both the choice of a hyperbolic bulk and its discretization by a regular tilings determine the geometrical features of the boundary, which can be seen to have a \emph{fractal} structure in two different ways: First, the question of which boundary regions are necessary for reconstructing information at a point $x$ in the bulk leads to the insight that not the entirety of the boundary whose entanglement wedge contains $x$ is needed; in fact, in the continuum an infinite number of subregions of decreasing size can be ``punched out'' of the boundary while still being reconstructable from the properties of the code. This means that from an operator algebra perspective, only operators on a fractal subset of this boundary are required to reconstruct the bulk, a property called \emph{uberholography} \cite{Pastawski2017}.
In the regular tiling discretization, this ability to remove pieces of the boundary with impunity is greatly restricted, but replaced by another notion of fractal geometry: Cutting off the tiling after a finite number of \emph{inflation steps} (discussed in more detail below) leads to a boundary whose  boundary geometry is inherently \emph{quasi-regular} \cite{Boyle:2018uiv,GluzaBoundary}, meaning it has self-similar geometric structures resembling those of a fractal. More generally, it has been proposed that tensor networks on regular hyperbolic geometries naturally encode the symmetry transformations of \emph{quasi-regular conformal field theory} (qCFT), a theory with discretely broken conformality, possessing properties distinct from continuum CFTs \cite{Jahn:2020ukq}.

\subsection{Majorana dimer codes}

What kind of \emph{physical boundary states} do holographic quantum error-correcting codes lead to? In the case of the hyperbolic pentagon code and its generalizations, this is not a question that can be studied in a straightforward manner: Tensor contraction of large-scale version of this code is, unlike the MERA, computationally inefficient. However, we can use the code properties of this model to study its correlation and entanglement structure in an almost completely analytical manner.

In the case of entanglement, these properties follow directly from the application of the greedy algorithm \cite{Pastawski2015}: For boundary regions $A$ and their complement $A^\text{C}$ that both reduce to the \emph{same} minimal cut $\gamma_A$ through application of the greedy algorithm on either region, the property of maximal entanglement of each individual tensor enforces maximal entanglement entropy across the cut. That is, the entanglement follows a discrete Ryu-Takayanagi formula
\begin{equation}
    S_A = |\gamma_A| \log \chi \ ,
\end{equation}
where $|\gamma_A|$ is the number of edges composing $\gamma_A$ and $\chi$ is the constant bond dimension of each perfect tensor bond, e.g.\ $\chi=2$ for the hyperbolic pentagon code (as the $[[5,1,3]]$ code is a spin code).
While this reduction of $A$ and $A^\text{C}$ to the same discrete bulk geodesic $\gamma_A$ is possible for most regions $A$ (even disjoint ones), there exist pathological cases where a \emph{residual bulk region} separates $\gamma_A$ from $\gamma_{A^\text{C}}$, affecting the resulting entanglement entropy.

\begin{figure}
\centering
\includegraphics[height=0.28\textheight]{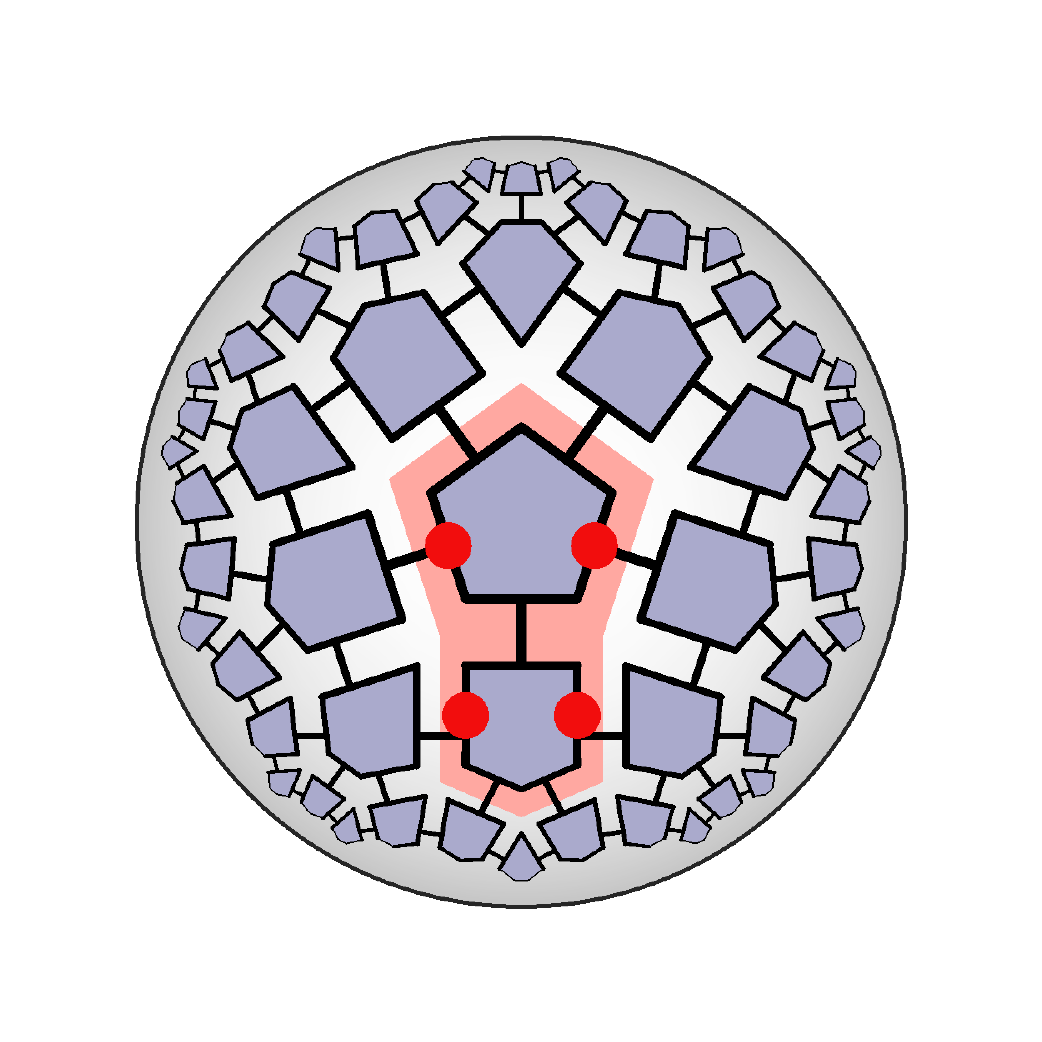}
\hspace{0.8cm}
\includegraphics[height=0.28\textheight]{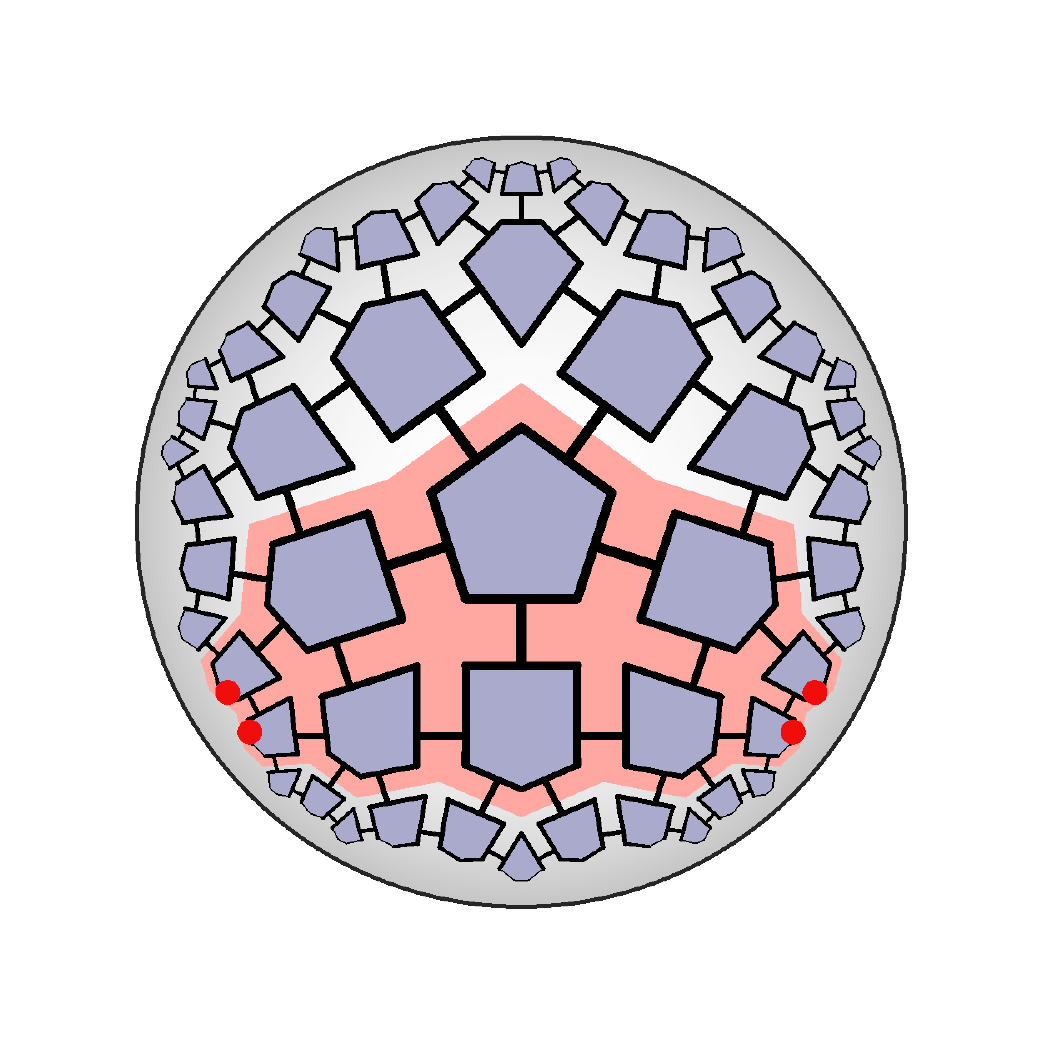}
\caption{\textsc{Left:} A non-reconstructable bulk region (red-shaded region) created by the insertion of four operators (red dots). The greedy algorithm will be unable to reach this region from any starting region.
\textsc{Right:} By pulling the four operators to the boundary along a discrete geodesics, the non-reconstructable region divides the tensor network in two. Expectation values of the four operators can be nonzero.
}
\label{FIG_WEDGES2}
\end{figure}

Beyond entanglement, one can consider two-point correlation functions, which appear to be simple at first. In the spin picture of the hyperbolic pentagon code, for example, consider a two-point function $\langle X_j X_k \rangle$ between Pauli $X_k$ operators on two boundary sites $j$ and $k$. If the operators act on edges of separate pentagons, then the effect of each operator corresponds to a correctable error, as the logical bulk state on each pentagon can still be reconstructed from its four other physical sites. This remains true even if both operators act on the same pentagon, as we can reconstruct the logical state from merely three physical sites as well.
We thus conclude that all two-point functions between single Pauli operators must vanish. Does this hold for more complicated two-point functions as well? Fig.\ \ref{FIG_WEDGES2} illustrates that this is not the case. By carefully choosing pairs of neighbouring sites, we can produce entire bulk regions stretching between two boundary regions on which the logical state that can no longer be reconstructed. This implies that two-point functions of such boundary operators with two-site support are generally nonzero.

\begin{figure}
\centering
\begin{align*}
    \begin{gathered}
    \includegraphics[height=0.26\textheight]{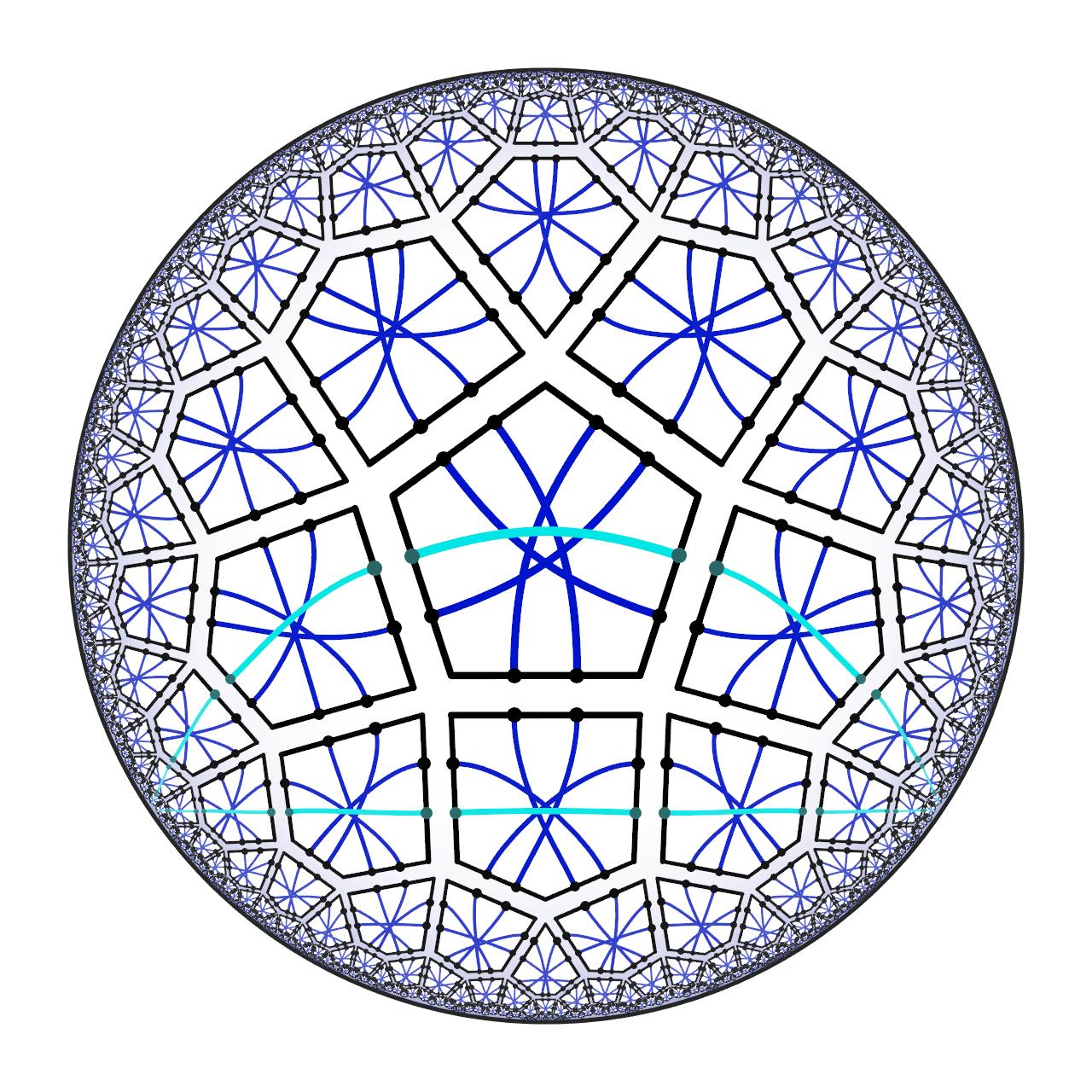}
    \end{gathered} 
    \quad \scalebox{2.0}{$\to$} \quad
    \begin{gathered}
    \includegraphics[height=0.26\textheight]{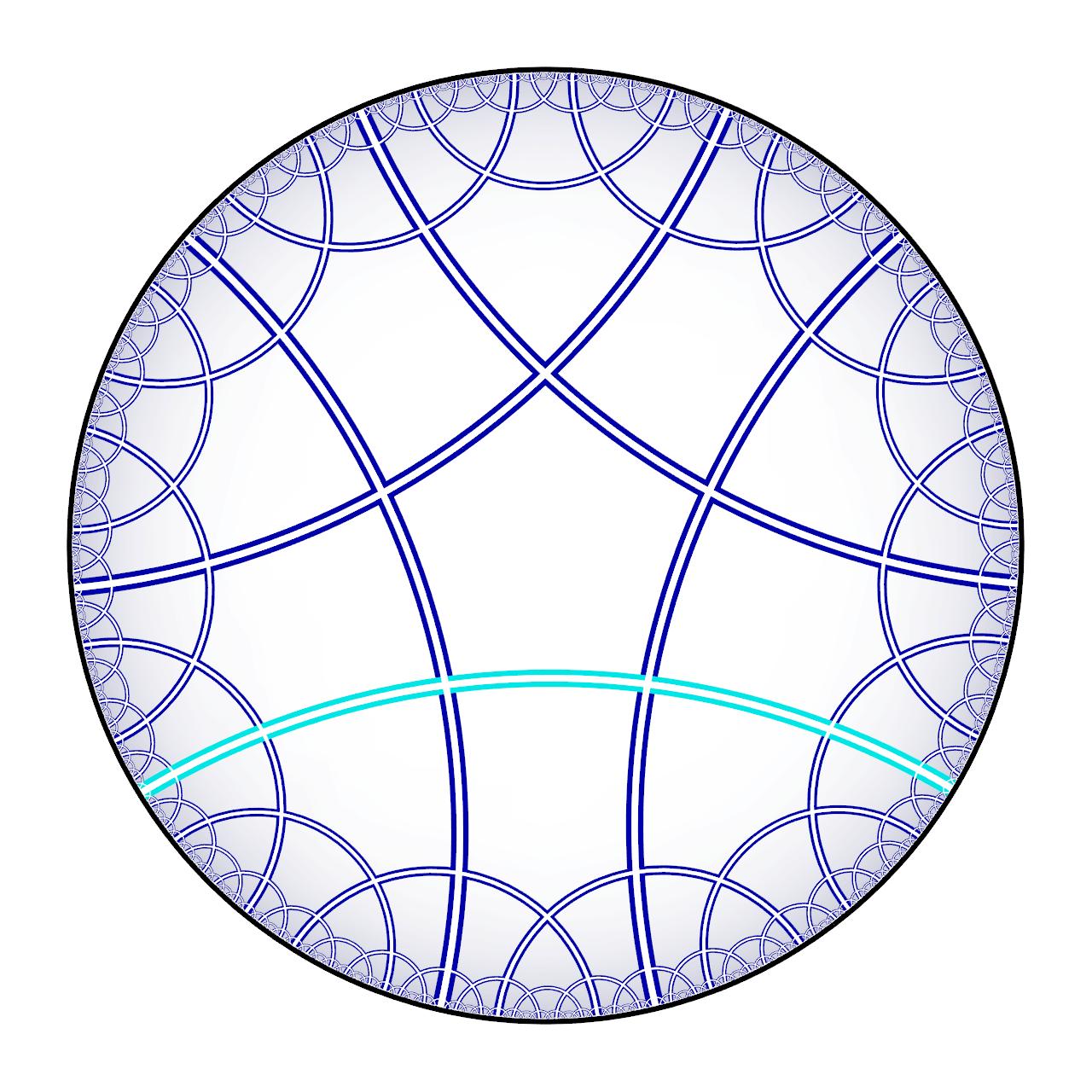}
    \end{gathered}
\end{align*}
\caption{The $\{5,4\}$ HaPPY code in terms of Majorana dimers for a local $\bar{0}$ input on all tiles, shown for the uncontracted states on each pentagon (left) and the full contraction (right). The full contraction contains only paired dimers, an example pair and its constituent dimer parts in the contracted system are highlighted.
}
\label{FIG_HAPPY_CONTR}
\end{figure}

Fig.\ \ref{FIG_WEDGES2} also suggests than these non-zero correlations are associated with endpoints of discrete geodesics of the tiling, an intuition that can be clarified by formulating the pentagon code not in the language of spins, but that of \emph{Majorana fermions}. We already mentioned in Summary 4 that the code states $\bar{0}$ and $\bar{1}$ of the 5-qubit code simplify in a fermionic setting; explicitly, an operator transformation of the form
\begin{align}
\m_{2k-1} &= {\sigma_z}^{\otimes (k-1)} \otimes \sigma_x \otimes \id^{\otimes (N-k-1)} \ , \\
\m_{2k} &= {\sigma_z}^{\otimes (k-1)} \otimes \sigma_y \otimes \id^{\otimes (N-k-1)} \ ,
\end{align}
that maps Pauli operators on $N=5$ spins to $2N$ Majorana operators $\m_k$ that obey anticommutation relations $\{\m_j,\m_k\} = 2\delta_{j,k}$. The four stabilizers $S_k$ of the $[[5,1,3]]$ code (and its fifth permutation $S_5$) then take the form
\begin{align}
    S_1 = \sigma_x \sigma_z \sigma_z \sigma_x \id &= -\i \m_2 \m_7 \ , \\
    S_2 = \id \sigma_x \sigma_z \sigma_z \sigma_x &= -\i \m_4 \m_9 \ , \\
    S_3 = \sigma_x \id \sigma_x \sigma_z \sigma_z &= -\i \Par \m_1 \m_6 \ , \\
    S_4 = \sigma_z \sigma_x \id \sigma_x \sigma_z &= -\i \Par \m_3 \m_8 \ , \\
    S_5 = \sigma_z \sigma_z \sigma_x \id \sigma_x &= -\i \Par \m_5 \m_{10} \ .
\end{align}
Here $\Par$ is the total parity operator expressed as $\sigma_z^{\otimes 5}$ in spin and $-\i \m_1 \m_2 \dots \m_{10}$ in Majorana operators. We immediately notice that up to parity, these stabilizers are products of only two Majorana operators: This implies that the ground state space of the stabilizer Hamiltonian $H= - \sum_k S_k$ is spanned by two parity eigenstates that are \emph{Gaussian}, i.e., each correspond to a non-interacting fermionic model.\footnote{Note that these two fermion Hamiltonians now contain five (rather than four) independent terms, as each stabilizes a single basis state.} This immensely simplifies working with these states: Gaussian states are entirely characterized by their two-point correlations, as the lack of interaction terms in the Hamiltonian of which they are ground states implies that all higher-order correlations are computable from Wick's theorem. For example, for any fermionic Gaussian state vector $\ket\psi$ the four-point Majorana correlation function is given by
\begin{align}
    \langle \m_i \m_j \m_k \m_l \rangle = \langle \m_i \m_j \rangle \langle \m_k \m_l \rangle - \langle \m_i \m_k \rangle \langle \m_j \m_l \rangle + \langle \m_i \m_l \rangle \langle \m_j \m_k \rangle \ ,
\end{align}
where we have denoted $\langle O \rangle:= \sandwich{\psi}{O}{\psi}$.
Note that minus sign appearing at any odd permutation of the operators due to their anticommutation relations.
Gaussianity allows us not only to describe quantum states using quadratic instead of exponential memory but also to contract tensor networks of Gaussian tensors (i.e., tensors describing Gaussian states) efficiently \cite{Valiant2002}.
For the HaPPY code, this implies that if we fix the bulk legs on each pentagon to a basis state input --- either $\bar{0}$ or $\bar{1}$ locally --- then the resulting tensor network is efficiently computable. While this can be performed with methods suitable for general Gaussian tensors \cite{Jahn:2017tls}, the HaPPY code tensors have additional structure than simplifies their contraction to simple graphical rules.
Let us begin with a single tensor describing a logical basis state $\bar{0}$ or $\bar{1}$ on one pentagon, i.e., on ten Majorana modes. We can visualize both state vectors with the diagrams
\begin{align}
\ket{\bar{0}} \;&=\;
    \begin{gathered}
    \includegraphics[height=0.14\textheight]{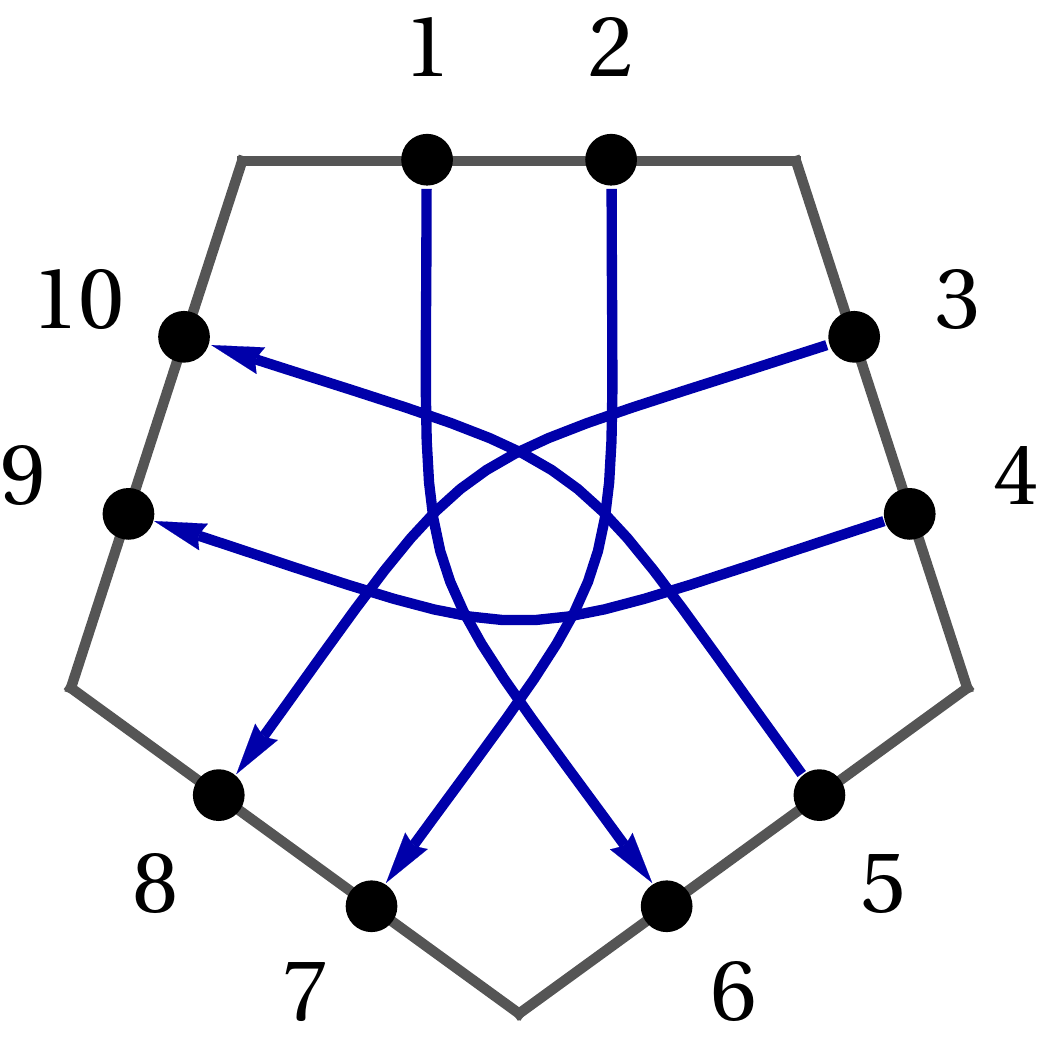}
    \end{gathered} \ ,
&
\ket{\bar{1}} \;&=\;
    \begin{gathered}
    \includegraphics[height=0.14\textheight]{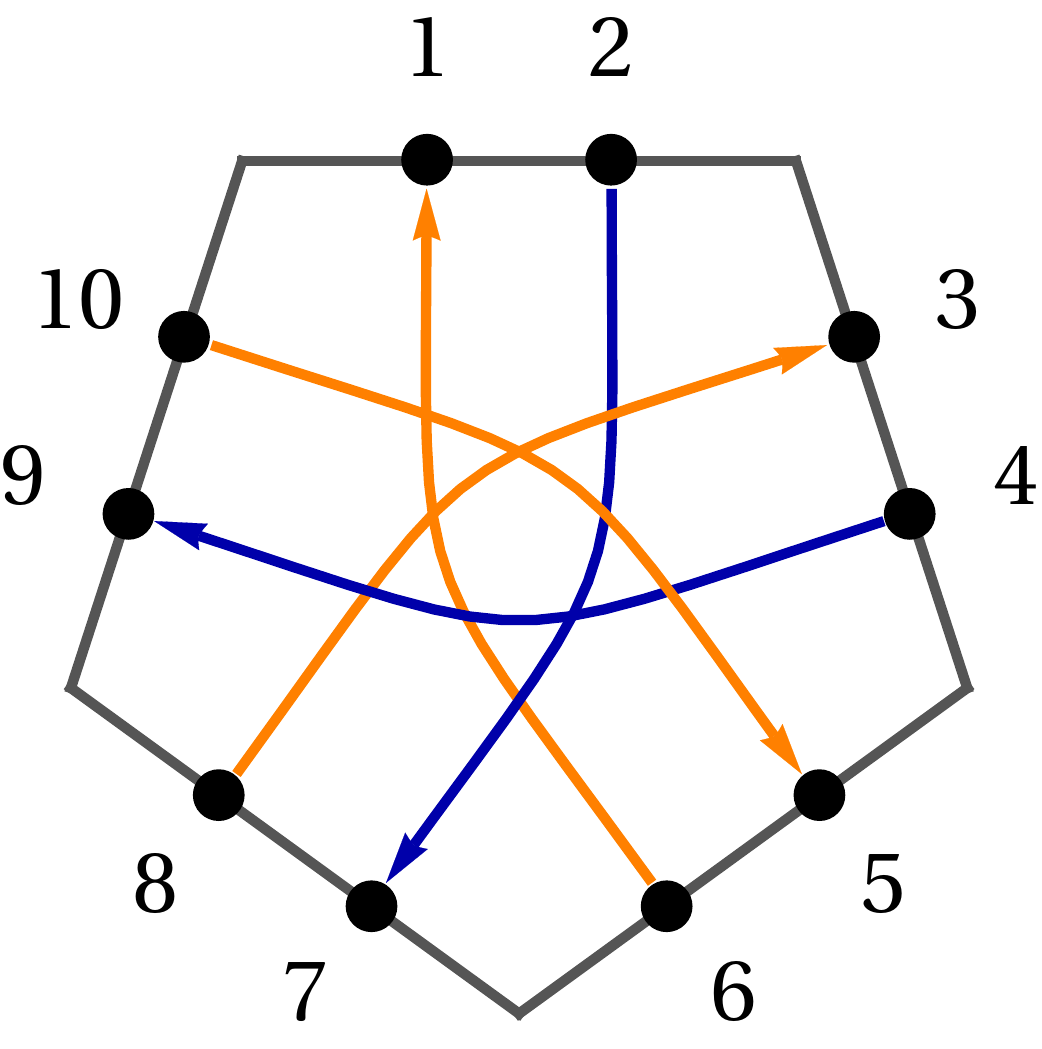}
    \end{gathered} \ .
\end{align}
Here each arrow from Majorana mode $j$ to $k$, called a Majorana dimer \cite{PhysRevB.94.115127,Jahn:2019nmz, GluzaBoundary}, denotes a stabilizer term $\i \m_j \m_k$. The states are uniquely defined by these diagrams up to a complex phase, as the Hamiltonians admits no ground state degeneracy. Alternatively, we can associate with each dimer an operator $\m_j + \i\m_k$ that annihilates the state; intuitively, we can thus picture each dimer as a delocalized fermion shared between sites $j$ and $k$. Note that the orientation of a dimer matters, as the Majorana operators do not commute.
In the visualization, dimer arrows from a mode $j$ to $k$ are shaded in blue if $j<k$ and in orange if $j>k$. As $\m_j + \i\m_k \propto \m_k - \i\m_j$, we can think of each ``inverted dimer'' as a fermionic hole. Each dimer can thus be associated with a \emph{dimer parity} $p_{j,k}$ for a dimer from $j$ to $k$, given by
\begin{equation}
    p_{j,k} = 
    \begin{cases}
    +1 & \text{if } j<k \\
    -1 & \text{if } j>k
    \end{cases} .
\end{equation}
The name ``parity'' comes from the observation that the \emph{total} parity $P_\text{tot}$ (the eigenvalue of $\Par$) of a Majorana dimer state is simply \cite{Jahn:2019nmz}
\begin{equation}
    P_\text{tot} = (-1)^{N_c} \prod_{(j,k)} p_{j,k} \ ,
\end{equation}
where the product runs over all dimers $(j,k)$ in the state and $n_c$ is the number of crossings of dimers in the diagram (which can only be changed by an even number when deforming it).
The Majorana dimer picture becomes particularly useful when performing contractions: For example, contracting two tensors corresponding to two adjacent pentagons encoding a $\bar{0}$ and $\bar{1}$ state, respectively, leads to the new state
\begin{align}
    \begin{gathered}
    \includegraphics[height=0.14\textheight]{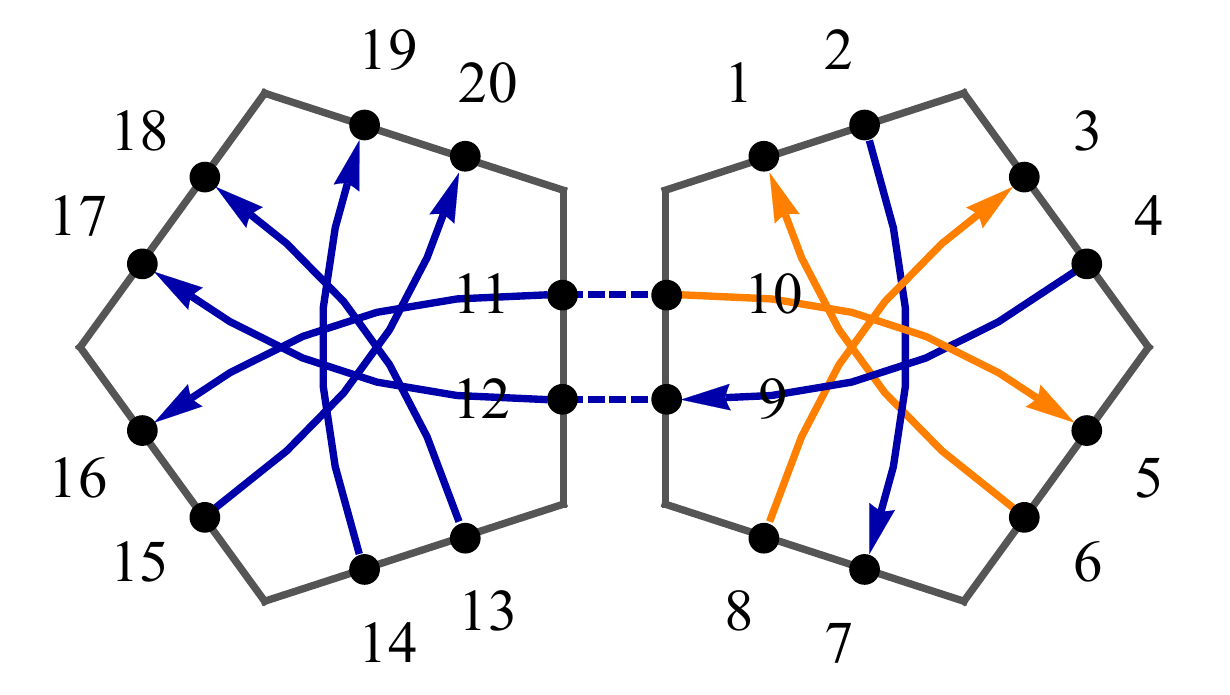}
    \end{gathered} 
    \; = \;
    \begin{gathered}
    \includegraphics[height=0.14\textheight]{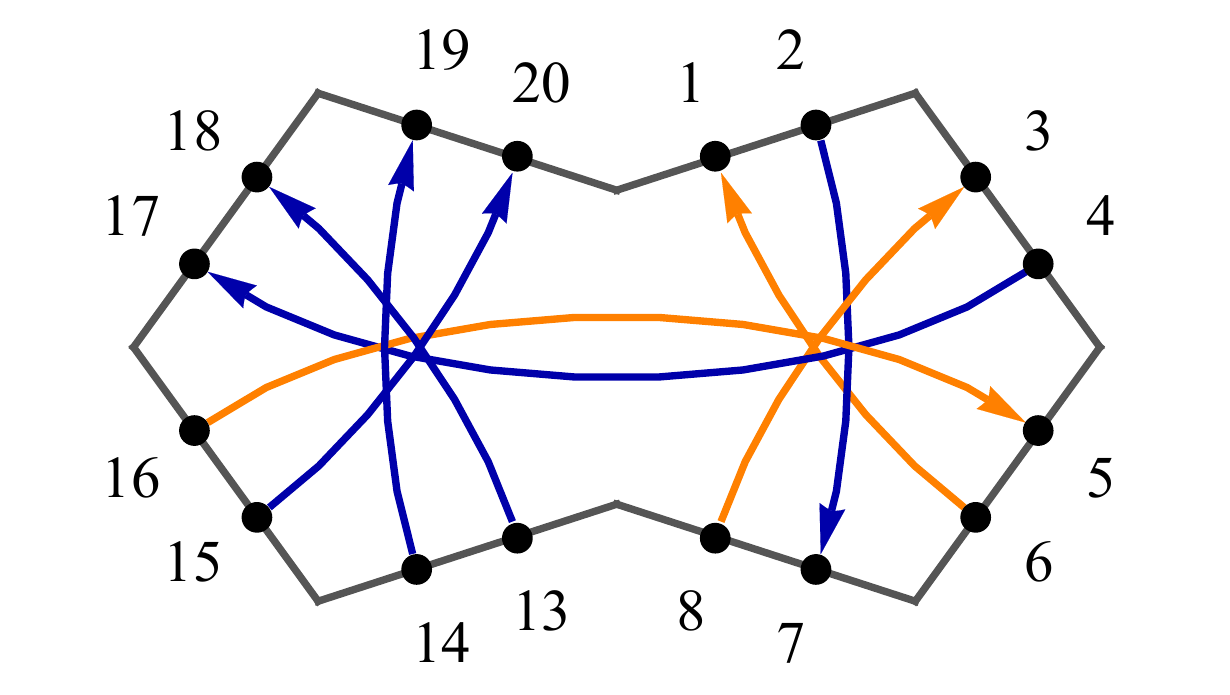}
    \end{gathered} \ .
\end{align}
Here the two dashed lines symbolizing contraction between the two pentagon edges (i.e., two indices) are intentionally suggestive of the result of the contraction of Majorana dimer state: Two pairs of dimers on either side of the contraction are paired up into two new dimers whose dimer parity is the product of the original two dimer parities (up to crossing terms appearing in self-contractions) \cite{Jahn:2019nmz}.
Equipped with this graphical method of contracting code basis states of the $[[5,1,3]]$ code, we can immediately evaluate the boundary states of the entire HaPPY code for fixed local bulk input. For a $\bar{0}$ input on every pentagon, the result can be visualized in a single picture shown in Fig.\ \ref{FIG_HAPPY_CONTR}.
For the full HaPPY code of infinitely many pentagons, the dimers meet up in pairs at the (asymptotic) boundary of the hyperbolic disk. These pairs of dimers trace the discrete geodesics of the $\{5,4\}$ hyperbolic tiling, which shows that the dimers produce the non-zero correlation along these geodesics that we deduced in Fig.\ \ref{FIG_WEDGES2} from the code properties. This statement can now be quantified: Fermionic two-point correlations $G_{j,k} = \frac{\i\,}{2} \langle \m_j \m_k - \m_k \m_j \rangle$ are simply zero if no dimer connects boundary site $j$ and $k$, and $-p_{j,k}$ otherwise. The average falloff of correlations between two large boundary regions is thus given by a histogram of the number of dimers plotted over the boundary distance $d$ between the two sites that they connect. Such a histogram shows that correlations fall off with $\propto 1/d$ \cite{Jahn:2017tls}, just as one would expect in a critical spin model (such as the $c=\frac{1}{2}$ Ising CFT) under a Jordan-Wigner transformation to fermions. Note that due to the contraction rules of Majorana dimers, logical basis state input other than a global $\bar{0}$ input only changes the signs of these correlations but not their structure. It is also possible to show that for \emph{arbitrary} logical input, i.e., superpositions of $\bar{0}$ and $\bar{1}$ logical states, the resulting non-Gaussian boundary state still has two-point correlation functions that follow the dimer pattern \cite{Jahn:2019mbb}.

\begin{figure}
\centering
\begin{align*}
&\begin{gathered}
\textbf{Residual bulk region (wedge picture)} \\
\includegraphics[height=0.28\textheight]{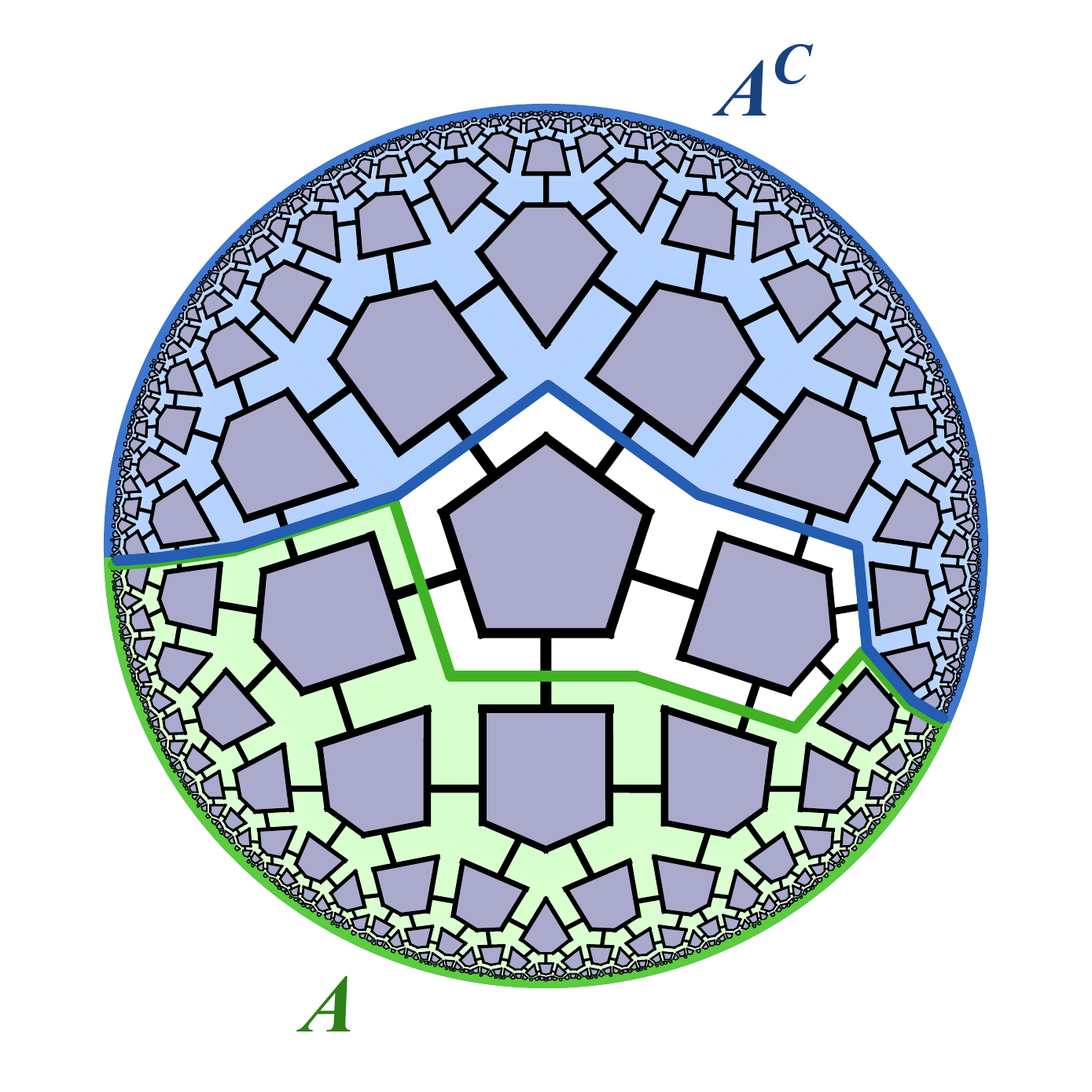}
\end{gathered}
&
&\begin{gathered}
\textbf{Residual bulk region (dimer picture)} \\
\includegraphics[height=0.28\textheight]{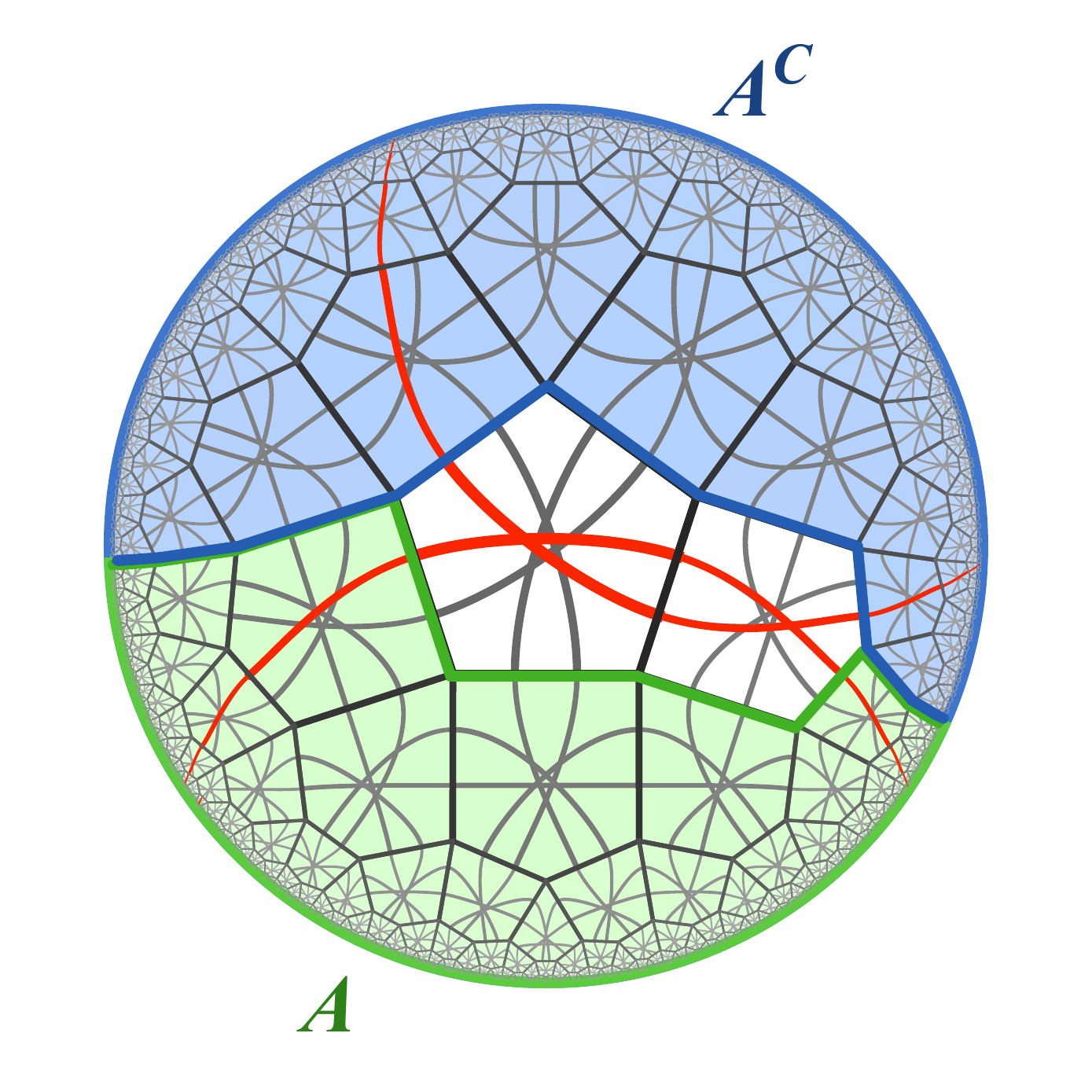}
\end{gathered}
\end{align*}
\caption{\textsc{Left:} A residual bulk region formed when the greedy algorithm for a boundary region $A$ and its complement $A^\text{C}$ do not converge to the same discrete bulk geodesic $\gamma_A$.
\textsc{Right:} Even though moving along geodesics themselves, certain dimers (shaded red) passing through a residual bulk region can pass through $\gamma_A$ twice, reducing the entanglement entropy of region $A$.
}
\label{FIG_WEDGES3}
\end{figure}

The Majorana dimer picture also explains the HaPPY code's entanglement structure without recourse to the greedy algorithm: As each dimer carries an entanglement entropy of $\frac{1}{2} \log 2$, computing the entanglement entropy of a compact\footnote{If define entanglement entropy in the spin picture, as is relevant for discussing the original HaPPY code, non-compact regions do not preserve operator locality after a transformation to fermions.} boundary region $A$ amounts to counting dimers between $A$ and its complement region $A^\text{C}$. From this the geodesic structure of dimers immediately implies a discrete version of the Ryu-Tayanagi formula: A discrete cut $\gamma_A$ through the bulk with the same endpoints as $A$ and which is of minimal length, i.e., discretizes a bulk geodesic, can only cut through such a dimer once. This is because no two shortest paths through a network (nor two continuous geodesics) can meet at more than one point. This leads to the formula
\begin{equation}
    S_A = |\gamma_A|  \log 2 \ ,
\end{equation}
where $|\gamma_A|$ is the number of edges in $\gamma_A$.
The only possible loophole to this argument requires a geometry where geodesics are not unique; these cases correspond precisely to the residual bulk regions unreachable by the greedy algorithm. Such a scenario is shown in Fig.\ \ref{FIG_WEDGES3}: For a pathological boundary region $A$, the greedy algorithm gets stuck at two adjacent pentagons which cannot be passed, as each only presents two edges which are insufficient to reconstruct the state. The tensor network thus cannot be a full isometry between $A$ and its complement region $A^\text{C}$. 
The Majorana dimer picture (Fig.\ \ref{FIG_WEDGES3}, right) resolves this issue: For a basis-state input, the entanglement entropy acquires a correction from the dimers passing through $\gamma_A$ twice, leading to $S_A = (|\gamma_A|-1) \log 2$. For a general input that corresponds to a superposition of dimer states, this correction depends on the exact logical input on the two pentagons in the residual bulk region \cite{Jahn:2019nmz}.

\begin{figure}
\centering
\begin{align*}
&\begin{gathered}
\bf 
\{5,4\} \textbf{ vertex inflation} \\
\includegraphics[height=0.28\textheight]{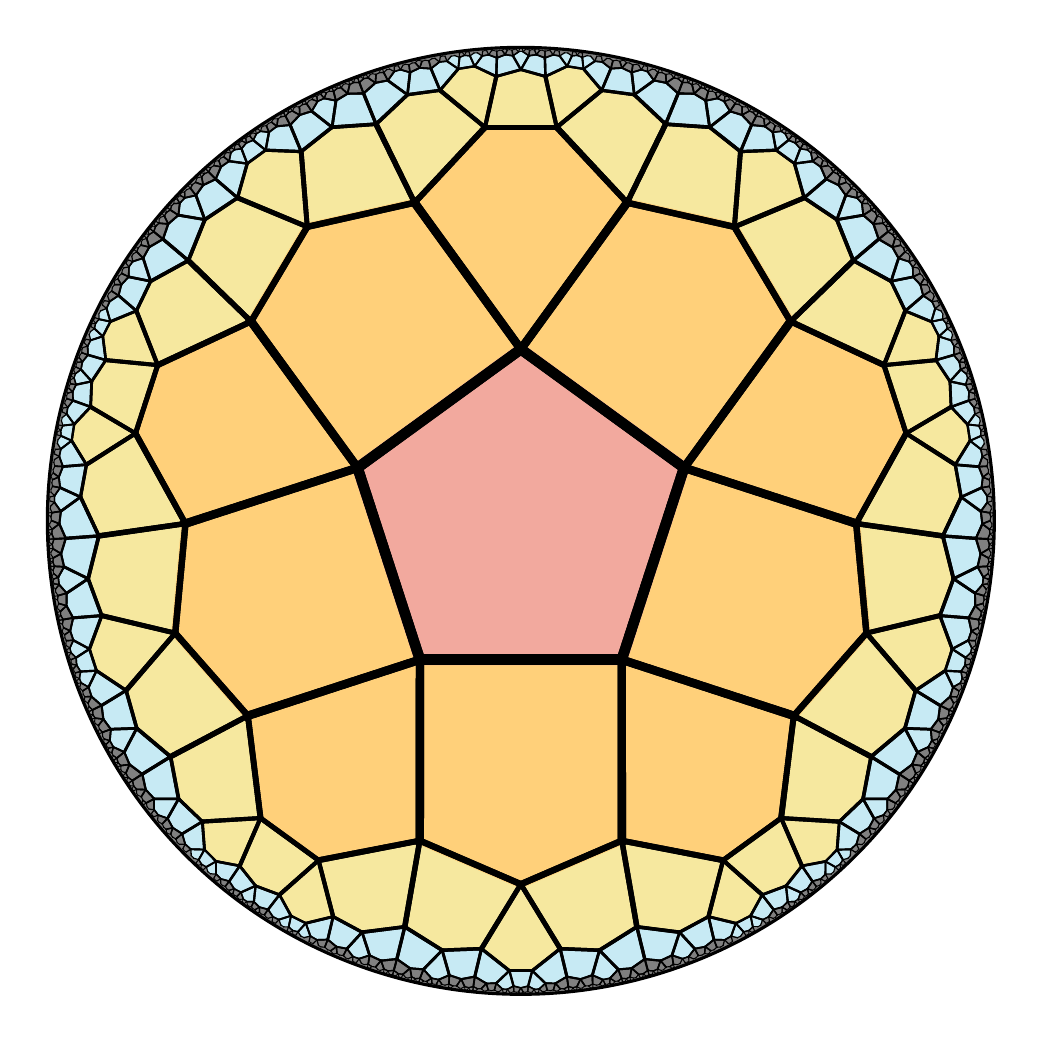}
\end{gathered}
&
&\begin{gathered}
\bf 
\{5,4\} \textbf{ edge inflation} \\
\includegraphics[height=0.28\textheight]{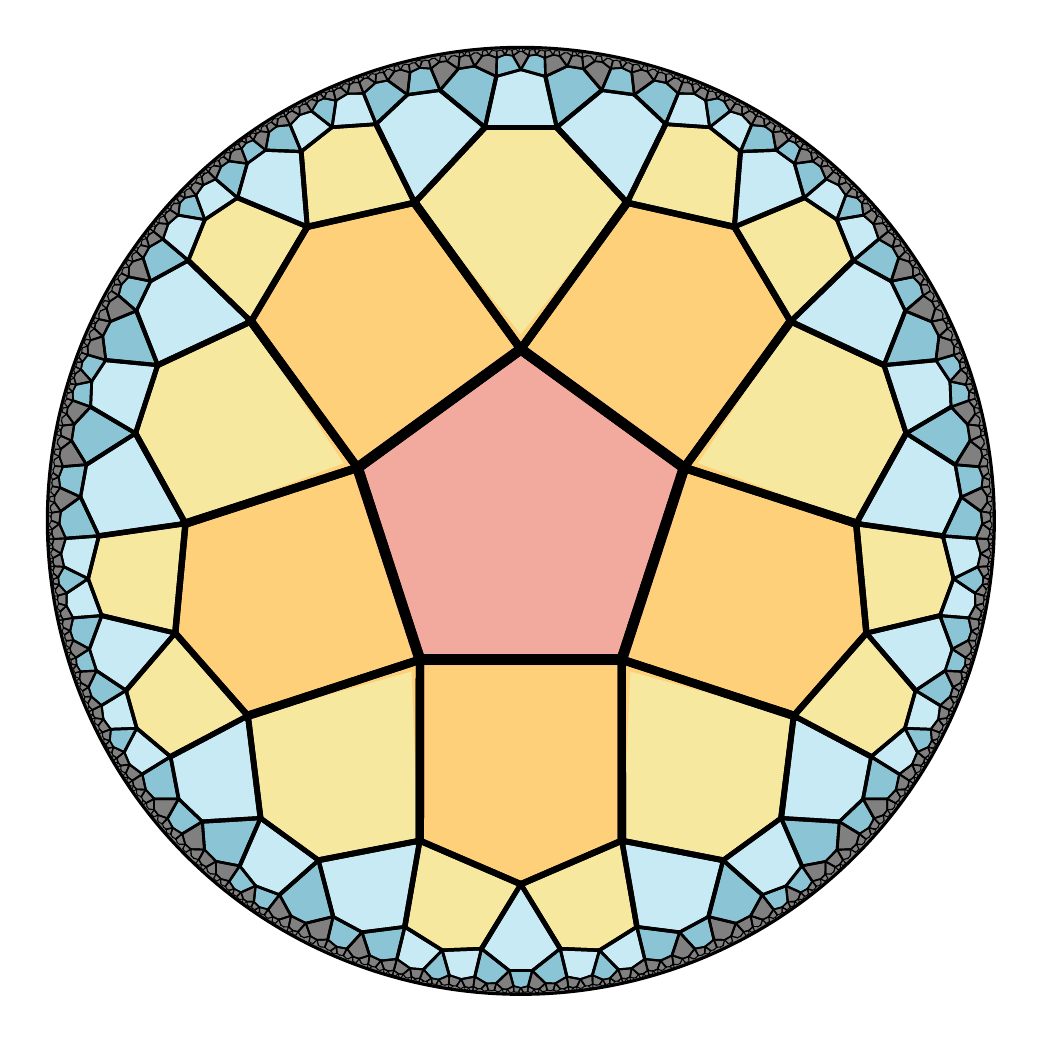}
\end{gathered}
\end{align*}
\caption{Inflation methods for the $\{5,4\}$ tiling, with layers in the inflation color-coded. \textsc{Left:} Vertex inflation, where each inflation step adds polygons vertex-adjacent to the previous layer, forming a closed band.
\textsc{Right:} Edge inflation, where each inflation step adds polygons edge-adjacent to the previous layer. 
}
\label{FIG_INFLATION_METHODS}
\end{figure}

The discrete Ryu-Takayanagi-like behaviour of entanglement entropy $S_A$ implies that, up to the small corrections produced by residual bulk regions, the HaPPY code's boundary states follow the logarithmic scaling \eqref{EQ_SA_CRIT} expected of ground states of conformal field theories on a circle. But what is the effective central charge $c$ of these states, i.e., the exact coefficient of entanglement scaling? 
Perhaps surprisingly, its value depends on the geometrical construction of the HaPPY code's tiling. As it consists of infinitely many tiles, properly defining a HaPPY boundary state requires a cutoff procedure or equivalently, a renormalization group (RG) step that iteratively extends the tiling. Such a step is not unique: Depending on the choice of \emph{inflation rule} chosen to grow the tiling (two of which are shown in Fig.\ \ref{FIG_INFLATION_METHODS}), the asymptotic boundary is approached differently. 
This leads to different effective RG steps of the boundary states. Again, under the assumption of basis state input the Majorana dimer picture helps quantify these statements: Expressed in terms of dimers, each inflation step acts as a fermionic analogue of an RG step of the \emph{strong-disorder renormalization group} (SDRG) \cite{StrongDisorder}, leading to an aperiodic correlation and entanglement structure that can be exactly solved \cite{Jahn:2019mbb}. 
For the two inflation methods shown in Fig.\ \ref{FIG_INFLATION_METHODS}, this leads to two different scaling factors of entanglement entropy and thus to different analytical values of the corresponding central charge,
\begin{align}
c_{\{5,4\}_\text{v}} &= \frac{9 \log 2}{\log\left(2 + \sqrt{3}\right)} \approx 4.74 \ ,
&
c_{\{5,4\}_\text{e}} &= \frac{6 \ln 2}{\ln\frac{3 + \sqrt{5}}{2}} \approx 4.32 \ ,
\end{align}
where the subscripts denote either vertex or edge inflation. Extensions to tilings other than the $\{5,4\}$ one are also possible \cite{Jahn:2019mbb}.

\subsection{Generalizations of HaPPY codes}
The conditions for a holographic quantum error-correcting code following the HaPPY proposal is quite constrained, as the perfect tensor property allows only for few solutions. For example, the nonexistence of absolutely maximally entangled states on four spins \cite{Goyeneche:2015fda,2017PhRvL.118t0502H} forbids a HaPPY code embedded into a $\{4,k\}$ tiling for a bond dimension $\chi=2$ (i.e., for a spin model), though constructions at larger bond dimension are possible.
Fortunately, it has been shown that many properties of HaPPY codes are preserved in a tensor network model with \emph{block-perfect} tensors \cite{PhysRevA.98.052301}.
While a perfect tensor corresponds to an isometry for any bipartition of indices, a block-perfect one is only isometric for bipartitions into adjacent sets of indices, i.e., if neither $A$ nor $A^\text{C}$ are disjoint regions.
This weaker constraint allows one to build holographic codes based on the more widely studied   \emph{Calderbank-Shor-Steane} (CSS) codes \cite{PhysRevA.98.052301},
a special and particularly important class of 
quantum error correcting codes that are constructed from pairs
of classical error correcting codes that share a number of desirable properties.
The specific construction made use of is based on the
\emph{7-qubit Steane code}. For any such holographic quantum error correcting code, the question of a suitable
\emph{decoding} appears --- as for any quantum error correcting code --- so that of 
a classical algorithm that based on the syndrome arising from local measurements would assign the likely error that has actually occurred. This line of thought of exploring decoders specifically for holographic quantum error correcting codes has been explored in Ref.\
\cite{PhysRevA.102.062417},
by suggesting an integer optimization decoder.

Block-perfect generalizations of the $[[5,1,3]]$-based holographic pentagon code can also be constructed in the Majorana dimer setting. To ensure that connected subregions are maximally entangled, basis states of such an $n$-gon code have to consist of dimers connecting modes $i$ and $(i+n) \mod 2n$ on opposite ends. Ensuring that not only the basis states but also their superposition have the same entanglement structure leads to conditions on the dimer parities that can be fulfilled only if $n=4i+1, i \in \mathbb{Z}$ \cite{Jahn:2019nmz}. The first non-HaPPY example of this construction is a ``nonagon code'' with basis state vectors
\begin{align}
\ket{\bar{0}} \;&=\;
    \begin{gathered}
    \includegraphics[height=0.15\textheight]{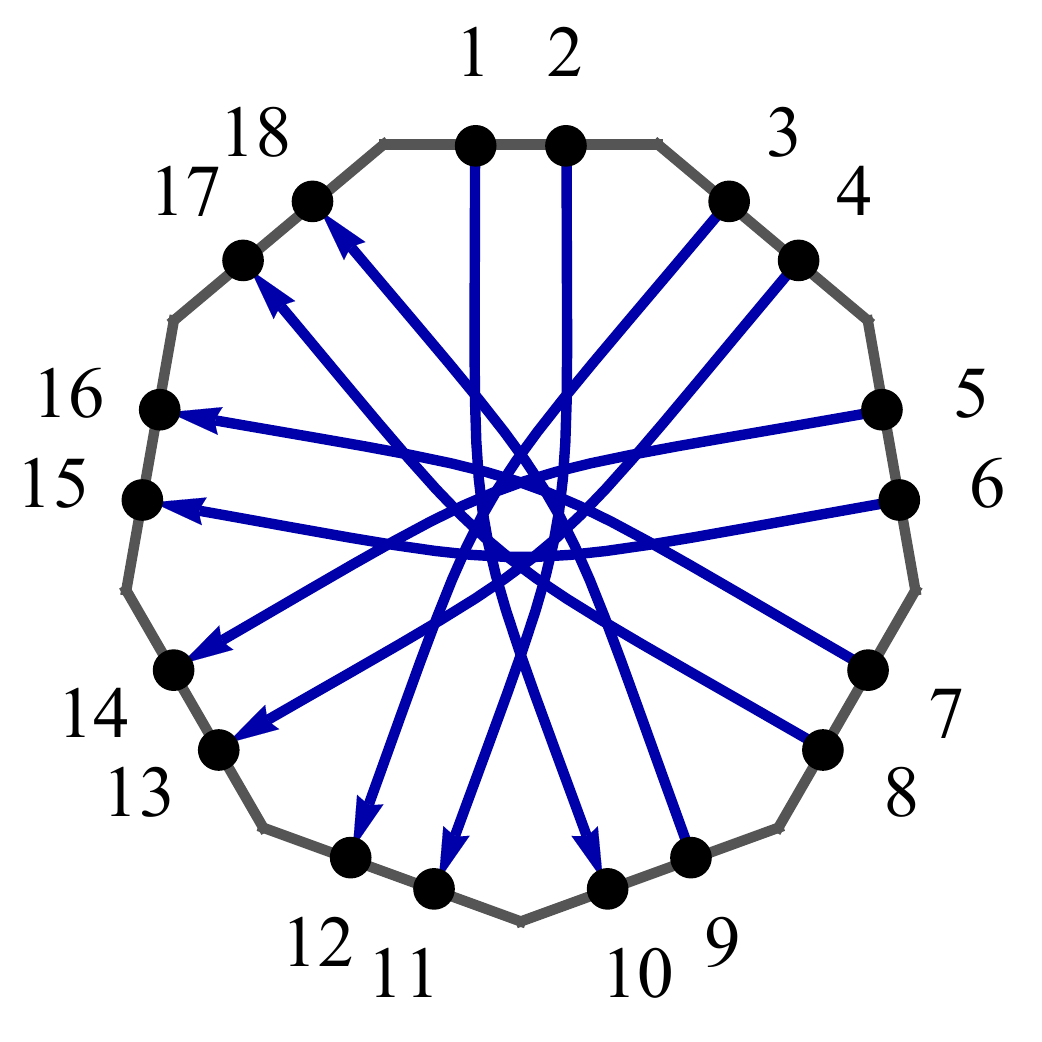}
    \end{gathered} \ ,
&
\ket{\bar{1}} \;&=\;
    \begin{gathered}
    \includegraphics[height=0.15\textheight]{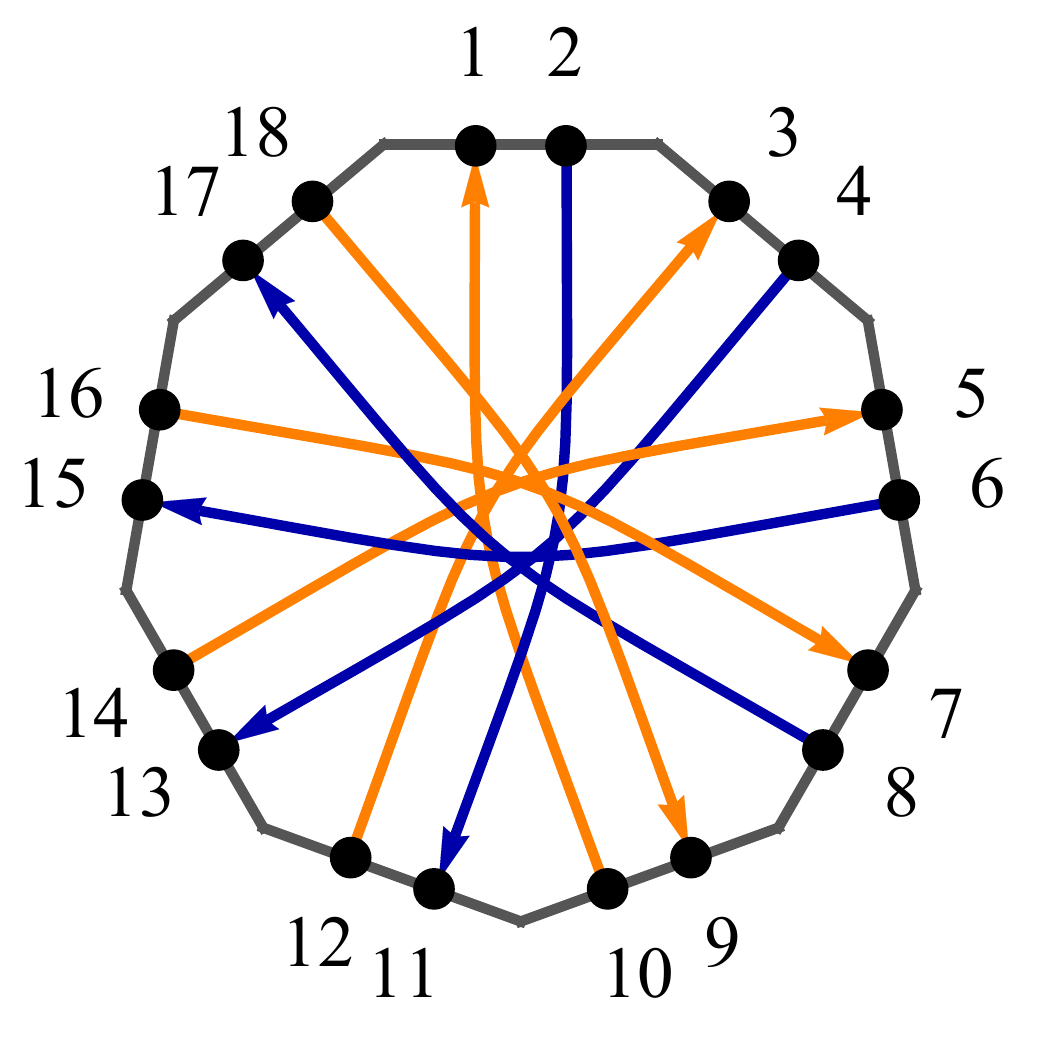}
    \end{gathered} \ .
\end{align}
Unfortunately, this code does not become more resilient to errors at larger $n$; for example, the operator $X_1 Z_{\frac{n+1}{2}} X_n$ (where $X_i$ and $Z_i$ are Pauli $\sigma_x$ and $\sigma_z$ operators acting on the $i$th site) has different eigenvalues on the basis states and therefore corresponds to a phase flip error of weight $3$, regardless of $n$.
While such codes are thus not useful in practical applications, they can be used as a tool to study holographic codes: For example, it can be shown that the effect of residual bulk regions becomes negligible in generalized Majorana dimer codes as $n$ becomes large \cite{Jahn:2019mbb}.

The generalization to more general types of tensors as such, however, is by no means the only line of thought to generalize
the HaPPY prescription. One can also introduce degrees of freedom on the edges 
of an associated tensor network, connected to 
further copies of the HaPPY code by an appropriate isometry. This mindset leads to
a generalization of HaPPY holographic quantum error-correcting code to provide 
toy models for bulk gauge fields or linearized gravitons \cite{Donnelly}.

Rather than studying static properties of holographic codes, one may also wonder if they admit \emph{dynamics}. Such a line of research has first been carried out \cite{Osborne:2017woa} for the particularly symmetric limit of holograpic codes on $\{n,k\}$ tilings were $k \to \infty$, the so-called \emph{ideal regular tilings}. The boundary symmetries of such tilings are encapsulated by Thompson's group $T$ which bears some similarities to the conformal group. The corresponding bulk symmetries are then given by the Ptolemy group $Pt$ describing a form of discretized diffeomorphisms, generated by \emph{Pachner moves} than re-arrange edges within the bulk, breaking regularity.
The resulting bulk/boundary dynamics differs from continuum notion of time evolution in some ways, specifically in that there is no well-defined Hamiltonian. The Pachner moves also act highly non-locally, as each edge in an ideal tiling stretches all the way to the asymptotic boundary. This makes it difficult to define notions of locality in the IR, which would be required for defining a discrete analogue of a particle in the bulk. Similarly, the tree tensor network structure of such models makes it difficult to produce boundary states with the entanglement structure of physical CFTs, though such geometries do appear in \emph{p-adic} models of AdS/CFT \cite{Heydeman:2016ldy}.
However, further studies along these lines in more general geometries may lead to a discrete bulk/boundary dictionary of holographic codes.

Another approach to boundary dynamics may be provided by the SDRG picture that arises from describing the HaPPY code in terms of Majorana dimers: Most traditional SDRG models, whose RG steps is formulated in terms of spin singlets rather than dimers, result from a strong disorder limit of certain Hamiltonians with only nearest-neighbour coupling. For example, consider the \emph{Fibonacci XXZ model} with Hamiltonian
\begin{align}
H &= \sum_i J_i \left(X_i X_{i+1} + Y_i Y_{i+1} +\Delta Z_i Z_{i+1} \right) \ ,
\end{align}
where the $X_i,Y_i,Z_i$ are Pauli spin operators acting on the $i$th site and the $J_i$ are coupling terms that vary along the sites according to an aperiodic Fibonacci sequence. This model is non-Gaussian for $\Delta \neq 0$ and generally difficult to solve for generic $J_i$. However, in the case of strong disorder, i.e., when the value of the couplings changes significantly with the aperiodic sequence, its ground state is approximately given by a configuration of singlets than can be recursively computed with the SDRG approach \cite{JuhaszZimboras2007,Igloi_2007}. 
Given that the boundary states of the HaPPY code can be produced in a similar SDRG process relying on fermions rather than spins \cite{Jahn:2019nmz}, it is thus plausible to speculate that the resulting boundary states are also the approximate ground state of a Hamiltonian with only local couplings, according to which one could define local time dynamics. Note that the HaPPY boundary states are of course ground states of a non-local Hamiltonian: For basis-state input, this is simply the free Hamiltonian coupling the endpoints of each Majorana dimer configuration.

The question of dynamics of holographic codes can also be approached from the stabilizer picture. Inflation of the hyperbolic tiling embedding the HaPPY code can be associated with a mapping between isometries, each additional inflation layer adding both bulk and boundary degrees of freedom, that allows for an explicit construction of the resulting stabilizers at each layer \cite{Gesteau:2020hoz}. The form of the resulting stabilizers implies that most long-range correlations (in the spin picture) vanish, which is equivalent to the sparseness of correlations we already saw in the Majorana dimer picture. However, the stabilizer picture can be used make statements about finite-temperature dependence of the resulting boundary model.
This inflation process can also be equivalently described in terms of $C^\star$ algebras \cite{Gesteau:2020rtg}, from which boundary entanglement is more readily computable.

\begin{figure}
\centering
\begin{align*}
&\begin{gathered}
\textbf{Alternating square/hexagon tiling} \\
\includegraphics[height=0.28\textheight]{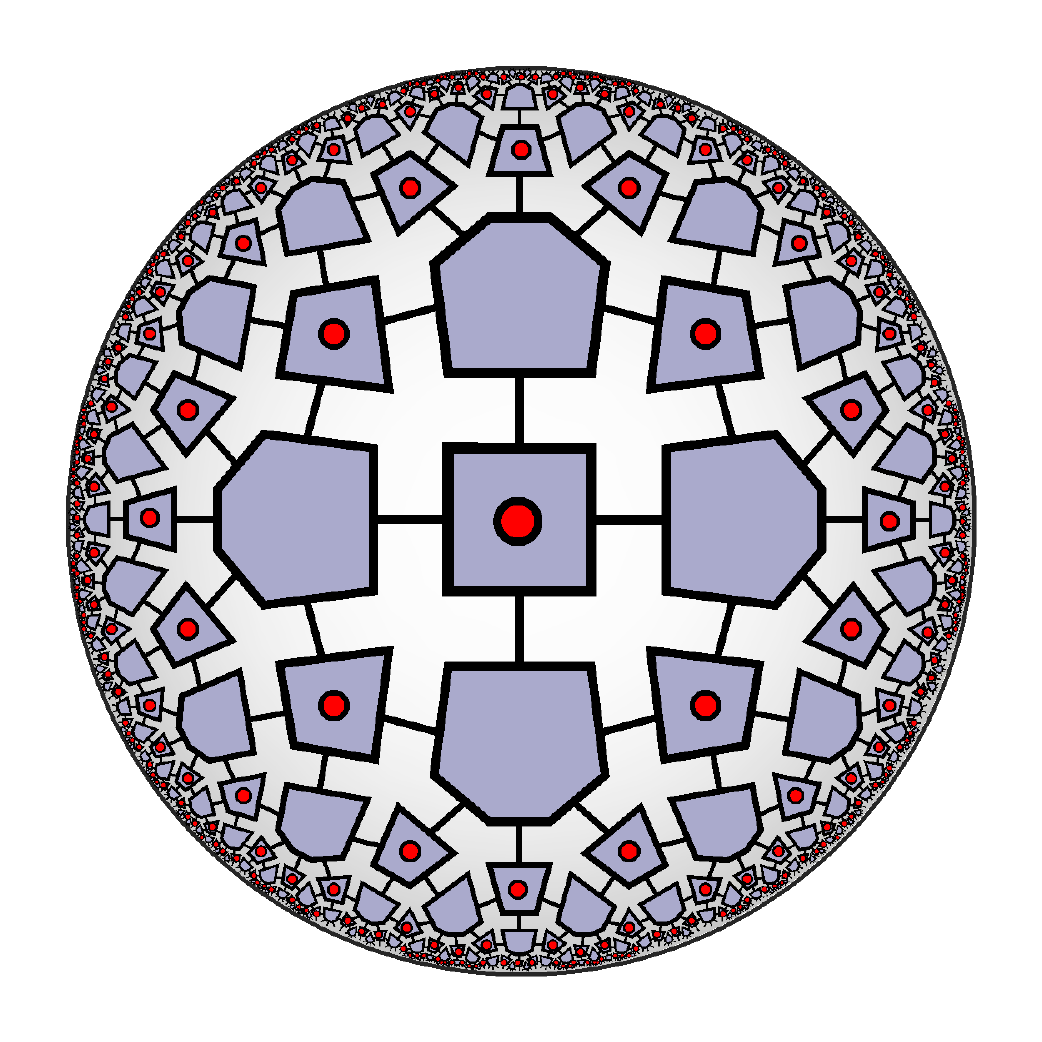}
\end{gathered}
&
&\begin{gathered}
\bf 
\{5,4\} \textbf{ HaPPY code with black hole} \\
\includegraphics[height=0.28\textheight]{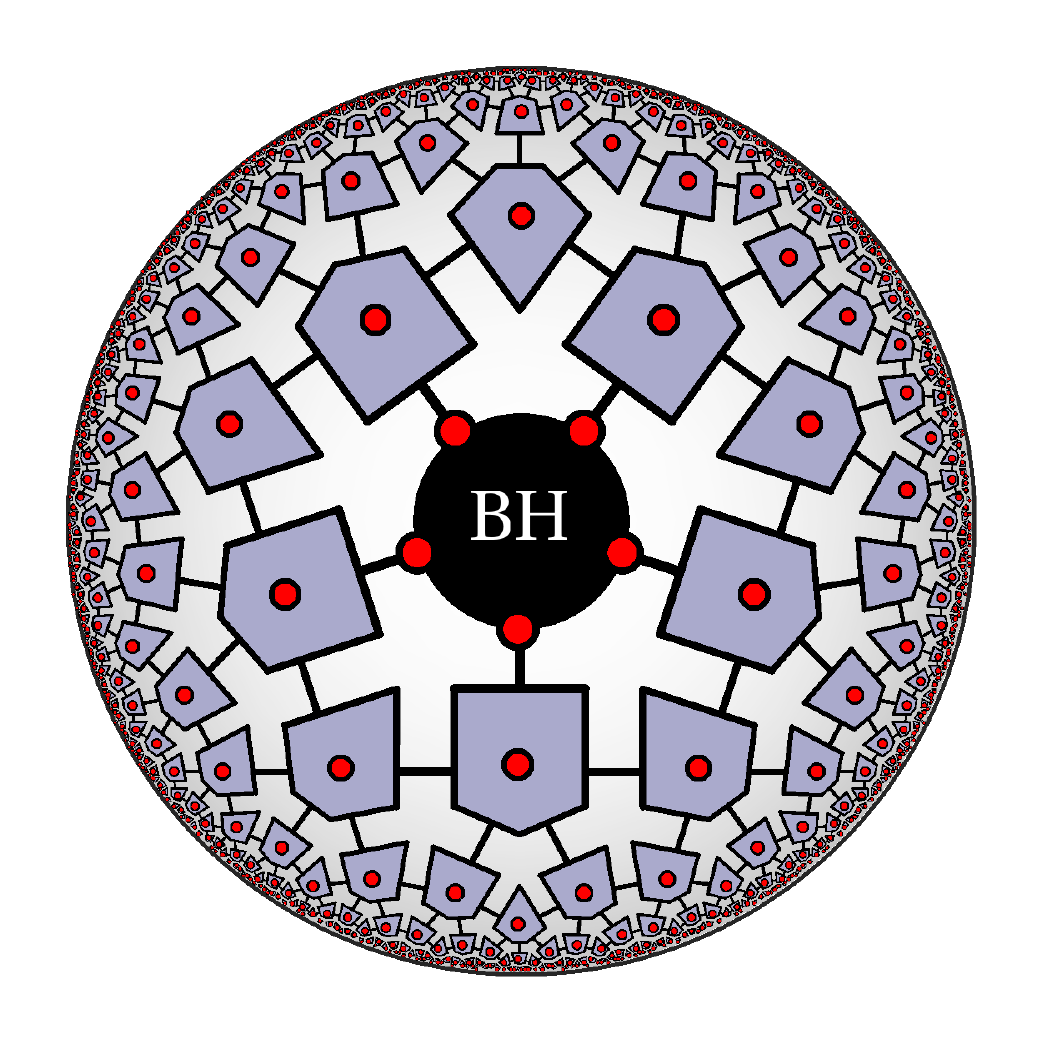}
\end{gathered}
\end{align*}
\caption{Tensor network generalizations of HaPPY codes, with logical states represented as red dots. \textsc{Left:} An alternating hyperbolic tiling of squares and hexagons used in Ref.\ \cite{Cao:2020ksw} with logical degrees of freedom encoded in a Bacon-Shor code on the squares.
\textsc{Right:} The black hole geometry from Refs.\ \cite{Pastawski2015,Kohler:2018kqk}, where a central tensor is removed, leading to additional logical ``horizon'' degrees of freedom on the remaining open edges. 
}
\label{FIG_HAPPY_EXTENSIONS}
\end{figure}

\subsection{Further approaches to holographic quantum error correction}

While the HaPPY model of a holographic code is already quite versatile and captures a number of holographic properties, it is not the only possible way to construct such a code.
Two plausible directions for constructing more general codes is to consider tensors describing more general encoding isometries and to consider tilings that are more complicated than simple regular $\{n,k\}$ ones.
One recent approach combining both directions is presented in Ref.\ \cite{Cao:2020ksw}, where the tensors are chosen to represent a \emph{Bacon-Shor code} which generalizes quantum error-correcting codes by including gauge degrees of freedom. This code is then embedded into an alternating hyperbolic tiling composed of squares and hexagons where the logical qubits are encoded only on the squares while the hexagons contain perfect tensors without bulk degrees of freedom. The resulting setup is visualized in Fig.\ \ref{FIG_HAPPY_EXTENSIONS} (left).
This construction inherits some of the issues of the original HaPPY model (such as residual bulk regions) but allows for a deformation to a \emph{skewed code} with only approximate error-correcting properties whose effect on entanglement wedges resembles a gravitational back-reaction of a massive bulk deformation.

Another approach to constructing holographic codes is to consider them as a mapping between local Hamiltonians, an idea developed in Ref.\ \cite{Kohler:2018kqk}: Such a mapping is indeed possible for a discretized hyperbolic bulk of three or more spatial dimensions (compared to two in the HaPPY model) in an approximate manner using \emph{perturbation gadgets}, a tool from Hamiltonian simulation theory. These higher-dimensional model require a generalization of the regular tilings to a tessellation with polytopes that follow the symmetries of a Coxeter group. Intriguingly, these models preserve locality both on the boundary and in the bulk, which makes it possible to consider a generic form of time evolution that reproduces certain aspects of black hole formation.
These constructions also reaffirm an idea present in the original HaPPY paper \cite{Pastawski2015} and developed in more detail in Ref.\ \cite{Yang_2016}: That low-energy bulk excitations should be describable by changes within the logical code space while high-energy ones (e.g.\ black holes, shown in Fig.\ \ref{FIG_HAPPY_EXTENSIONS}, right) explicitly break the code space and modify the bulk geometry into which the holographic code is embedded.

Previous tensor network models of holography can also be included into the framework of quantum error-correcting codes: Given the strong resemblance of MERA with discrete instances of hyperbolic geometries as featuring in the AdS/CFT correspondence \cite{PhysRevD.86.065007}, it comes as no surprise that the connection between quantum error correction and MERA has been explored. Indeed, MERA serves as an example to solidify the idea of creating quantum error-correcting codes arising from an encoding map from the bulk theory to the boundary theory. Ref.\ \cite{Kim2016} explores the connection between quantum error correction and MERA, in fact guided by a bold motivation:
Here, it is not only attempted to realize some specific holographic quantum error correction code.
Instead, the point is made that if there is a unitary equivalence of conformal field theory and a quantum theory of gravity in AdS space-time, one should be able to explain holographic codes as emerging directly from properties of the underlying CFT.

That said, in order to establish such a connection in the framework of Ref.\ \cite{Kim2016}, notions of quantum error correction have to be slightly weakened to an \emph{approximate quantum error correction}.
An erasure of a given region is correctable --- appropriately modified in an approximate version thereof --- if and only if that region does not contains any logical information, and hence, if and only if that region is uncorrelated with the purifying space for all the code words. Refining this insight,
a notion of \emph{local correctability} \cite{LocalCorrection} extends this to a connection between local correctability and the degree to which different subsystems are separated in correlations. In Ref.\ \cite{Kim2016}, these notions are applied to and put into the context of MERA codes. Specifically, assume that $A$ is a simply connected region ``shielded'' by a region $B$ such that $AB \equiv A\cup B$ contains all sites within a distance $x$ from $A$, and that the MERA contains sufficiently layers $s$ so that $|AB| < 2^s$. 
If we further denote with $C$ the complement of $AB$, then it is shown that there exists a recovery map $\mathcal{R}$ reconstructing the region $AB$ from $B$ under the bound
\begin{equation}
   \| \mathcal{R}(\rho_{BCR})
   -\rho_{ABCR}\|_1
   \leq 
   c
   \left(
   \frac{|A|}{c}
   \right)^{\nu/2},
\end{equation}
for all purified code states $\rho_{ABCR}$ (involving a further purifying system $R$), where $c>0$ is a constant and $\nu>0$ is the scaling dimension of the isometries of the MERA.
$\|.\|_1$ here is the trace norm meaningfully quantifying the statistical distinguishability of quantum states.
This notion of recoverability provides a broader basis to the understanding that low energies of the critical systems should have a certain error correction properties \cite{Pastawski2016} without demanding tensors to obey such strict bounds as perfect tensors.
These results apply broadly to critical ground states within the extent that they can be approximated by the MERA.

Finally, rather than beginning the search for suitable holographic tensor network models with a fixed geometry and then exploring suitable choices of tensors, it is also possible to generate a tensor network geometry dynamically from boundary entanglement in a process of \emph{entanglement distillation} iteratively applied to the boundary state \cite{Bao:2018pvs,Bao:2019fpq}. As this approach leads to tensor networks that fulfill holographic entanglement entropy (for tree tensor networks) and entanglement of purification (for more general geometries) of holographic CFT states by virtue of the construction, tensor networks can be regarded not merely as toy models but as suitable representations of certain properties of continuum AdS/CFT. In the tree tensor network case, this approach naturally incorporates the operator reconstruction properties of holographic quantum error-correcting codes, though for more generic geometries that capture (sub-)AdS bulk locality, this connection remains to be worked out \cite{Bao:2018pvs}.
Such endeavors should also serve as a reminder that there is no uniquely ``correct'' tensor network geometry for discretizing AdS/CFT; for example, recent path integral studies of holography suggest that we may represent the same boundary state with different tensor network geometries corresponding to different slices through AdS space-time \cite{Miyaji:2015yva,Miyaji:2016mxg,Caputa:2017urj,Milsted:2018yur,Milsted:2018san,Caputa:2020fbc}.

\section{Outlook}
\label{SEC_OUTLOOK}

In this topical review, we have laid out some developments of a growing field of research at the 
interface of high energy physics and quantum information
theory that aims at fleshing out aspects of holography in a particularly transparent picture. The two main ingredients in this endeavour are on the one hand tensor networks 
that capture the natural underlying entanglement structure of quantum states. On the other hand, these are notions of quantum error corrections, concepts originally having arisen in the context of quantum computing in the presence of noise, but actually being closely intertwined with notions of holography. We have explored here the roots in string theory and high energy physics, made the connection to tensor networks, and moved on to explain the connection to quantum error correction. We have discussed in great detail holographic quantum error correction in tensor networks and toy models of holography in a stabilizer picture as well as their fermionic representation in terms of coupled Majorana modes, but also the various efforts made to generalize these approaches in the search of more comprehensive tensor network models of holographic quantum error correction.

And yet, many exciting questions remain open, and much of what has been said here can be seen as an invitation to pursue these steps. The study of the \emph{dynamics} of holographic models --- only hinted upon above --- is just beginning to unfold \cite{Osborne:2017woa,Gesteau:2020hoz,Kohler:2018kqk}.
Just as importantly, questions of meaningful \emph{continuum limits of tensor networks} relating discrete and continuous models of holography are being extensively pursued \cite{PhysRevLett.104.190405,PhysRevLett.105.260401,Nozaki:2012zj,PhysRevLett.110.100402,Magic}, with the particularly interesting potential of describing regimes of interacting quantum fields \cite{GcMERA,Cotler:2018ehb,Cotler:2018ufx,Fernandez-Melgarejo:2019sjo,Fernandez-Melgarejo:2020fzw}.
Notions of circuit and state complexity \cite{Nielsen:2006mn2} have also taken centre stage in recent discussions of holography \cite{ChapmanMarrochioMyers,JeffersonMyers,Complexity1,BigComplexity,ComplexityGrowth}, not the least due to the bold ``complexity equals volume''  \cite{PhysRevD.90.126007} and ``complexity equals action'' \cite{Brown:2015bva} conjectures due to Leonard Susskind, Douglas Stanford,  Adam R. Brown, Brian Swingle and others mentioned above.
The precise connection of complexity to tensor network models of holography is yet to be fully established but may be related to path integral approaches to holography \cite{Miyaji:2016mxg,Caputa:2017urj,Caputa:2017yrh,Boruch:2020wax}.
Holographic tensor network models can also be used to construct practical stabilizer codes with natural decoders \cite{Farrelly:2020mxf}, showing that insights from these models can become useful in an applied setting.
Applying insights from the theory of quantum error correction to holography reliably produces interesting new results, as can be seen in a recent extension of the Eastin-Knill theorem \cite{Eastin_2009} to holographic codes, showing that local boundary operations generally only implement Clifford gates on the logical qubits in the bulk \cite{Cree:2021rxi}. 
It is the hope that the present article, beyond merely giving an overview over current developments of an exciting field, will serve as a source of inspiration for further endeavours exploring the intricate connections between notions of holography, quantum error correction, and tensor networks.

\subsection*{Acknowledgements}
We would like to warmly thank numerous colleagues for stimulating discussions on the topics addressed in this review. A list of those colleagues includes but is by no means limited to
A.\ Altland,
P.\ Caputa,
J.\ Conrad,
B.\ Czech,
G.\ Evenbly,
J.\ C.\ M.\ de la Fuente,
M.\ Gluza,
L.\ F.\ Hackl,
D.\ Harlow,
M.\ Heller, 
R.\ C.\ Myers,
F.\ Pastawski,
J.\ Prior,
S.\ Singh,
M.\ Steinberg,
T.\ Takayanagi,
G.\ Vidal,
M.\ Walter,
C.\ Wille,
H.\ Wilming,
X.-L.\ Qi, 
B.\ Yoshida, and 
Z.\ Zimboras. 
This work has been supported by the DFG (CRC 183 and EI 519/15-1) and the FQXi.
Parts of it were produced under a Visiting Graduate Fellowship at the Perimeter Institute for Theoretical Physics. Research at Perimeter Institute is supported by the Government of Canada through the Department of Innovation, Science, and Economic Development, and by the Province of Ontario through the Ministry of Research and Innovation.
This review includes excerpts of the doctoral thesis submitted by AJ at the Free University of Berlin.

\bibliographystyle{ieeetr_href}
\small
%\bibliography{References}

\end{document}